%% file: barcode_rss.tex
\documentclass[fleqn,usenatbib]{mnras_custom}

\usepackage{newtxtext,newtxmath}

\usepackage[T1]{fontenc}
\usepackage{ae,aecompl}

\usepackage{amsmath}	
\usepackage{amssymb}	
\usepackage{bm}       
\usepackage{tikz}     
\usepackage{mathrsfs} 
\usepackage{stackengine}
\usepackage{subcaption}    
\usepackage{mathtools} 
\usepackage{anyfontsize}  

\usepackage{soul}  
\usepackage{xcolor} 

\newcommand{\barcode}{\textsc{barcode}}
\newcommand{\bs}[1]{{\bm{#1}}}
\newcommand{\bx}{\bs{x}}   
\newcommand{\bq}{\bs{q}}   
\newcommand{\bv}{\bs{v}}   
\newcommand{\bk}{\bs{k}}   
\newcommand{\hatx}{\hat{\bs{x}}} 
\newcommand{\displacescalar}{\Psi}
\newcommand{\displace}{\bs{\displacescalar}}
\newcommand{\divdisp}{\vartheta}
\newcommand{\zspace}{\bs{s}}
\newcommand{\displacez}{{\displace_s}}
\newcommand{\twodcf}{\xi(r_\perp, r_\parallel)}
\newcommand{\Deloneq}{\Delta^{(1)} (\bq)}
\newcommand{\msun}{\mbox{\({\rm M}_{\odot}\)}}
\newcommand{\mpc}{\mathrm{Mpc}}
\newcommand{\gpc}{\mathrm{Gpc}}
\newcommand{\mpch}{~h^{-1} \mpc}
\newcommand{\hmpc}{\mpch}

\newcommand{\gpchcube}{~h^{-3} \gpc^3}
\newcommand{\likeli}{\mathscr{L}}
\newcommand{\lnL}{\ln\likeli}
\newcommand{\prior}{\mathscr{P}}
\newcommand{\lnP}{\ln\prior}
\newcommand{\hamHMC}{\mathscr{H}}
\newcommand{\potHMC}{\mathscr{E}}
\newcommand{\kinHMC}{\mathscr{K}}
\newcommand{\masHMC}{\mathscr{M}}
\newcommand{\posHMC}{\breve{q}}
\newcommand{\momHMC}{\breve{p}}
\newcommand{\pdif}[2]{\frac{\partial\ #1}{\partial\ #2} }
\newcommand{\bnabla}{\bs{\nabla}}
\newcommand{\one}{{(1)}}
\newcommand{\two}{{(2)}}
\newcommand{\obs}{\mathrm{obs}}

\newcommand{\ua}{\underline{a}}
\newcommand{\ub}{\underline{b}}
\newcommand{\uc}{\underline{c}}
\newcommand{\ud}{\underline{d}}
\newcommand{\krodel}{\delta^{\rm K}}
\newcommand{\delone}{\delta^{(1)}}
\newcommand{\deloneq}{\delta^{(1)}(\bq)}

\newcommand{\Delone}{\Delta^{(1)}}
\newcommand{\Deltwo}{\Delta^{(2)}}
\newcommand{\ft}{\mathcal{F}}
\newcommand{\dft}{\ft_D}

\newcommand*\circled[1]{\tikz[baseline=(char.base)]{
            \node[shape=circle,draw,inner sep=1pt] (char) {#1};}}

\newcommand{\changed}[1]{#1}
\newcommand{\moved}[2]{#2}

\title[Bayesian cosmography from redshift space maps]{Bayesian cosmic density field inference from redshift space dark matter maps}

\author[E. G. P. Bos et al.]{
E. G. Patrick Bos,$^{1,2}$\thanks{E-mail: \href{mailto:p.bos@esciencecenter.nl}{p.bos@esciencecenter.nl}}
Francisco-Shu Kitaura$^{3,4}$
and Rien van de Weygaert$^2$
\\
$^1$Netherlands eScience Center, Science Park 140, 1098XG, Amsterdam\\
$^2$Kapteyn Astronomical Institute, University of Groningen, PO box 800, 9700AV, Groningen\\
$^3$Instituto de Astrofisica de Canarias (IAC), Calle Via Lactea s/n, 38200, La Laguna, Tenerife, Spain \\ 
$^4$Departamento de Astrof\'{\i}sica, Universidad de La Laguna (ULL), E-38206, La Laguna, Tenerife, Spain
}

\date{Accepted XXX. Received YYY; in original form ZZZ}

\pubyear{\changed{2019}}

\begin{document}
\label{firstpage}
\pagerange{\pageref{firstpage}--\pageref{lastpage}}
\maketitle

\input{inc/abstract_keywords}



\input{inc/sec_intro}
\input{inc/sec_method}
\input{inc/sec_input+qual_validation}
\input{inc/sec_results}
\input{inc/sec_discussion}
\input{inc/sec_conclusions}
\input{inc/sec_acknowledgments}



\bibliographystyle{mnras}
\bibliography{bib/egpbib}



\appendix

\input{inc/ap_barcode_summary}
\input{inc/ap_hamiltonian_likelihood_force}

\input{inc/ap_no_rsd_model_problem}

\input{inc/ap_mcmc_performance}

\input{inc/ap_2D_corr_fct}
\input{inc/ap_density_est}
\input{inc/ap_2lpt_sc}


\bsp	
\label{lastpage}
\end{document}

%% file: inc/abstract_keywords.tex

\begin{abstract}
We present a self-consistent Bayesian formalism to sample the primordial density fields compatible with a set of dark matter density tracers after cosmic evolution observed in redshift space.
Previous works on density reconstruction \changed{did not self-consistently consider redshift space distortions or} included an additional iterative distortion correction step.
We present here the analytic solution of coherent flows within a Hamiltonian Monte Carlo posterior sampling of the primordial density field.
We test our method within the Zel'dovich approximation, presenting also an analytic solution including tidal fields and spherical collapse on small scales.
Our resulting reconstructed fields are isotropic and their power spectra are unbiased compared to the true field defined by our mock observations.
Novel algorithmic implementations are introduced regarding the mass assignment kernels when defining the dark matter density field and optimization of the time step in the Hamiltonian equations of motions.
Our algorithm, dubbed \barcode, promises to be specially suited for analysis of the dark matter cosmic web \changed{down to scales of a few Megaparsecs.
This large scale structure is} implied by the observed spatial distribution of galaxy clusters --- such as obtained from X-ray, SZ or weak lensing surveys --- as well as that of the intergalactic medium sampled by the Lyman alpha forest or perhaps even by deep hydrogen intensity mapping.
In these cases, virialized motions are negligible, and the tracers cannot be modeled as point-like objects.
It could be used in all of these contexts as a baryon acoustic oscillation reconstruction algorithm.
\end{abstract}


%% file: inc/sec_intro.tex

\section{Introduction}
\label{sec:intro_barcode_rsd}

The Cosmic Web of the Universe arises from the gravitational instability caused by tiny primordial density perturbations, which presumably have their origin in quantum fluctuations.
At initial cosmic times, perturbations are linear and the statistics describing them is extremely close to Gaussian, though some deviations from Gaussianity could be expected depending on the inflationary phase the Universe has probably experienced after its birth \citep{1980PhLB...91...99S,2018arXiv180706211P}.
Therefore, the Cosmic Web encodes the information to understand nonlinear structure formation
\citep{bond96,weygaert96,cautun14,2014MNRAS.440L.106A}
and disentangle the interplay between dark matter and dark energy
\citep{parklee07,bos12,vogelsberger14}.

One of the great challenges of the coming years is to rigorously test these theories using observations \citep{werner08,2016ApJ...817..160L,2018arXiv180504555T,2018ApJ...867...25C}.
However, observations are based on biased tracers, which are affected by their proper motions, as they are measured in so-called redshift space \citep{jackson72,sargentturner77,kaiser87,hamilton98}.
To test our theories, we need to make the step from these biased tracers to the full underlying matter density field that they probe.
This requires us to make completely explicit the entire process by which the biased observations were generated from the physical world that our theories stipulate.
In this, we face three main challenges: one is the action of gravity linking the linear primordial fluctuations to the final nonlinear density field, the second one is modeling the bias of the dark matter tracers and the third one is solving for redshift space distortions.
In this paper we present a novel approach to the latter problem in the context of Bayesian reconstruction of primordial density fields given biased observations of clusters of galaxies at $z=0$.

\subsection{\changed{Reconstruction of primordial density fields}}

The topic of reconstruction of primordial density fields underwent great progress in the late 80s and early 90s.
\citet{peebles89} first proposed a least action method linking the trajectories of particles from initial to final cosmic times.
\citet{weinberg92} proposed a rank-ordering scheme, realizing the different statistical nature between the initial and final density fluctuations.
\citet{nusser92} proposed a time reversal machine based on the Zel'dovich approximation \citep{zeldovich70}.
Since these pioneering works, however, there have recently been dramatic developments.
The minimization of an action led to two approaches: the FAM method \citep{nusser00,2002MNRAS.335...53B} and the MAK method \citep{2003MNRAS.346..501B,2006MNRAS.365..939M,lavaux10}.
Time reversal machines based on higher order corrections have been proposed \citep{1993ApJ...405..449G,1999MNRAS.308..763M,kitauraangulo12}.
The same concept has been proposed to enhance signatures encoded in the primordial density fluctuations, such as the Baryon Acoustic Oscillations (BAOs) \citep[see][]{2007ApJ...664..675E}, as this linearizes the density field transferring information from the higher order statistics to the two point correlation function \citep[see][]{2015PhRvD..92l3522S}.
Constrained Gaussian random field simulations \citep{bertschinger87,hoffmanribak91,vdwbert96} have been applied to study the Local Universe \citep{clues10}, but it is hard to find a configuration of constrained peaks that give rise to a desired Cosmic Web configuration after non-linear gravitational evolution.
The reverse Zel'dovich approximation can provide a first order correction to this limitation of constrained simulations \citep{doumler13,2018MNRAS.476.4362S}.
The problem of these methods is that they are limited by the  approximation of reversing gravity, which breaks down with so-called shell crossing \citep{2016IAUS..308...69H}.

Bayesian approaches have permitted the incorporation of forward modeling, which solves this latter problem, at a considerably higher computational cost \citep{kitaura13,jaschewandelt13,wang13}.
We will follow here this approach and more specifically the setting described in \citet{wang13} based on the combination of a Gaussian prior describing the statistical nature of the primordial fluctuations and a Gaussian likelihood minimizing the squared distance between the reconstructed density and the true one.
Nonetheless, we would like to remark that the algorithmic approach of connecting the primordial density field with the final one followed by \citet{wang13} is based upon the one described in \citet{jaschewandelt13}, as we also do here.

\subsection{\changed{Self-consistent redshift space treatment}}

Our novel contribution to the Bayesian reconstruction formalism consists of including coherent redshift space distortions (RSD) in an analytical way, within the Hamiltonian Monte Carlo sampling of the posterior distribution function (resulting from the product of the prior and the likelihood).
\changed{\citet{1994ApJ...421L...1N} and \citet{1995MNRAS.272..885F} have also taken the approach of a self-consistent formalism for reconstructing real space densities from the biased redshift space tracers that are galaxies.
Both derive their formalism in linear theory, which allows \citet{1994ApJ...421L...1N} to use the Zel'dovich approximation to correct for redshift space distortions, while for \citet{1995MNRAS.272..885F} it allows an analytical one-to-one transformation of the radial harmonics of the field from redshift to real space via a matrix inversion.}
While \citet{jaschewandelt13} do not give a treatment of RSD, \citet{wang13} rely on an iterative RSD correction applied prior to the Bayesian reconstruction method \citep{wang09,wang12}.
In a recent study in the context of Bayesian reconstructions of the primordial fluctuations, \citet{2018arXiv180611117J} apply a redshift space transformation to the dark matter field, in which the bias is defined with respect to redshift space.
Our formalism entails a more natural approach, and includes a biasing scheme with respect to Eulerian space \citep[see also][]{thesis}.
This implies additional terms in the Hamiltonian equations, and has the advantage that it admits natural real space bias models that couple the large scale dark matter distribution to the galaxy population \citep[see e.g.][and references therein]{wang13,kitaura14,2014MNRAS.441..646N,ata15,2016MNRAS.456.4156K,2017MNRAS.472.4144V}.
The only restriction to our approach is to include a treatment for the virialized motions, as we are not able at this moment to include a random term within the analytical posterior sampling expression.
Virial motions can, in general, be corrected for with fingers-of-god \citep[FoGs;][]{1978IAUS...79...31T} compression \citep{tegmark04,2015AJ....149..171T,2016A&A...588A..14T,2017MNRAS.469.2859S}.
Within a Bayesian context, it is possible to include a virialized cluster term following the approach of \citet{kitaura13}, which includes a likelihood comparison step object by object and not cell by cell as in \citet{jaschewandelt13} or \citet{wang13}.
In the object based approach, the likelihood comparison can be done in redshift space, including virialized motions as was demonstrated in \citet{hess13}.
This method requires two steps within an iterative Gibbs-sampler to handle the reconstruction problem, one to transform dark matter tracers from redshift space at final cosmic times to real space at initial cosmic times, and a second one to make the link between the discrete distribution to the continuous Gaussian field.
It is, in fact, also possible to include an additional Gibbs-sampling step in the Bayesian formalism to self-consistently sample peculiar motions and solve for RSD \citep{kitaura16rsd,ata17}.

\begin{figure*}
\centering
\includegraphics[width=\textwidth]{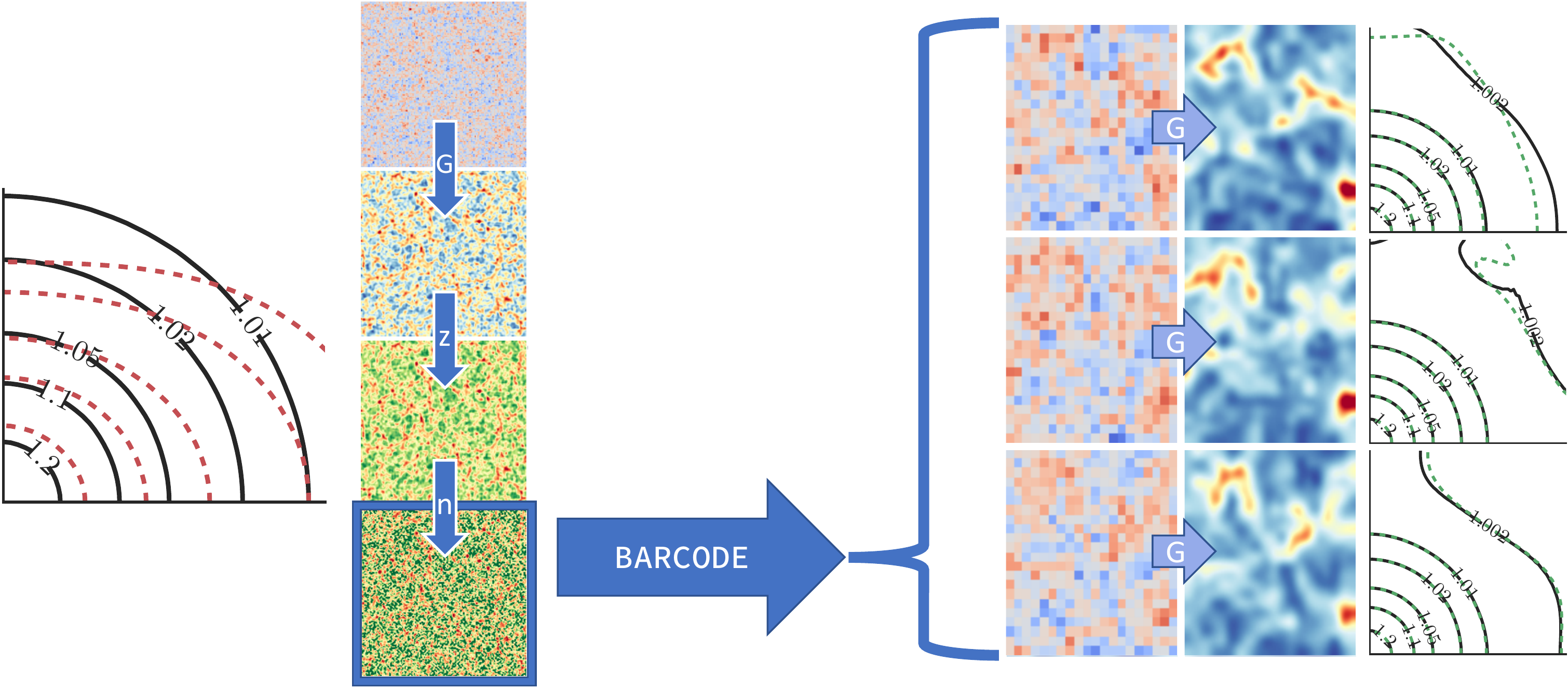}
\caption{
A high level overview of the \barcode\ workflow and pipeline.
\\
Left-most panel: illustration of the challenge posed when applying a reconstruction algorithm on redshift space distorted input data.
The two-dimensional correlation function of a (statistically) isotropic density field is a nearly perfectly circular pattern.
This is what we see in the solid black lines that represent the expected (``true'') field that we want to reconstruct.
When comparing the two-dimensional correlation function of the real --- non-redshift --- space dark matter density fields without correcting for the RSDs, the result is a strong anisotropic deviation (red dashed lines) from this expected pattern.
\barcode\ has been specifically designed to deal with this circumstance, and samples primordial density fluctuation field realizations that are constrained to evolve into a present-day mass distribution in accordance with the input set of observations. 
\\
\barcode\ seeks to reconstruct the underlying real-space dark matter density field from a set of observational input data.
The procedure is statistically sampling the allowed realizations under these observational constraints.
\changed{A crucial distinguishing feature of} \barcode\ is its ability to directly process observations as they are obtained in redshift space without the need for an intermediate redshift distortion correction step.
Also implicit in the formalism is that it takes into account the \changed{stochastic nature of the relation between the observational input data and the underlying density field}.
The four left-hand side panels illustrate this sequence of complications.
Top panel: the primordial density fluctuations that we sample.
Second panel: the corresponding dark matter density field in real space, gravitationally evolved from the primordial field.
Third panel: the field in redshift space.
Bottom panel: the redshift space field with ``observational'' noise added.
\\
The \barcode\ procedure described in this work produces likely reconstructions of the density field, on the basis of the physical and statistical assumptions and prescriptions of the Bayesian model relating the observed mass distribution to the primordial density perturbations.
It takes into account the gravitational evolution of the mass distribution, stochastic processes and observational effects.
\\
The right-hand side panels show zoom-ins on to three realizations of the \barcode\ procedure: primordial fields on the left, corresponding gravitationally evolved real space fields on the right.
Note that we sample primordial fields, which can in turn be used as initial conditions for evolution studies of the large scale structure.
The images reveal the substantial level of variation between the resulting reconstructions of the underlying dark matter distribution.
The differences between the reconstructions reflect the uncertainties emanating from the biased and noisy observational input data, as well as those induced by the approximate nature indigenous to the physical modeling of gravitational evolution and biasing.
Despite the intrinsic variations between permissible distributions of the dark matter distribution, the success of the \barcode\ procedure in correcting for the redshift distortions in the input observational data is evidenced by the three far right-hand side panels.
For three different input fields the algorithm has produced near perfect isotropic correlation functions in real space, clear evidence for the removal of anisotropies due to presence of redshift distortions.
}
\label{fig:barcode_overview}
\end{figure*}

In this work, we propose a self-consistent transformation of the \emph{data model} into redshift space, allowing to sample directly from the posterior distribution function \emph{in one step} within a Hamiltonian Monte Carlo sampler.
Thus, this represents the first self-consistent (coherent) redshift space data based primordial density fluctuation reconstruction code based on one iterative sampling step, which can also naturally deal with masks.
Our implementation of this approach is called \barcode.
A high level overview of the method and the problem it solves is presented in figure~\ref{fig:barcode_overview}.
In this code, we transform from Lagrangian coordinates $\bq$ to Eulerian coordinates $\bx$ using a structure formation model and within the same equation also from $\bx$ to the corresponding redshift space $\zspace$.
To implement the redshift space transformation into the Hamiltonian Monte Carlo formalism, we need to derive the corresponding analytical derivatives.
For our structure formation model, we will restrict our studies to the Zel'dovich approximation, as we seek the most efficient, simple formalism, which has been shown to be accurate \changed{on scales down to a few Megaparsecs} when it is fully treated \changed{\citep[see e.g.][]{2013MNRAS.433.2389M,white15}}.

\subsection{\changed{Astrophysical motivation}}

\barcode\ has already been referred to in various works giving details of its implementation \citep{bos16tallinn,thesis}.
One of our main motivations for developing \barcode\ is the analysis of galaxy clusters for which the mass has been measured in X-ray \citep{sarazin86,mulchaey00,ikebe96,nulsen10}, Sunyaev-Zel'dovich cluster surveys \citep[e.g.][]{1999PhR...310...97B,2015ApJS..216...27B,2016A&A...594A..24P,2018ApJS..235...20H} and weak lensing studies of clusters (e.g.\ \citet{1993ApJ...404..441K,1997ApJ...489..522L,2010arXiv1002.3952U,laureijs11,dejong12,2013SSRv..177...75H,2016MNRAS.459.1764S,2017A&A...598A.107R,2018arXiv180901669T}).
This frees us from having to consider galaxy bias, contrary to many similar studies in the literature \citep[see e.g.][]{2019MNRAS.483.5267B}.
\changed{
    Rather, in this work we deal with cluster bias, which is very different from the classical galaxy bias.
    In the case of galaxies we have tracers and a bias factor (in the linear case) to roughly relate galaxy density and the underlying full density field.
    In our case we want to identify the peaks in the density field and ``move them back in time'' to recover the full underlying surrounding primordial density field; this is, in fact, the central focus of this work.
    The peaks of the density field are biased tracers according to the background split picture introduced by \citet{1984ApJ...284L...9K}.
    This implies that having unbiased estimates of the density distribution of a cluster from e.g. X-ray or lensing measurements, will not lead to unbiased estimates of the whole dark matter field, unless the density field in the intra-cluster region is properly reconstructed.
    This means that we are facing two different problems in terms of bias.
    One is to solve for the bias within the cluster and another is to solve for the bias introduced from having density estimates only in cluster regions.
    The first problem can be solved either by preparing the data beforehand \citep[see e.g.][]{wang09} or with a self-consistent deterministic and stochastic bias description within a Bayesian framework \citep{ata15}.
    The second problem can be also solved with a deterministic and stochastic bias description or alternatively with a mask.
    In the latter approach the regions where no cluster data is available can be set to have a zero response function, whereas the regions where the cluster has been observed can be set to one in the response function \citep[see e.g.][]{zaroubi95,kitaura08,jaschekitaura10,wang13}.
    We have taken this latter approach.
    Once the full dark matter field is reconstructed the results are unbiased (see results in Section~\ref{sec:results_solution}).
}

More details to the context and motivation of our focus on clusters have been given in \citet{thesis}.
We note that in such a context, the approach proposed here is potentially more appropriate than the ones cited above modeling tracers as point-like objects, because the shapes and orientations of clusters play a decisive role in their tidal effects on their environments \citep{bond96,vdwbert96}.
\changed{In practice, we will in most cases be forced to limit our orientation constraints to the two coordinates on the sky, the projected axes of the full ellipsoid.
In rare cases, rotation measure tomography \citep{2011A&A...525A.104P} or gravitational lens reconstruction \citep[e.g.][]{2019MNRAS.485.3738L} may provide additional information about the radial structure.}
We leave a more detailed investigation of this to later work.

Broader cosmological applications are also envisioned.
Extending this work to make growth rate sampling is trivial, and can be done in a similar fashion to \citet{granett15}, but including a more complex and accurate algorithm, as presented here.
In fact, redshift space distortions can also be leveraged as a proxy for constraining the nature of gravity and cosmological parameters \citep[e.g.][]{berlind01,zhang07,jainzhang08,guzzo08,nesseris08,songkoyama09,songpercival09,percivalwhite09,mcdonaldseljak09,white09,song10,zhao10,song11}.
To this end, many recent studies focus on the measurement of redshift space distortions \citep{cole95,peacock01b,percival04,daangela08,okumura08,guzzo08,blake11,jennings11,2012ApJ...748...78K,2012MNRAS.420.2102S,2012MNRAS.426.2719R,2012JCAP...11..014O,2013MNRAS.429.1514S,2013PhRvD..88j3510Z,2013MNRAS.436.3089B,2013A&A...557A..54D,2014MNRAS.439.3504S,2014MNRAS.440.2692S,2014A&A...563A..37B,2014MNRAS.440.2222T,2014JCAP...05..003O,2014MNRAS.443.1065B}.

\subsection{\changed{Outline}}

This paper is structured as follows.
First we present the methodology, giving core high level details of the analytical implementations (see Section~\ref{sec:method}), while leaving most mathematical derivations to the appendices.
In particular we focus on the novel redshift space formalism implemented in \barcode.
In Section~\ref{sec:input_data}, we set the stage for our analyses by describing and illustrating our mock observations and the reconstructions that \barcode\ yields from them.
 
The core results of our investigation are described in Section~\ref{sec:results_solution}.
The reconstructions including the redshift space model demonstrate the major benefits of our novel approach.
For contrast, we show what would happen if one used \barcode\ without our redshift space extension on observational data in redshift space coordinates in Appendix~\ref{ap:results_problem}.

We close with a discussion of the results in Section~\ref{sec:discussion_barcode_rsd} and conclusions in Section~\ref{sec:conclusions}.

Additional material is available in the appendices.
We first briefly go into more detail than in Section~\ref{sec:method} about the most important \barcode\ concepts and equations in Appendix~\ref{ap:barcode_summary}.
Then in Appendix~\ref{ap:hamiltonian_force_zspace} we present the full derivation of the necessary formulas for implementing our redshift space model in the HMC formalism.
The astronomically oriented results of Section~\ref{sec:results_solution} are followed by a more technical discussion on the performance aspects of \barcode\ with this new model in Appendix~\ref{ap:mcmc_performance}.
There we evaluate whether the added stochasticity that redshift space transformations add cause the MCMC chain to evolve differently than without our redshift space model.
The two dimensional correlation function, which is used extensively in the results sections, is defined in Appendix~\ref{ap:2D_corr_fct}.
An important technical aspect in \barcode\ is the use of kernel density estimation, the details of which we discuss in Appendix~\ref{ap:density_est}.
Finally, in Appendix~\ref{ap:2lpt_sc}, we give a complete derivation of the equations necessary for configuring \barcode\ to use a second-order Lagrangian Perturbation Theory (2LPT) structure formation model, optionally including the spherical collapse terms of the ``ALPT'' model.

%% file: inc/sec_method.tex

\section{Method}
\label{sec:method}
The Bayesian reconstruction algorithm connecting the evolved dark matter density to the primordial fluctuation field has already been presented in several works \citep{kitaura12,kitaura13,jaschewandelt13,wang13,bos16tallinn}.
We summarize the \changed{core concepts} used in \barcode\ --- our C\texttt{++} implementation of this algorithm --- in \changed{Section~\ref{sec:barcode} and list and derive central equations in} Appendix~\ref{ap:barcode_summary}.

\changed{The central tenet of this paper is that now} we consider the Hamiltonian Monte Carlo calculations in redshift space $\zspace$ (equation~\ref{eqn:redshift_space}) instead of Eulerian coordinates $\bx$.
Densities are then functions of $\zspace$, e.g.\ $\rho^\obs(\zspace)$.
For convenience, in this section we write $\posHMC_i$ instead of $\delta(\bq_i)$ for the initial density field (the sample in the MCMC chain), where $i$ is a grid location (i.e.\ a component of the sample vector).

\changed{In Section~\ref{sec:redshift_space_coordinates}}, we will define what we mean by redshift space coordinates.
Next, in Section~\ref{sec:rsd_in_lpt}, we give a brief overview of the different kinds of non-linearities involved in the redshift space transformation and our Lagrangian Perturbation Theory models of structure formation.
In particular, we discuss what this means for the representation of non-linear cosmic structure.

\subsection{\barcode}
\label{sec:barcode}
\changed{Before we describe our redshift space formalism within the Hamiltonian Monte Carlo reconstruction method, we briefly recap the basic idea behind \barcode, our implementation of this method.}
\moved{Appendix~\ref{ap:barcode_summary}}{
  \barcode\ is a code that aims at reconstructing the full Lagrangian primordial density perturbation field corresponding to a configuration of observed objects.
  It has been designed specifically for clusters of galaxies as principal constraining objects.
}

\moved{}{
  We want this reconstruction to reflect the inherently stochastic nature of both observations and the non-linear structure formation process.
  This stochasticity can be encoded in a \emph{probability distribution function} (PDF), or posterior.
  The posterior describes the complete physical process leading to an observed density field; starting from the initial density perturbations, going through gravitational collapse and finally including observational effects.
}

\moved{}{
  \barcode\ uses an MCMC method called Hamiltonian Monte Carlo \citep{neal93,taylor08,jaschekitaura10,neal12} to \emph{sample} this posterior, comparing the results to the actual observed density field and in this way finding the initial density distributions that match the data.
  This yields not just one reconstruction, but an ensemble of possible reconstructed Lagrangian density fields.
  The variation in these realizations reflects the full array of stochastic effects as included in the model.
}

\changed{
  To compare Lagrangian space primordial density fields (our ``sample variable'') to the observations at $z=0$, \barcode\ evolves the sampled Lagrangian fields forward in time using a structure formation model.
  Then, in the likelihood (the Bayesian posterior being the product of likelihood and prior), the difference of the evolved sampled field to the observed field is quantified.
  The likelihood effectively ``scores'' the sample by its similarity to the observations.
}

\moved{}{
  In the HMC method, the PDF is represented by a Hamiltonian (Section~\ref{sec:hmc_terminology}), where the PDF's stochastic variable (the signal we sample) is treated as a position analogue.
  The HMC algorithm then samples the posterior PDF by repeating the following procedure:
  \begin{enumerate}
    \item \emph{Initialize the signal:} define the ``starting position'' in the parameter space (either a manually supplied initial guess or the position from the previous sample).
      Our ``signal'' $\delta(\bq)$ is defined as
      \begin{equation}
        \delta(\bq) \equiv D_1 \Deloneq \,,
        \label{eqn:delta_definition}
      \end{equation}
      where $D_1$ is the first order growing mode linear growth factor and $\Deloneq$ is the spatial part of the first order term of the LPT primordial density field expansion.
    \item \emph{Stochastic step:} draw a random ``momentum'' vector (we use a Gaussian field momentum distribution with the Hamiltonian mass as the covariance matrix).
    \item \emph{Dynamical step:} solve the Hamiltonian equation of motion using a leap-frog solver (Section~\ref{sec:leap_frog}) to find the next ``position'', i.e.\ the next sample \emph{candidate}.
    \item \emph{Accept or reject} the candidate based on the acceptance criterion (Section~\ref{sec:hmc_acceptance_criterion}).
  \end{enumerate}
  This process can continue indefinitely.
  Usually, the user supplies some maximum number of desired samples.
}

\changed{
  In Appendix~\ref{ap:barcode_summary} details can be found on our HMC formalism, its terminology and the algorithmic steps listed above.
  For the main discussion of this paper, it suffices to mention that the so-called \emph{Hamiltonian likelihood force} plays a central role in the algorithm.
  Without going into too much detail here, this quantity, which is defined as the derivative of the likelihood to the signal, is used to find the next sample in our HMC Markov chain.
  This term is mathematically non-trivial, which is why we devote Appendix~\ref{ap:hamiltonian_force_zspace} to its derivation.
  It is also computationally quite intensive, which calls for a highly efficient implementation.
  However, it is of critical importance, since it provides HMC with the information from both the data, its stochastic properties and our assumed structure formation model.
  HMC uses this information to efficiently ``orbit'' the high-dimensional space of signals (primordial density fields) and especially to converge quickly on the region in this space of highest posterior probability, i.e.\ of fields that best match the data.
}

\subsection{Redshift space coordinates}
\label{sec:redshift_space_coordinates}

From a practical perspective we will describe redshift space as follows.
Redshift space coordinates $\zspace$ can be defined as
\begin{equation}
  \zspace \equiv \bx + \frac{1}{Ha} \left( (\bv-\bv_\obs) \cdot \hatx \right) \hatx \equiv \bx + \displacez \,,
\label{eqn:redshift_space}
\end{equation}
where $\zspace$ is a particle's location in redshift space, $\bx$ is the Eulerian comoving coordinate, $\bv$ is the particle's (peculiar) velocity and $\bv_\obs$ is the velocity of the observer, and  $\hatx$ is the comoving unit vector.
$a$ is the cosmological expansion factor, which converts from physical to comoving coordinates, and $H$ is the Hubble parameter at that expansion factor, which converts the velocity terms to distances with units $\mpch$.
We call $\displacez$ the \emph{redshift space displacement field}, analogously to the displacement field $\displace$ that transforms coordinates from Lagrangian to Eulerian space in LPT (equation~\ref{eqn:lagrangian_to_eulerian}).

One important point to stress is that in equation~\ref{eqn:redshift_space} (and all equations henceforth), the comoving coordinates are defined such that the observer is always at the origin.
The unit vector $\hatx$, and thus the entire redshift space, changes completely when the origin changes.

From redshift measurements $z$, we can derive the radial component of the total velocity $\bs{u}$ of galaxies and other objects:
\begin{equation}
  \bs{u} = \bv_H + \bv \,,
\end{equation}
where $\bv_H$ is the Hubble flow velocity and $\bv$ is the peculiar velocity.
Redshift space (comoving) coordinates $\zspace$ can be written in terms of the total velocity as
\begin{equation}
  \zspace \equiv \frac{\bs{u}}{Ha} \cdot \hatx \hatx \,.
\label{eqn:redshift_space_total_velocity}
\end{equation}
We can compare such observational redshift space positions directly with theoretical models by transforming our models to redshift space as given in equation~\ref{eqn:redshift_space}.

\subsection{Redshift space distortions in Lagrangian Perturbation Theory}
\label{sec:rsd_in_lpt}

\changed{For the results presented in this paper, we used first order Lagrangian Perturbation Theory --- better known as the Zel'dovich approximation (ZA) --- as our model of structure formation (see Appendix Section~\ref{sec:lag2eul}).}

Lagrangian Perturbation Theory (LPT) is accurate in its position and velocity estimates up to (quasi-)linear order.
Note that since our objective is to use them in a model based on data at low redshift, and we are dealing with strong non-linearities, LPT is not a fully accurate model.

To clarify, there are two types of non-linearity involved in redshift space distortion theory \changed{\citep{1991ApJ...372..380Y,white12}}.
\citet{kaiser87} and others --- when talking about linear redshift space \changed{distortions ---} are dealing with non-linearities in the mapping from Eulerian to redshift space.
One marked example of this non-linear mapping is the creation of caustics in the triple-value regions.
In the centers of clusters, we also have to take into account that the velocity field itself is non-linear, which, for instance, leads to FoGs.

Since we make no approximations in the redshift space mapping, we are dealing with its full non-linear form.
The (quasi-)linear velocity field of LPT, on the other hand, will become increasingly inaccurate as we venture into the non-linear density and velocity-field regime, leading to increasingly inaccurate redshift space representations.

Large scale (linear regime) coherent flows are still well modeled in our LPT-based framework.
\changed{The large scale velocity field} $\bv$ in LPT \changed{scales} linearly with the displacement field $\displace$:
\begin{equation}
  \bv = a H \sum_{i=1}^{o} f^{(i)} \displace^{(i)} \,,
\end{equation}
where $i$ is summed up to the used perturbation order $o$ ($o = 1$ for the Zel'dovich approximation and $o=2$ for 2LPT) and $f$ is the linear velocity growth factor:
\begin{equation}
  f^o \equiv \pdif{\ln D^o}{\ln a} \,,
\end{equation}
In linear theory, the displacement is directly proportional to the gravitational acceleration, which follows directly from the initial density field.
On large scales, structure still resides mostly in the linear regime, so that LPT manages to describe the velocity field fairly accurately.
The squashing effect on large scales will therefore be well represented in our model.

Cluster in-fall is modeled poorly in Lagrangian Perturbation Theory.
Virialized and multi-stream motion are not at all included in LPT.\@
This means we do not model FoGs nor the triple value regime within the turnaround radius \citep[see e.g.][]{haarlemvdw93} in this work.
Any coincidental triple value regions cannot be trusted in our approach, even though there might be FoG-like features around clusters in LPT as well.
Since LPT breaks down there, they cannot be considered as true features.
\changed{
  In any case, only about 30\% of the FoGs can be attributed to coherent motions, the rest being the result of random motions \citep{2016JCAP...08..050Z}.
  Even with a prohibitively more expensive structure formation model that could reproduce such motions, a statistical description of FoGs is necessary if no peculiar velocity information is available, as the phase space dynamics in the Vlasov equation can not be solved.
}
We shortly come back to this in the discussion in Section~\ref{sec:discussion_rsds_and_lpt}.

\changed{
In this work, we use a plane parallel approximation of redshift space.
Effectively, we put the observer at an infinite distance.
This greatly simplifies the equations derived for the Hamiltonian likelihood force in Appendix~\ref{ap:hamiltonian_force_zspace}.
}

%% file: inc/sec_input+qual_validation.tex

\section{Input data}
\label{sec:input_data}

Our novel approach towards integrating redshift space in the Hamiltonian solver code needs to be validated.
The question of whether the code converges correctly towards the true Lagrangian density and in what way it diverges from it (if differently than in the regular Eulerian space treatment of previous works) will be answered in Section~\ref{sec:results_solution} and Appendix~\ref{ap:mcmc_performance}.
To this end, we sampled a set of reconstructions based on mock observed density distributions.
We describe and illustrate the mocks and reconstructions in the rest of this section.

\begin{figure*}
\centering
\includegraphics[width=\textwidth]{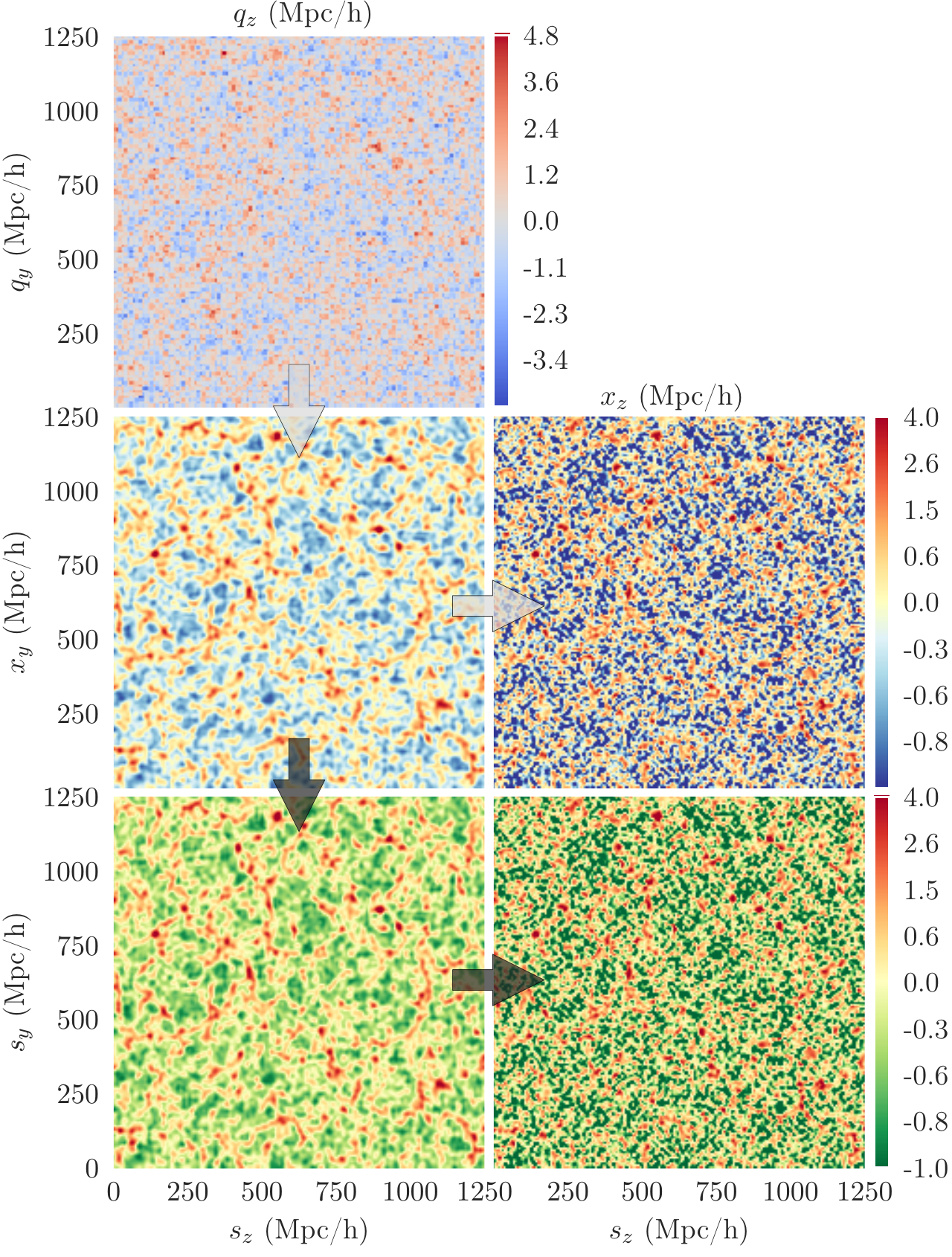}
\caption{
  From true Lagrangian field (\emph{top left}) to true Eulerian field (\emph{center left}).
  From there we can go two ways: either directly to mock observational input field in Eulerian space as we did in \citet{thesis} (\emph{center right}), or via true redshift space field (\emph{bottom left}) to \changed{mock} in redshift space (\emph{bottom right}).
}
\label{fig:img_results_01_lag_eul_rss_nobs}
\end{figure*}

In our numerical experiments we use four categories of parameters, detailed in the following subsections:
\begin{enumerate}
  \item cosmological (\S~\ref{sec:cosmological_parameters}),
  \item statistical (\S~\ref{sec:statistical_parameters}),
  \item astronomical (\S~\ref{sec:astronomical_parameters}),
  \item numerical (\S~\ref{sec:numerical_parameters}).
\end{enumerate}

\subsection{Cosmological parameters}
\label{sec:cosmological_parameters}
For the cosmological parameters we use the maximum likelihood results from the WMAP 7-year observations \citep{komatsu11}\footnote{
    For more up to date parameters, see \citet{2018arXiv180706209P}.
}.
The relevant parameters for \barcode\ are given in table~\ref{tab:cosmo_parameters}.
\begin{table}
  \caption{Cosmological parameters used in this work. These are results from the WMAP 7-year observations \citep{komatsu11}, except for the box size.}
  \label{tab:cosmo_parameters}
  \centering

  \begin{tabular}{l|cc}

  \hline
  & \textbf{value} & \textbf{description} \\
  \hline
  $\Omega_m$      & $0.272$          & Matter density parameter \\
  $\Omega_b$      & $0.046$          & Baryonic matter density parameter \\
  $\Omega_\Lambda$& $0.728$          & Dark energy (DE) density parameter \\
  $w$             & $-1$             & DE equation of state parameter \\
  $n_s$           & $0.961$          & Primordial power spectrum power-law index \\
  $w_a$           & $0$              & DE linear time dependency parameter \\
  $\sigma_8$      & $0.807$          & Density field fluctuation rms at $8\mpch$ \\
  $h$             & $0.702$          & Hubble parameter (units of $100~\mpc~\mathrm{km}^{-1} s$) \\
  $L$             & $200$            & Comoving box size in$\mpch$ \\
  \hline

  \end{tabular}
\end{table}
The cosmological parameters also determine the power spectrum.
We use CAMB \citep{camb} to generate the power spectrum which is used for the prior computation in \barcode.
We consider cubical volumes of $1.95 \gpchcube$.
The Zel'dovich structure formation model is applied to transform from Lagrangian to Eulerian coordinates at $z=0$.

\subsection{Statistical parameters}
\label{sec:statistical_parameters}
\barcode\ has a number of parameters to tune HMC and some choices of statistical models.
The results in this paper use the following statistical parameters:

\emph{Leap-frog step size $\epsilon$:}
We use an adaptive step size (Appendix~\ref{sec:leap_frog_time_step_precision}).
It is bounded to be between $0$ and $2$, but within those limits it is adapted to give an acceptance rate of $0.65 \pm 0.05$ (calculated over the $50$ previous steps).

\emph{Number of leap-frog steps per iteration $N_\epsilon$:}
The number of leap-frog steps is randomly drawn from a uniform distribution of integers between $1$ and $256$.
For Section~\ref{sec:optimizing_number_leap_frog_steps} the maximum was increased to $4096$ to assess performance as a function of this parameter.

\emph{Hamiltonian mass:}
\changed{
  In HMC, the \emph{Hamiltonian mass} in the kinetic energy of equation~\ref{eqn:kinetic_term} is used as the inverse of the correlation matrix of the Gaussian distribution from which the Hamiltonian momenta are drawn (see Appendix~\ref{ap:barcode_summary} for technical details).
  A smart choice of mass enhances Markov chain performance, because it will tune the magnitude of momentum components to their respective \emph{Hamiltonian position} dimensions \citep{neal93}.
  Recall in this that Hamiltonian position equals our N-dimensional signal, i.e.\ the gridded primordial density field.
}
We use the inverse correlation function type mass (see \citet{taylor08,thesis} for more details):
\begin{equation}
  \hat{\masHMC} = 1/P(k) \,.
\end{equation}
\changed{
  This makes sure that the scale of momentum fluctuations matches that of the primordial density field, though only on a statistical level, i.e.\ without taking into account the actual spatial structure of the sampled field at that point in the Markov chain.
}

\emph{Likelihood:}
We use a Gaussian likelihood as described in equation~\ref{eqn:log_likelihood}.
For this work, we set \changed{the parameters of that equation} $\sigma_i$ and $w_i$ to $1$ for all cells, effectively removing these parameters from the algorithm.
\changed{
  When needed, these parameters can be used to quantify per cell $i$ respectively the amount of uncertainty in the observed density and a window function to weigh regions of the observed density field more or less strongly.
  As described in the next subsection, we generate the noise (right-hand panels of figure~\ref{fig:img_results_01_lag_eul_rss_nobs}), so $\sigma_i$ is given.
}

\subsection{Astronomical/observational parameters}
\label{sec:astronomical_parameters}

\emph{Mock observed density field:}
The algorithm is based on the comparison of an observed density field with sampled fields from the posterior PDF.
For the runs in this study, we generated the mock observed fields in redshift space (instead of Eulerian space as \changed{in most previous works \citep{jaschewandelt13,wang13,thesis}}) as follows.
First, we generate a Gaussian random field given the cosmological power spectrum.
This Lagrangian space field is mapped to an Eulerian space density field at $z=0$, using the assumed structure formation model\changed{, the Zel'dovich approximation}.
\changed{
  Without fundamental changes, the formalism can be upgraded to higher order approximations like 2LPT, as shown in Appendix~\ref{ap:2lpt_sc}.
  At the cost of increased complexity of mainly the Hamiltonian force (Section~\ref{sec:barcode}), an N-body method could be used as well, as first shown in \citet{wang14}.
  Here we restrict ourselves to the easily interpreted and implemented Zel'dovich approximation.
}
The Eulerian space field is, in turn, transformed to redshift space on the basis of the ballistic velocities according to the Zel'dovich approximation.
Following this, to simulate actual data, we add random noise to the redshift space density field.
The noise is drawn from a Gaussian with zero mean and $\sigma_i$ standard deviation.

Note that since our likelihood is Gaussian, we must also allow negative densities to be produced by the noise.
In \citet{thesis}, we truncated negative values so as not to generate unphysical negative masses.
This turned out to be the cause of the power deficiency in the reconstructions we discussed at length in that work.
In this work, we show the potential of the formalism, including its ability to converge on its target, which is why we allow unphysical negative densities in order to satisfy our model choice of a Gaussian likelihood.
In a more realistic use case, using real observed data, one would have to choose an error model that does not allow negative values, like a gamma or log-normal distribution.

The lower right panel of figure~\ref{fig:img_results_01_lag_eul_rss_nobs} displays the resulting noisy mock observed field (the panel above that describes the end result of the mock field used in a model where density is defined in regular comoving coordinates instead of redshift space).

For our comparison study, two types of runs were set up, in addition to the runs from \citet{thesis}  (in this work referred to as ``regular'' runs, i.e.\ runs with data and the model in regular comoving space):
\begin{enumerate}
  \item In the first, we used the same algorithm as the runs in \citet{thesis}, not taking redshift space into account; we refer to these as ``obs\_rsd'' runs.
  \item For the second set of runs, the redshift space formalism from Section~\ref{sec:method} was used; we call these the ``rsd'' runs.
\end{enumerate}

\begin{figure}
\centering
\includegraphics[width=\columnwidth]{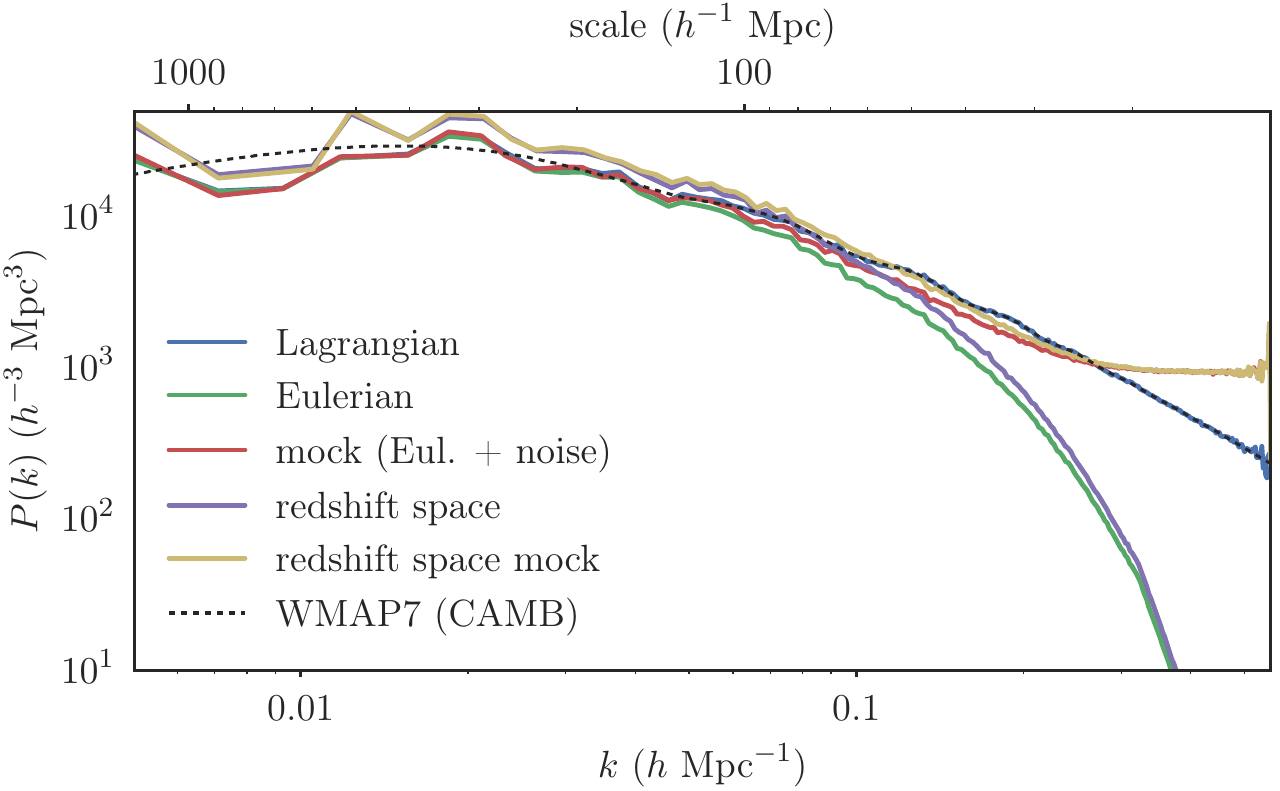}
\includegraphics[width=\columnwidth]{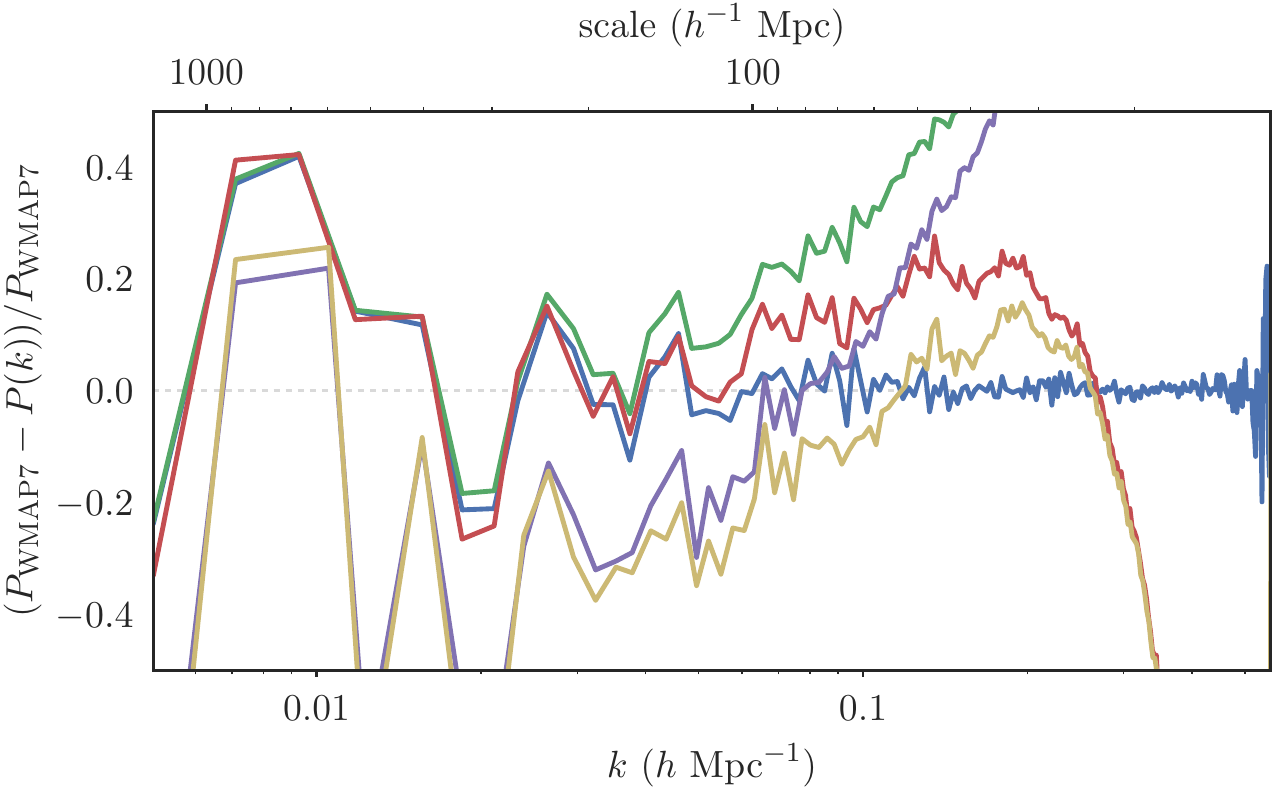}
\caption{
  Power spectra of density field in Lagrangian space, Eulerian space and redshift space and of the mock observational density fields that are used as input (in Eulerian or redshift space).
  \changed{The bottom panel shows the fractional deviation of the spectra from the WMAP7 (CAMB) spectrum that was used to sample the true Lagrangian density field from.}
  The redshift space spectrum peaks above the Eulerian space one, by a factor called the Kaiser factor \citep{kaiser87}.
  The mock observational density fields have a high-$k$ tail that represents the Gaussian noise that is put on top of it.
}
\label{fig:img_results_13_ps_true_lag_eul_rss}
\end{figure}

In figure~\ref{fig:img_results_13_ps_true_lag_eul_rss} we show the power spectra of the density fields from figure~\ref{fig:img_results_01_lag_eul_rss_nobs}.
The redshift space spectrum clearly reflects the Kaiser effect: a boost of power at low $k$ and a deficiency at higher $k$.

\emph{Window / mask:}
The window parameter $w_i$ in the likelihood was not used in this work, i.e.\ it was set to $1$ for all cells $i$.

\emph{Starting configuration:}
To start the MCMC random walk, we need to provide a starting point, or ``initial guess'', in the $N_x^3$-dimensional sampling space.
We supply the code with a non-informative guess about the cosmic web structure: a field with value zero in every grid cell.
This is also the way to run the code using real observations.

\subsection{Numerical parameters}
\label{sec:numerical_parameters}

We run with a resolution of $L/N_x = 9.766 \mpch$, corresponding to a grid of $N_x = 128$ cells in each direction and a volume side of $L = 1250 \hmpc$.
The total number of cells is $N = 2097152$.
In \citet{thesis} we studied computations with $L/N_x=3.125$.
The Zel'dovich approximation as a structure formation model correlates well with a full N-body simulation above scales of $\sim10\mpch$, which is why for this work we chose a higher grid cell size.

We used three seeds to initialize the random number generator in the code.
Each of these seeds was used for a separate code run, giving separate, differently evolving chains.

We output the field realizations every 50 steps of the chain.
This assures that the saved samples are sufficiently far apart, although it is a bit overcautious, because when the chain performs well, the HMC algorithm should already make sure that subsequent samples are uncorrelated (indeed the likelihood panel in figure~\ref{fig:perform_reg_rsd} shows that this is the case).
In the statistics of the following sections (means and standard deviations of reconstructions) we use 101 samples, the range from sample 500 to 5500.

We used the SPH density estimator (Section~\ref{ap:density_est}).
We set the SPH scale parameter equal to the cell size $L/N_x = 9.766 \mpch$.

\subsection{Redshift space reconstructions illustration}

\changed{
  To illustrate the accuracy of the algorithm, we show and discuss qualitatively the reconstructed density fields.
  Quantitative results are presented in Section~\ref{sec:results_solution}.
  The goal of the algorithm is to recover primordial density fields.
  However, these are Gaussian random fields with a power spectrum that makes it hard to visually discern large scale structures.
  Therefore, instead we compare the evolved density fields to the ``true'' density field, which illustrates more prominently the accuracy and deviations of the implied large scale structure.
}

\begin{figure*}
\centering
\includegraphics[width=\textwidth]{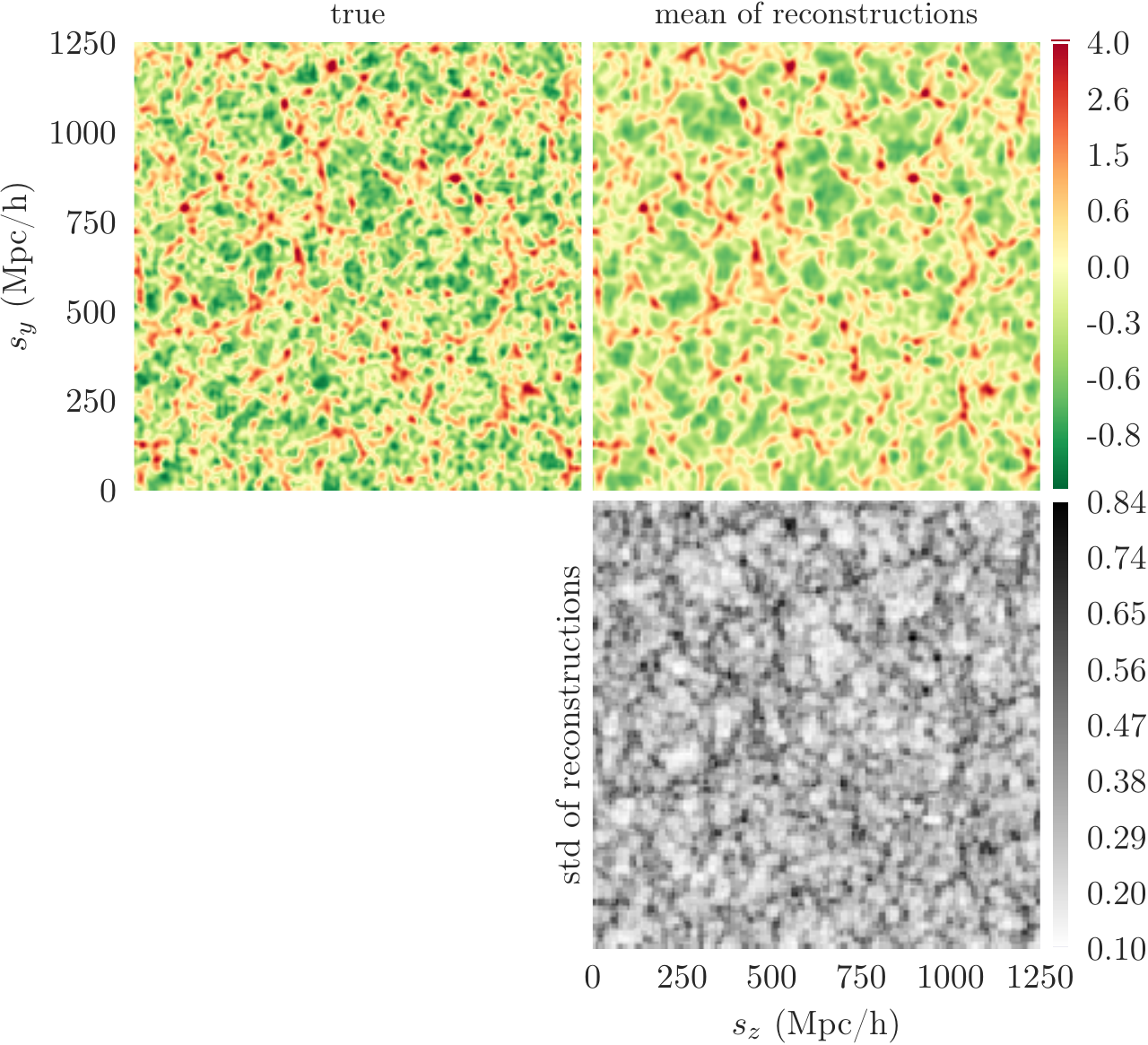}
\caption{
  Redshift space density fields.
  \emph{Top left:} the true field.
  \emph{Top right:} the mean of the sampled fields.
  \emph{Bottom:} the standard deviation of the samples.
}
\label{fig:true_mean_rss}
\end{figure*}

First, in figure~\ref{fig:true_mean_rss} we show the difference between the mean and true fields, but now in redshift space coordinates $\zspace$.
The mean reconstruction in redshift space shows a high degree of similarity to the true field's configuration, both in the density values and in the shapes of the web structures.
The variance around the mean (bottom panel) reflects the propagation of data, model and systematic errors embodied by the posterior model of the cosmic densities that we sample.
As is the case when sampling densities in regular comoving coordinates (see e.g.\ \citet{thesis}, figures~3.14 and~3.15), the highest variance can be found in the high density regions, reflecting the higher number statistics of the particles that cluster to form these regions.
These same statistics lead the voids in reconstructions to be dominated by random fluctuations from the prior.
These fluctuations average out in the mean field.
The underdense voids, hence, contain less structure in the reconstructions' mean than in the true field.

We can further compare the redshift space results to the regular Eulerian space density shown in the top two panels of figure~\ref{fig:den_vstrue_eul_rsd}.
The coherent inflow part of the redshift space distortions causes expected features along the line of sight in redshift space: squashing and enhancement of density contrast for overdense large scale structures and a stretching of underdense voids\footnote{
These features may be hard to appreciate from static figures.
We provide an animated GIF at \url{http://egpbos.nl/rsd-eul__rsd-rss/} for more convenient comparison of the top right-hand side panels of figures~\ref{fig:true_mean_rss} and~\ref{fig:den_vstrue_eul_rsd}.
}.
These effects are quantified in the power spectrum, which in redshift space shows the Kaiser effect (figure~\ref{fig:img_results_13_ps_true_lag_eul_rss}; see also figure~\ref{fig:ps_vstrue_lag_obs} of Appendix~\ref{ap:results_problem}, which illustrates the imprint of the Kaiser effect on reconstructions that were not corrected for redshift space distortions), and the two-dimensional correlation function, further discussed in Section~\ref{sec:2Dcf_isotropy}.

%% file: inc/sec_results.tex

\section{Results}
\label{sec:results_solution}

In this section we will show the benefits of using our self-consistent redshift space (RSS) Zel'dovich model over a regular Zel'dovich model that does not self-consistently treat observations as being in redshift space, the latter of which is used in previous models \citep[e.g.][]{jaschekitaura10,jaschewandelt13,wang13}.
When properly dealt with, the anisotropies caused by RSDs can be modeled in the data and hence eliminated from the reconstruction.
This enables reconstruction of the mean density field based on redshift space data.
By directly incorporating RSDs in the formalism, ad-hoc corrections for RSDs are no longer necessary.

From an algorithmic point of view, the most pressing question is whether the code does converge correctly towards the true Lagrangian density.
Also, we want to characterize in what way the samples diverge from it.
\changed{
  This latter question is addressed in Section~\ref{sec:density_variation_around_mean}.
}

One might wonder what would happen if redshift space data are used without a redshift space model, i.e.\ ignoring the effect of redshift space distortions in the observations by not modeling them in the reconstruction algorithm, or more specifically: the posterior.
In Appendix~\ref{ap:results_problem} we illustrate the non-negligible impact of the redshift space effects in such a naive approach.
This clearly motivates the need for the model used in our approach.

\begin{figure*}
  \centering
  \includegraphics[width=\textwidth]{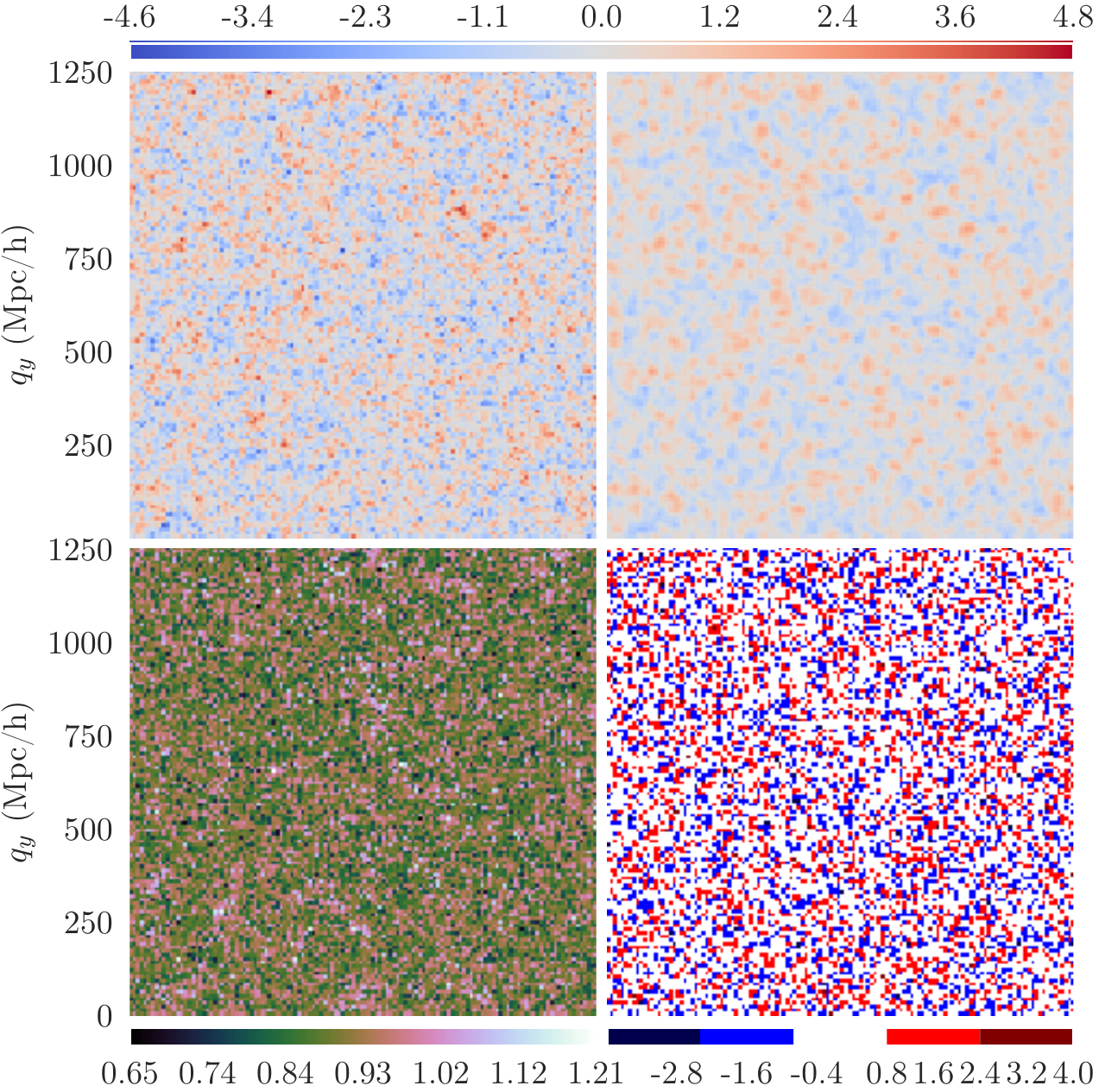}
  \caption{Mean sample \emph{Lagrangian} densities (top right) compared to true (top left), sample std (bottom left) and true minus mean (bottom right).
  }
  \label{fig:den_vstrue_lag_rsd}
  \end{figure*}

\subsection{Large scale structure reconstruction}

\begin{figure}
  \centering
  \includegraphics[width=\columnwidth]{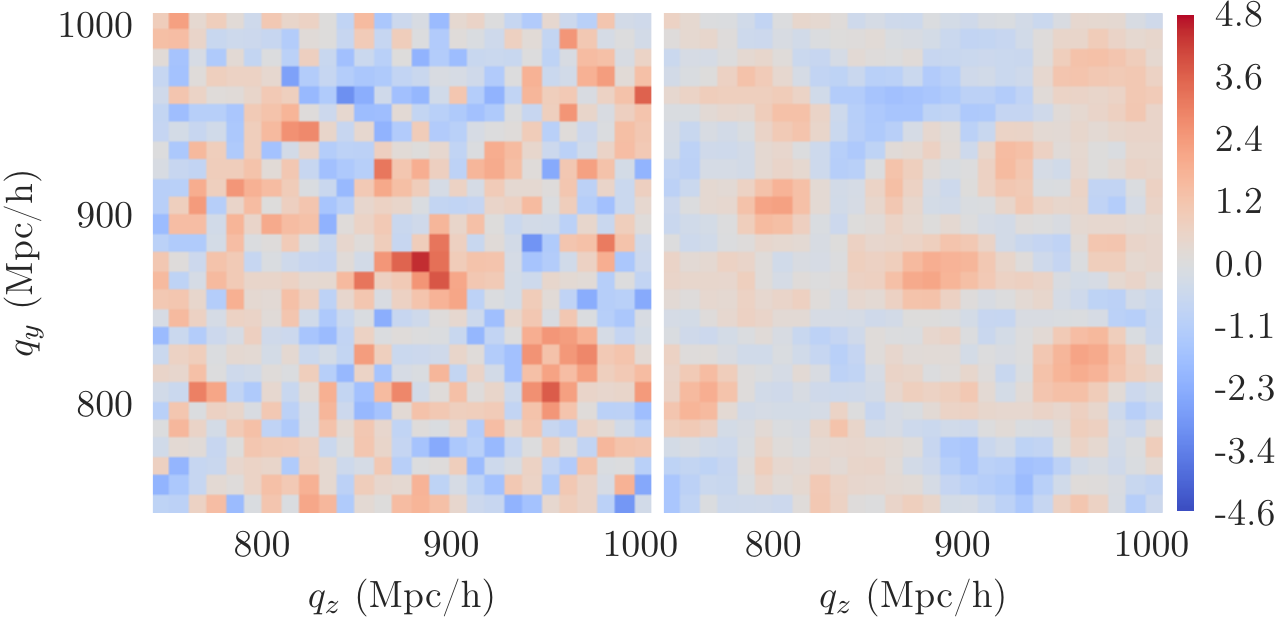}
  \caption{
    Zoom-in of figure~\ref{fig:den_vstrue_lag_rsd}.
  }
  \label{fig:den_vstrue_lag_rsd_zoomin}
\end{figure}

We first inspect the match of the Lagrangian density obtained by the chain samples, after burn-in, to the true underlying density.
Comparing the true field to the ensemble average in the top panels of figure~\ref{fig:den_vstrue_lag_rsd}, we find a good match in terms of the large scale structures.
Many high peaks and low troughs have their counterpart in the samples, even though their exact shapes and heights do consistently differ.
In some cases peaks or troughs are combined, in some they are split and in some cases random peaks or troughs are added that should not exist at all.
\changed{
  These differences and similarities can be appreciated even better in the zoom-in of figure~\ref{fig:den_vstrue_lag_rsd_zoomin}.
}
  
\begin{figure*}
  \centering
  \includegraphics[width=\textwidth]{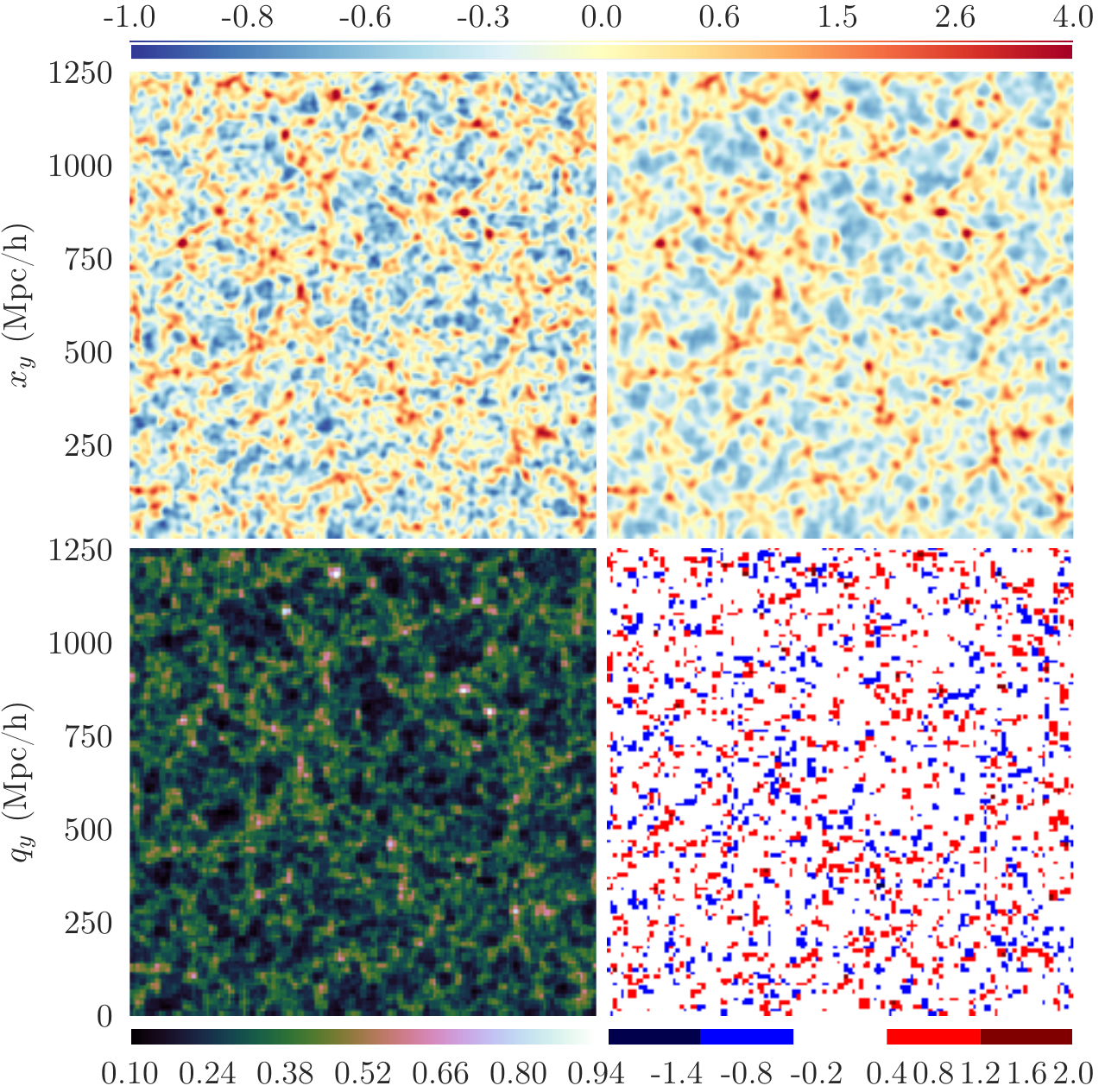}
  \caption{Mean sample \emph{Eulerian} densities (top right) compared to true (top left), together with std (bottom left) and the difference of the true and mean fields (bottom right); with RSD model.}
  \label{fig:den_vstrue_eul_rsd}
  \end{figure*}  

We should note that these kinds of fluctuations are to be expected.
In the individual samples, such fluctuations will be even more pronounced (as we discuss in Section~\ref{sec:density_variation_around_mean}).
In the mean sample, the algorithm is expected to fluctuate around the true density field to a great extend.
However, it is highly unlikely that the mean field will ever exactly match the true field.
This is prevented by the uncertainties that are inherent to all the processes involved in the posterior.

The algorithm not only manages to reconstruct the structures, also quantitatively it works well: the difference between true and reconstructed density is typically on the same order of magnitude as the values themselves.
It leads to a good reconstruction of the density distribution function, but still allows for significant variation.
The standard deviation of sampled densities is typically around 1 (bottom left panel of figure~\ref{fig:den_vstrue_lag_rsd}).
Using the RSS model gives a very comparable amount of variation to when one does not need it, as can be seen by comparing to figure~\ref{fig:den_vstrue_eul_obs}.
In principle, given the added uncertainty in spatial mass distribution configurations along the line of sight introduced by redshift space distortions, one might expect that the variation around the mean would be smaller when using the RSS model.
Stated differently, the degeneracy of the $z$-axis could lead to the ``leaking'' of the MCMC chain of some of its movement in the statistical coordinate space $\posHMC$ to velocity space.
Here we see no such effect, which may indicate that our chosen HMC parameters, in particular $N_\epsilon$, cause a sufficiently uncorrelated sampling process for both cases.
We mention this here explicitly, because in our previous work on higher resolution runs \citep{thesis} we did see a difference in sample variability between runs with and without RSS model.
The sample variation is further discussed in Section~\ref{sec:density_variation_around_mean} \changed{and in Appendix~\ref{ap:mcmc_performance}}.

Following gravitational evolution, the corresponding Eulerian density fields reveal an evolved web.
Our analysis therefore also allows a similarity inspection of the prominent web-like mass distribution features in the Cosmic Web.
On the basis of figure~\ref{fig:den_vstrue_eul_rsd}, it becomes clear that this model performs very well in reconstructing the large scale features that we are interested in.

As expected, there is still some difference, as seen in the bottom right panel.
With the redshift space distortions included in the reconstructions, the outstanding $z$-directionality or bi-polarity that can be noted in the uncorrected process (see figure~\ref{fig:den_vstrue_eul_obs} of Appendix~\ref{ap:results_problem}) have fully disappeared.
Moreover, the variation seems evenly spread out over the entire field.
If the variation were caused by redshift space distortions, one would expect a concentration of differences around clusters and ``great walls''.
As expected, compared to the Lagrangian case, the standard deviation of the samples is somewhat lower, but of the same order of magnitude.
\changed{
  We discuss this further below in Section~\ref{sec:density_variation_around_mean}.
}

\begin{figure}
  \centering

  \includegraphics[width=\columnwidth]{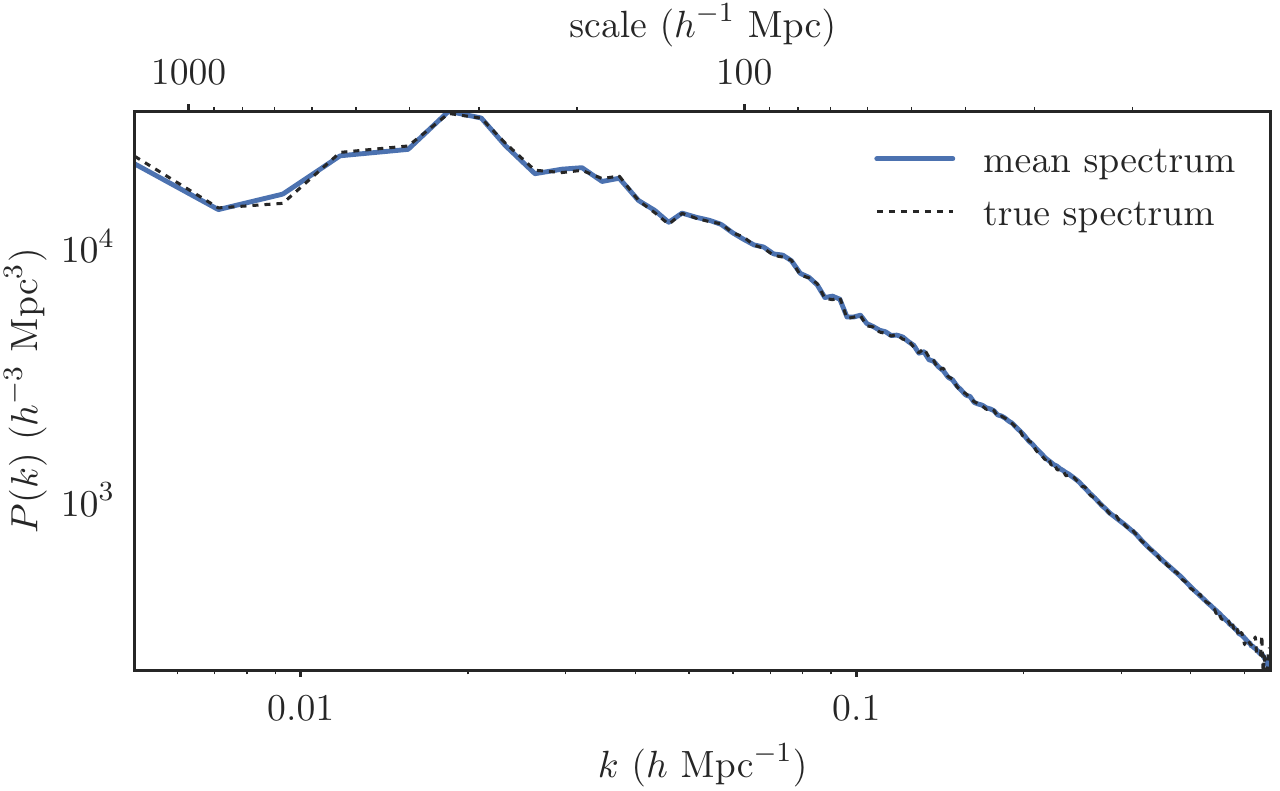}

  \includegraphics[width=\columnwidth]{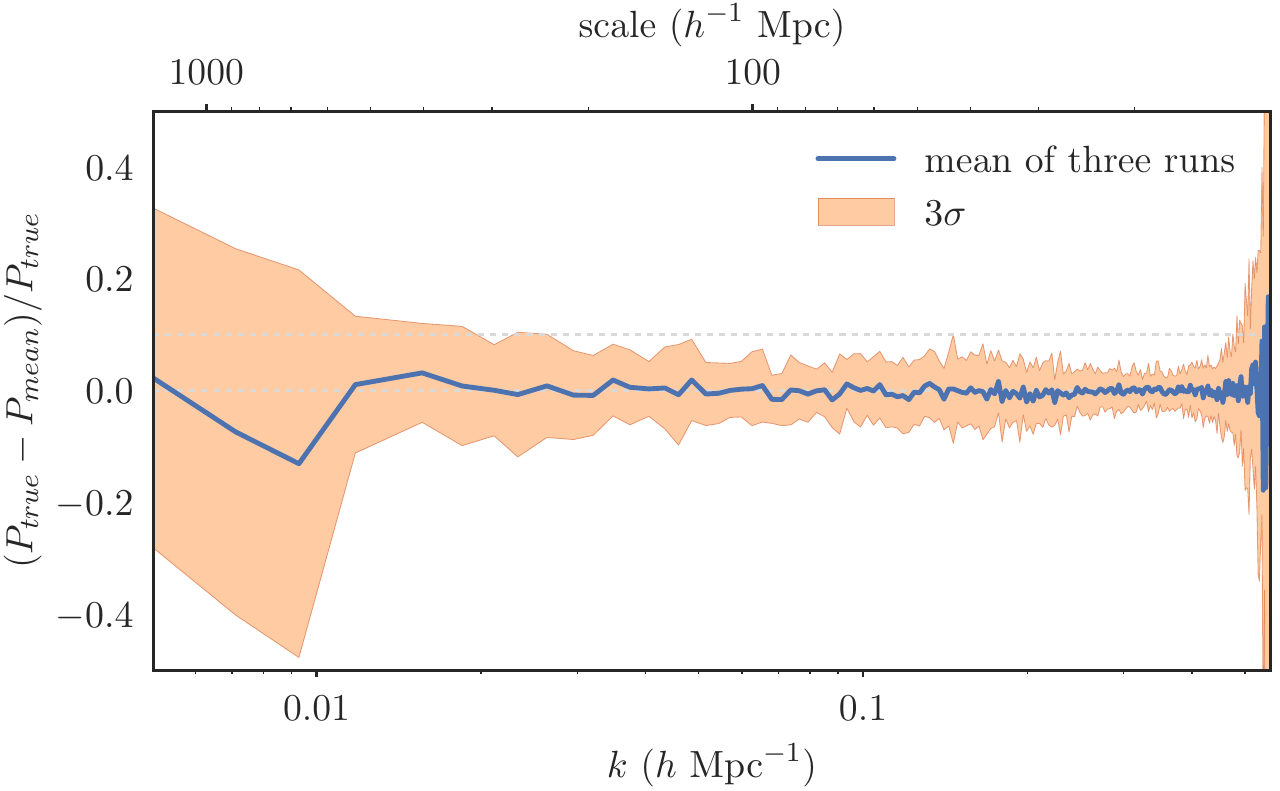}

\caption{
  True Lagrangian power spectrum compared to the mean and variance of the sampled density fields' power spectra \emph{with RSD model}.
  Note that the difference plot (bottom panel) uses samples from three different runs with different true fields, while the power spectrum plot itself (top panel) is only from one true field.
}
\label{fig:ps_vstrue_lag_rsd}
\end{figure}

Looking further at the sample power spectra in figure~\ref{fig:ps_vstrue_lag_rsd}, we find an excellent match to the expected true spectrum.
The difference of the mean to the true spectrum in the bottom panel shows that the Kaiser effect has been completely accounted for.
The deficiency in the reconstructed power, peaking at $k=0.2 h \mathrm{Mpc}^{-1}$, that we discussed in \citet{thesis} is no longer present here, due to allowing the mock observed density distribution to have values $\rho < 0$, so as to properly match the Gaussian likelihood (see Section~\ref{sec:astronomical_parameters}).

\subsection{Density variation around mean}
\label{sec:density_variation_around_mean}

As demonstrated above, the variation of samples around the mean seems similar to that obtained with a regular Eulerian space model \changed{(Appendix~\ref{ap:results_problem} and \citet{thesis})}.
One could have expected the algorithm to have somewhat more trouble evolving the MCMC chain in the redshift space case.
However, as argued above, it seems that in our situation, the leap-frog step size was sufficiently large for the chain to orbit similar volumes of posterior parameter space in both cases.

\begin{figure}
  \begin{subfigure}[b]{\columnwidth}
    \centering
    \includegraphics[width=\columnwidth]{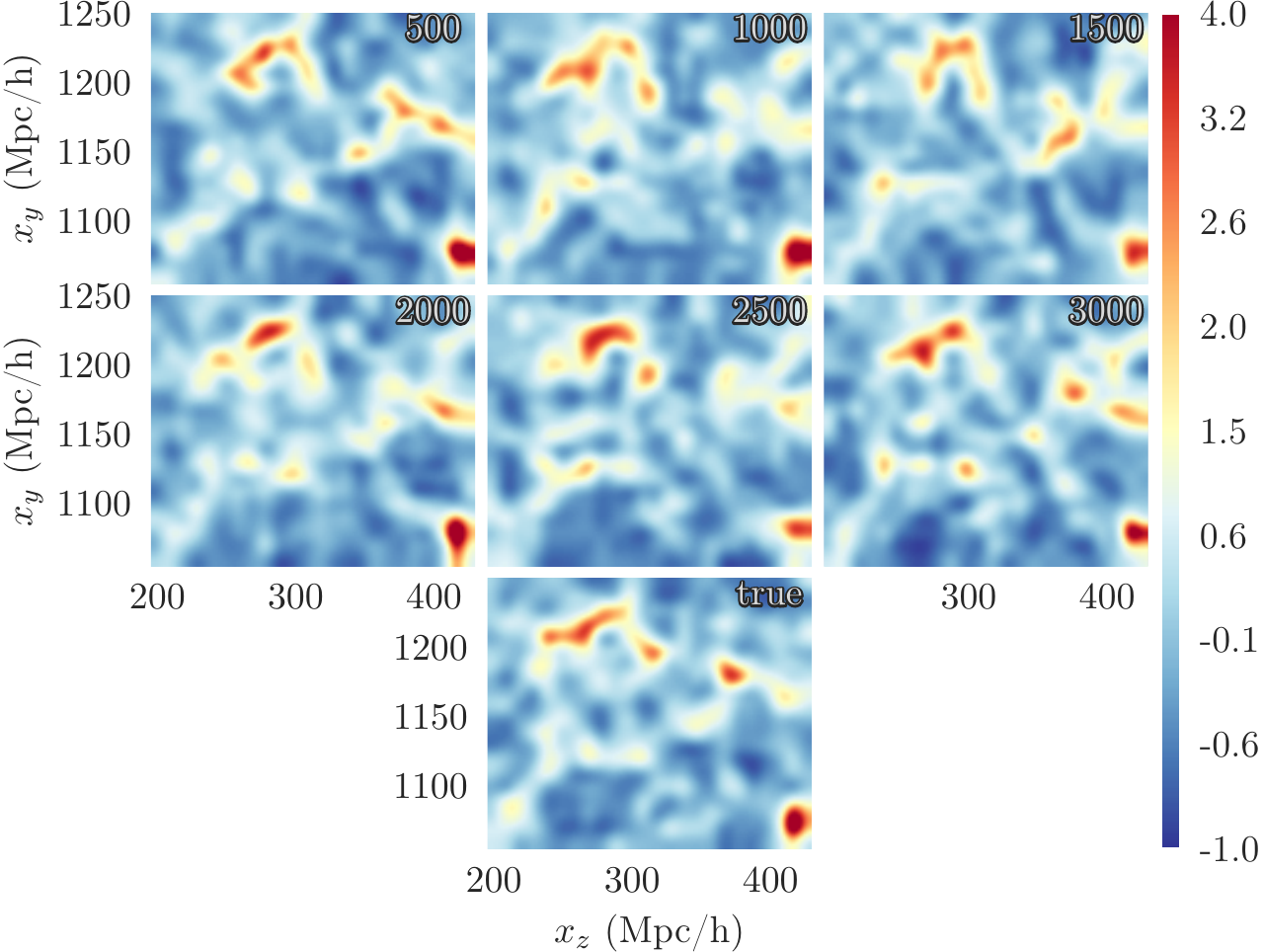}
    \caption{Real space, with RSD model}
    \label{fig:den_eul_zoomin_rsd}
  \end{subfigure}

  \begin{subfigure}[b]{\columnwidth}
    \centering
    \includegraphics[width=\columnwidth]{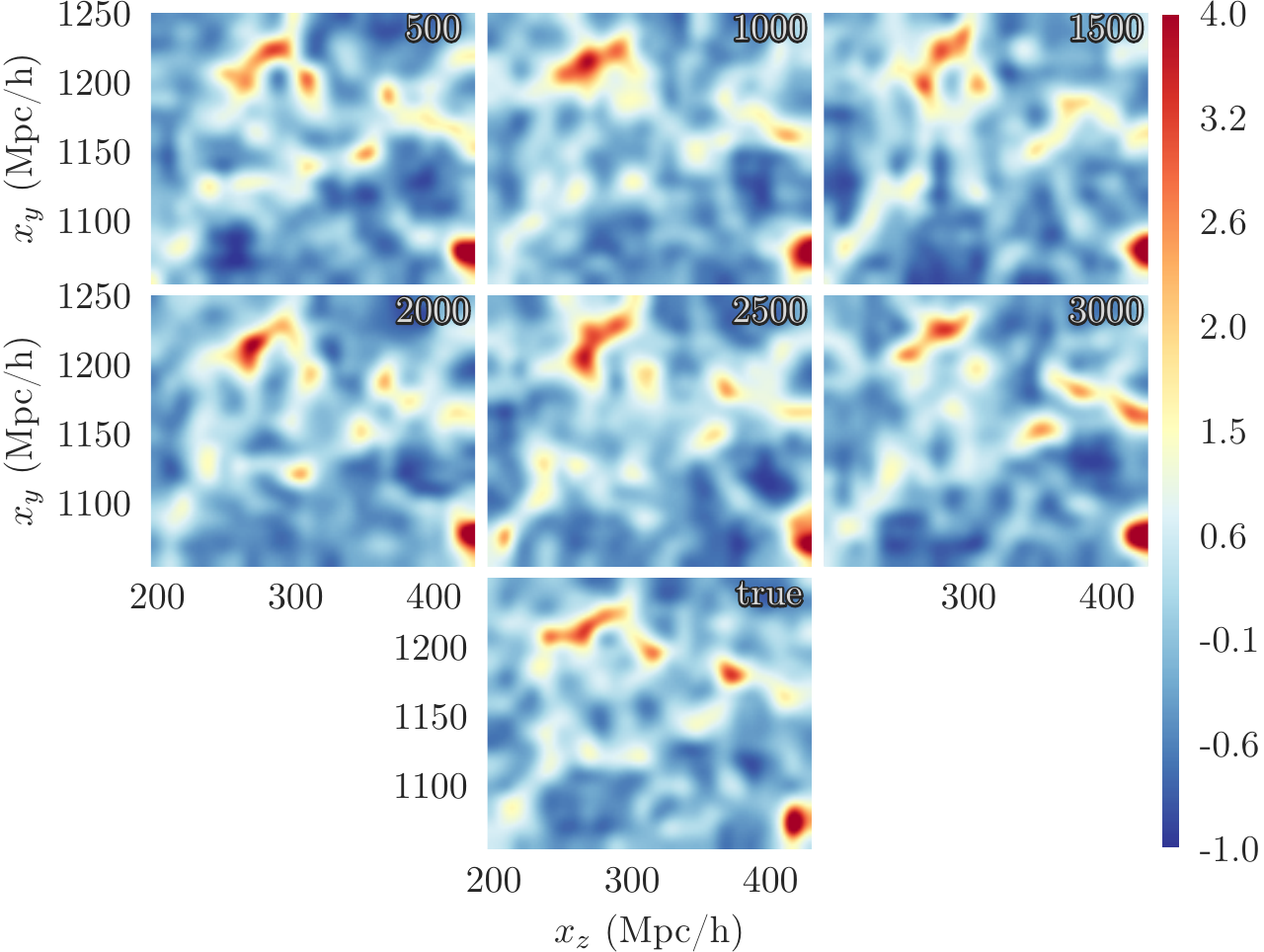}
    \caption{Real space, without RSD model}
    \label{fig:den_eul_zoomin_obs}
  \end{subfigure}

  \begin{subfigure}[b]{\columnwidth}
    \centering
    \includegraphics[width=\columnwidth]{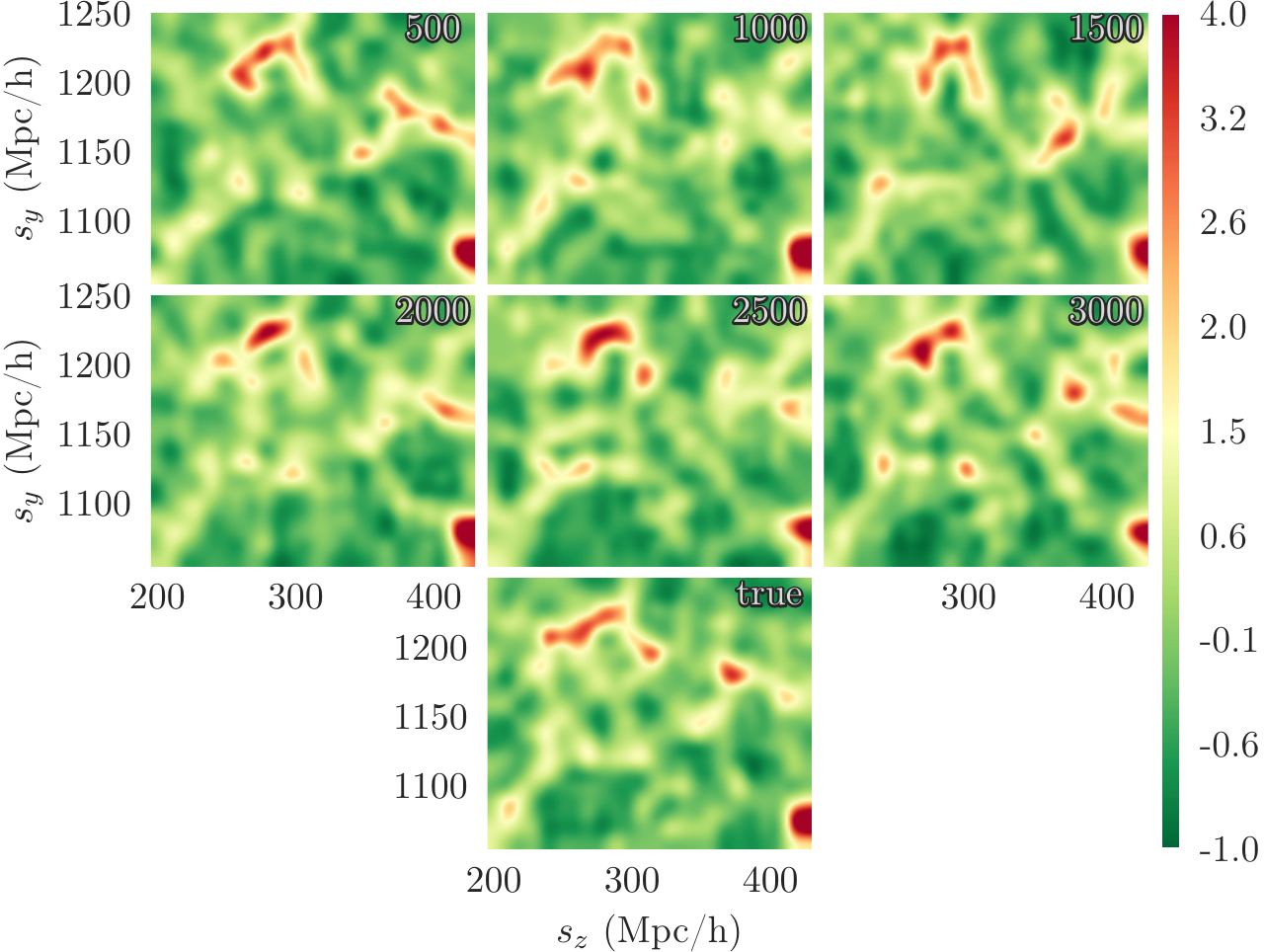}
    \caption{Redshift space, with RSD model}
    \label{fig:den_rss_zoomin_rsd}
  \end{subfigure}

\caption{Zoom-in on Eulerian real and redshift space density fields of iterations 500--3000 (top six panels) and the true field (bottom panel): with and without RSD model.}
\label{fig:den_zoomin}
\end{figure}

In figure~\ref{fig:den_zoomin} we take another close-up look at the evolution of the chain.
The same region of the reconstructed Eulerian density field is shown for six different iterations of the chain in figure~\ref{fig:den_eul_zoomin_rsd}.
These can be compared to the true density field in the bottom panel.
We see here the stochastic sampler at work.
The main large scale features can be identified in all samples, like the very massive peak in the bottom right corner and the filamentary structure in the top left.
The massive peak is rather stable, with some additional protrusions sometimes being formed at random.
The filaments, however, undergo some major variations in shape, orientation and connectedness.

We compare this behavior to that of a regular space run in figure~\ref{fig:den_eul_zoomin_obs} (see Appendix~\ref{ap:results_problem} for details on this run).
While there are clear differences with the redshift space corrected runs of figure~\ref{fig:den_eul_zoomin_rsd}, the results do not seem qualitatively different by eye in terms of amount of structural variation, which is in line with what we discussed above.
At the same time, we see deeper reds in clusters and blues in voids in figure~\ref{fig:den_eul_zoomin_obs} than in the corrected runs.
This is because the uncorrected runs reproduce features that look like redshift space distortion --- amongst which: enhanced density contrast --- but then in real space.
The enhanced contrast can also be seen in the redshift space zoom-ins of figure~\ref{fig:den_rss_zoomin_rsd}, which corroborates this fact.
The uncorrected runs are discussed in more detail in Appendix~\ref{ap:results_problem}.

\subsubsection{Density-density full field comparison}

\begin{figure}
\centering
\includegraphics[width=\columnwidth]{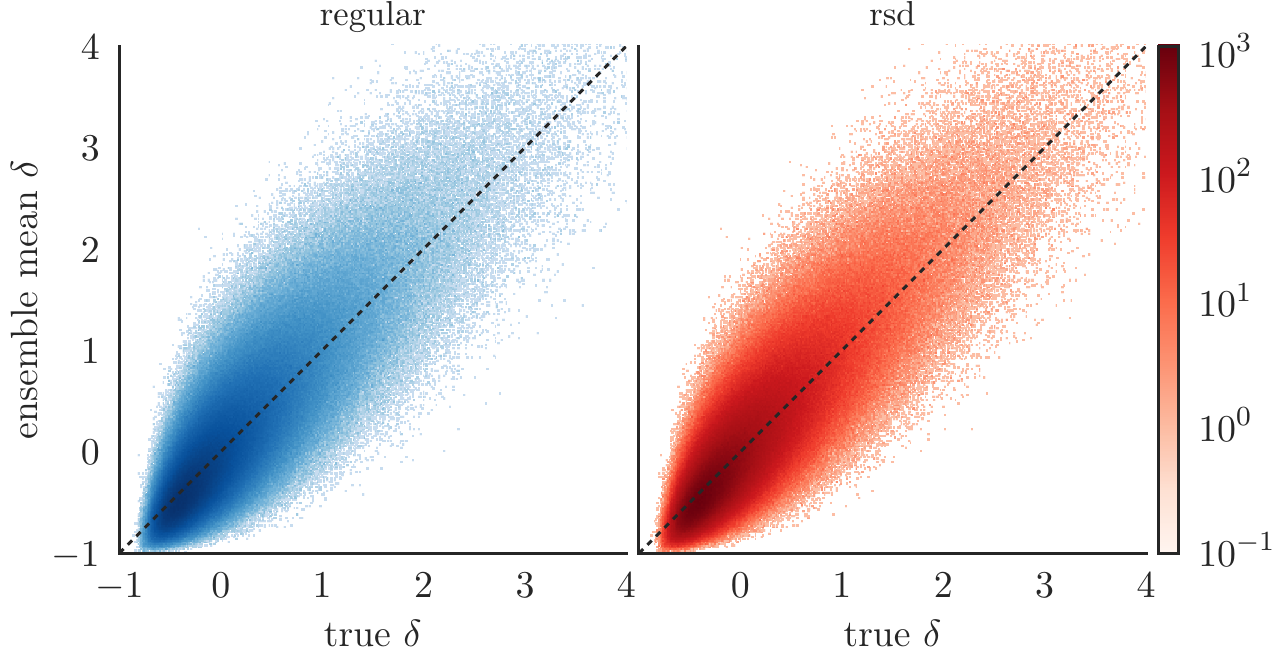}
\caption{
  Histogram of the density of the true Eulerian density field (horizontal axis) versus the corresponding density in the ensemble average of the chain (vertical axis), compared in each grid cell.
  In red the redshift space model and the regular model in blue.
}
\label{fig:denvsden_mean_rsd_reg}
\end{figure}

Figure~\ref{fig:denvsden_mean_rsd_reg} shows, for each grid cell, the density of the true Eulerian density field versus the corresponding density in the ensemble average of the chain.
The right panel shows the results for the redshift space model, the left panel for the Eulerian space model.
These plots corroborate some of our previous findings.
The variation (width of the distribution around the $x=y$ line) is more or less the same, corresponding to the similar variation in the samples we saw before.
Generally, the correspondence is good for both models.

\subsection{2D correlation function isotropy}
\label{sec:2Dcf_isotropy}

Because of the anisotropy of redshift space, RSD effects show up more clearly in directional statistics like the 2D correlation function $\xi(\sigma,\pi)$ (see Appendix~\ref{ap:2D_corr_fct} for more detail).
In $\xi(\sigma,\pi)$, the three spatial directions are collapsed into two.
One uses the directions along and perpendicular to the line of sight.
Anisotropies in the data will show as a deviation from a perfectly circular $\xi(\sigma,\pi)$.
RSDs will thus have a marked signature, see e.g. \citet{hawkins03,peacock01b,guzzo08}.

\begin{figure}
\centering
\includegraphics[width=\columnwidth]{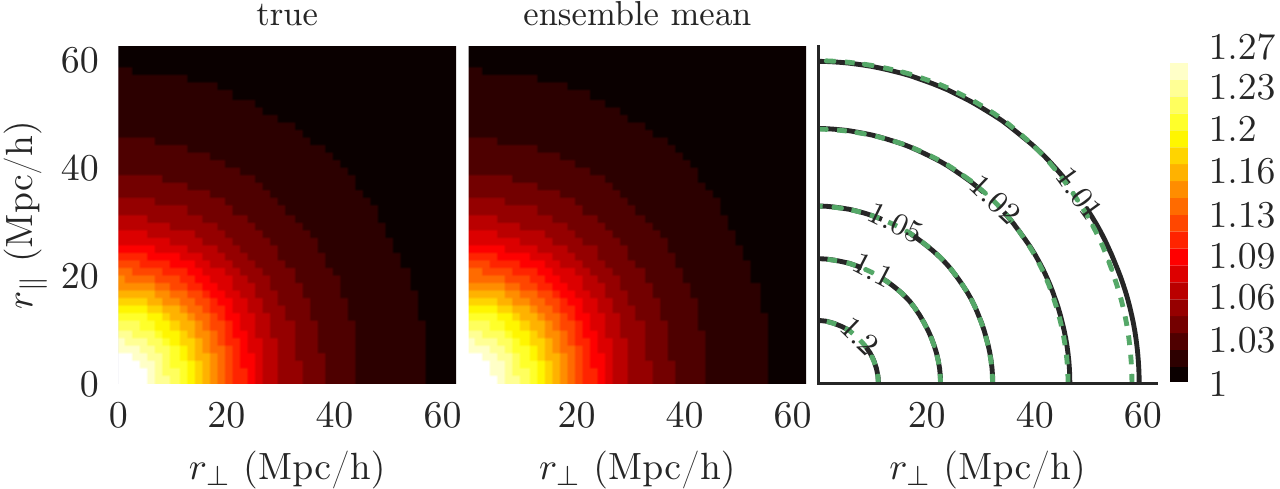}
\caption{
  Eulerian space 2D correlation functions: true (left) vs ensemble mean (center).
  The right-hand panel shows the two functions in contours: solid black for the true function and dashed green for the ensemble mean.
}
\label{fig:2DCF_vstrue_mean_rsd}
\end{figure}

The two-dimensional correlation function $\twodcf$ of the reconstructions matches the true one when using the redshift space model.
This indicates that the large scale redshift space anisotropies have been eliminated\footnote{
  We use the notation $\twodcf$ instead of $\xi(\sigma, \pi)$, replacing $\sigma = r_\perp$ and $\pi = r_\parallel$.
}.
In figure~\ref{fig:2DCF_vstrue_mean_rsd} we compare the correlation functions of the ensemble mean of chain samples and the true field.
The difference between them, $\twodcf_\mathrm{true}~-~\twodcf_\mathrm{mean}$, is maximally $\sim0.001$, with the mean difference being $\sim0.000013$.

\begin{figure}
\centering
\includegraphics[width=\columnwidth]{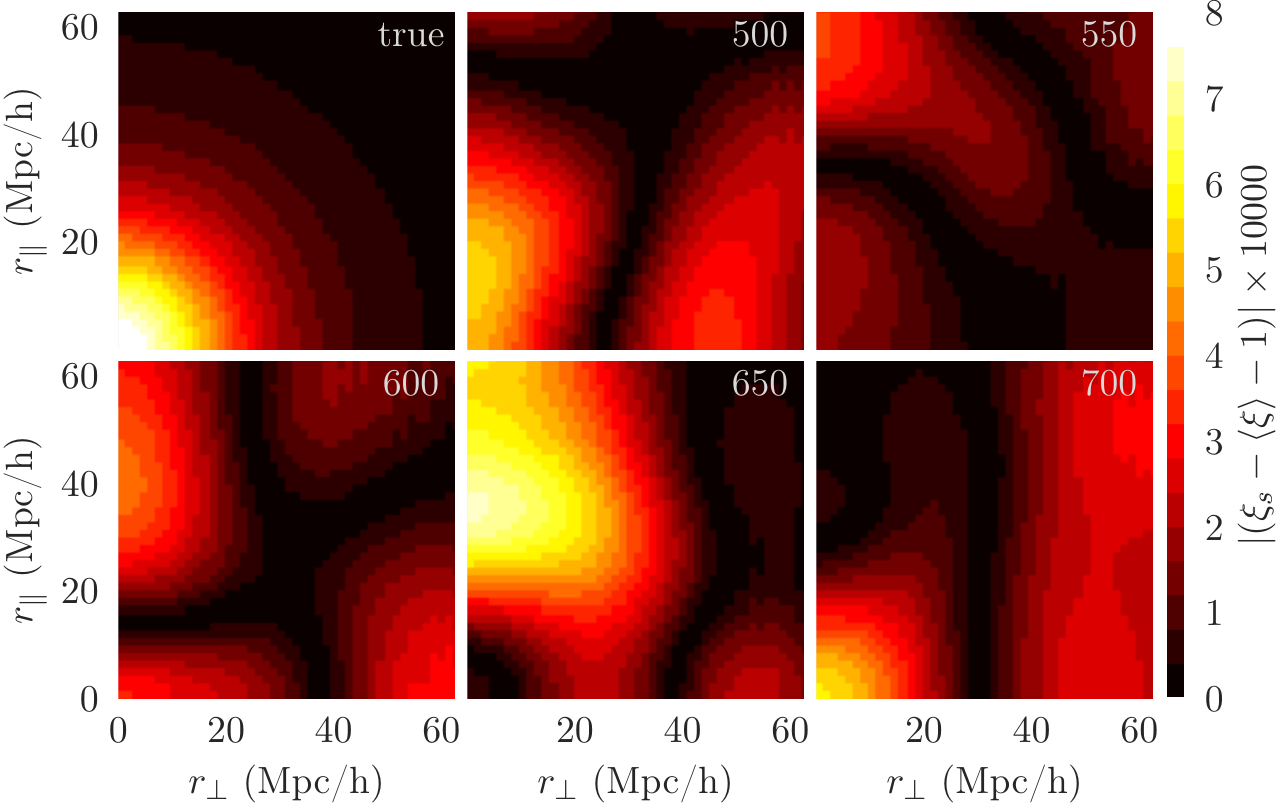}
\caption{Five sample Eulerian 2D correlation functions minus the mean of all reconstructions, compared to the true field in the top left panel.}
\label{fig:2DCF_vstrue_samplediff_rsd}
\end{figure}

\begin{figure}
\centering
\includegraphics[width=\columnwidth]{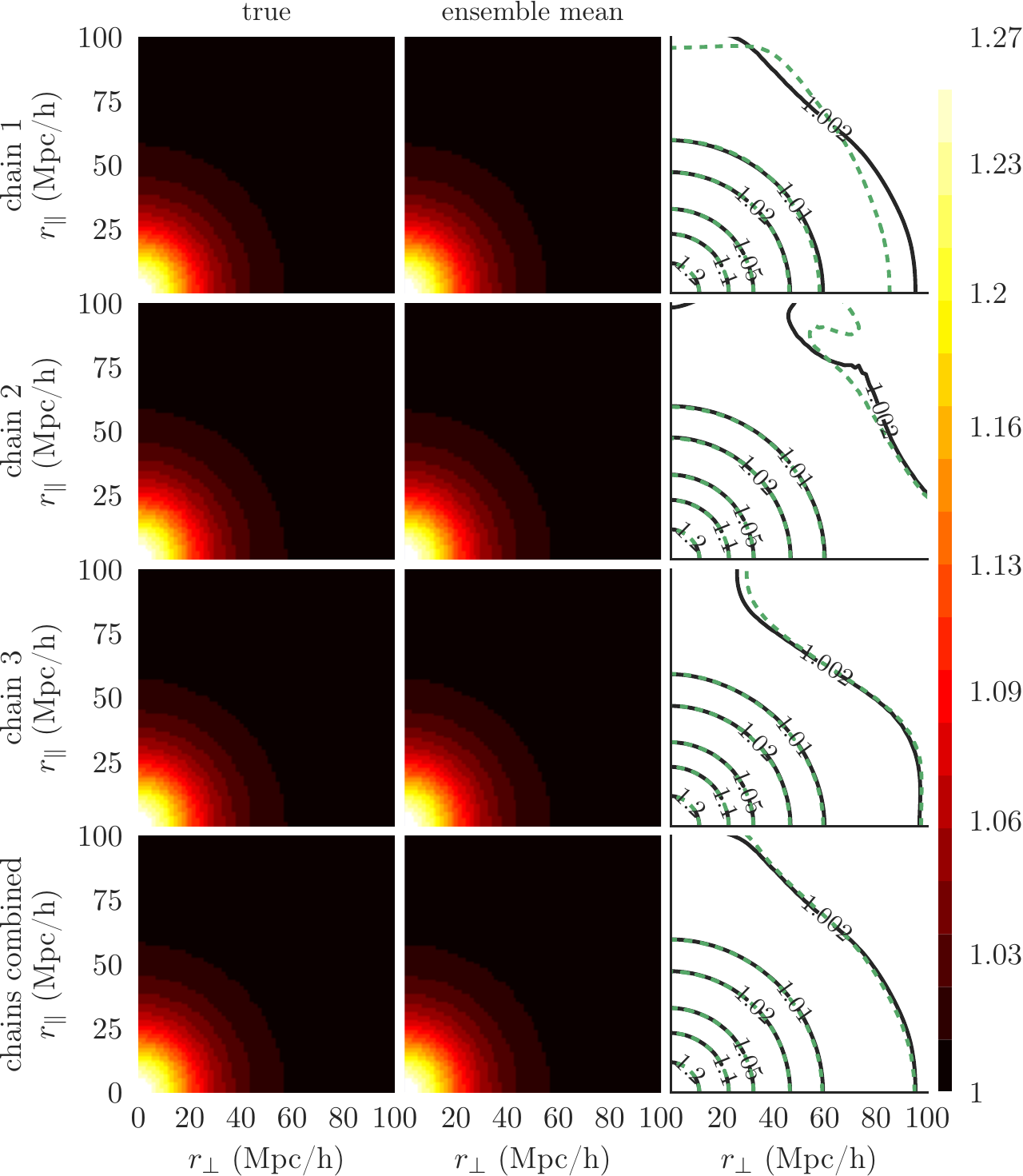}
\caption{
  Sample Eulerian 2D correlation functions compared to true: the first three rows are based on three different true Lagrangian density fields, while the last row contains the average of the three runs.
  The left-hand column shows the true fields, the middle row the mean of the ensemble and the right-hand row compares the two.
  The ensemble mean is represented in dashed green and the true field in solid black lines.
  This sample images clearly illustrate the role of cosmic variance in the anisotropy of the field.
}
\label{fig:2DCF_vstrue_cosvar}
\end{figure}

The nature of the MCMC sampling process itself introduces additional anisotropies in the samples.
No single realization will perfectly match the 2D correlation function of the true field.
This sampling effect is illustrated in figure~\ref{fig:2DCF_vstrue_samplediff_rsd}, where we show the difference of the correlation functions of 5 individual realizations with that of the mean of all realizations.
The differences are very small, on the order of $\delta \xi / \xi \sim 10^{-3}$.
To account for this case-by-case variation, we only look at sample ensembles instead of just at single samples.

Note that for reasons of cosmic variance, some amount of anisotropy is always expected.
A finite sized box will never be perfectly isotropic.
This fact is illustrated in figure~\ref{fig:2DCF_vstrue_cosvar}.
The first three rows are each based on a different true Lagrangian density field, generated with a different random seed.
It is immediately apparent from their deviation from perfect circularity on larger scales that the true fields (left column) contain a considerable amount of anisotropy.
To account for this effect, one should always compare to the true field and take its anisotropies into consideration.
Indeed, it can be seen that the true and reconstructed anisotropies match well.
Alternatively, one could run a large number of chains with different random input fields and average out their statistics.
For the entire Universe we expect a perfectly isotropic condition, reflecting the cosmological principle.
Even averaging over only three different chains does not suffice, as may be appreciated from the bottom row of figure~\ref{fig:2DCF_vstrue_cosvar}.

%% file: inc/sec_discussion.tex

\section{Discussion}
\label{sec:discussion_barcode_rsd}

We developed and tested a self consistent method for the reconstruction of an ensemble of likely primordial density fields based on mock observed data in redshift space.
We showed that this significantly improves the quality of the reconstructions.
They are more isotropic and redshift space artifacts are removed.
This forms a contrast to a naive application of an Eulerian space algorithm to redshift space data, which would retain these features.

The novelty of our method relates to the explicit treatment of redshift space.
We extend existing Hamiltonian MC methods \citep{jaschewandelt13,wang13} to a slightly more realistic data model.
The rationale is that pre-processing data should be kept to a minimum.
Properly modeling the data is always preferable.
This way, all the possible interacting complexities in the data will be properly handled, whereas pre-processing data might lead to additional noise and/or errors.
For instance, to artificially remove Fingers of God one must make a lot of simplifying assumptions, like sphericity of the clusters \citep[e.g.]{tegmark04,tempel12}.
Simple, automated FoG removers will inevitably generate false positives and also miss out on true positives.
In our integrated redshift space approach, no assumptions are needed at all.
This is not only considerably easier, but also seems like the only way that one can ever consistently model all the complexities in the data.

Note that redshift distortions have been treated in a different way by \citet{wang13}.
Their treatment of redshift distortions \citep{wang09} is based on a linear approximation of the velocity field, which they use to correct for the positions of dark matter haloes.
This is a pre-processing step that is done only once and in a unique way, before running their Hamiltonian sampler.
In so doing, the intrinsic uncertainty from using redshift as a proxy for distance is circumvented rather than accounted for.  

In the rest of this section, we will discuss some of the (astro)physical and astronomical applications of our formalism.
The method presented in this work is still not fully complete with respect to modeling all complexities of the cosmic structure formation process, which we will also touch upon.
In addition, a range of technical aspects and open questions will be addressed and we will discuss some further points pertaining to the novel redshift treatment.

\subsection{Regions/environment of applicability}

Using the \barcode\ framework, we aim to use cluster data as input and statistically study the Cosmic Web that forms in between.
All the stochastic effects involved in the evolution of large scale structure and in the observations thereof can be self-consistently included.
The analysis of the resulting ensemble of reconstructions should then be straightforward.
This matter was explored in Chapter~5 of \citet{thesis}.

One point of attention when using this code is the fact that it is biased, and is better tuned towards constraining high density regions than low density regions.
Because many Lagrangian $\bq$ locations will end up in clusters on the real-space $\bx$ or $\zspace$ grid, the real-space constraints in \barcode\ are biased towards more high density regions.
This means that even though you might have put constraints on $\rho^\obs(\zspace)$ in void-like regions, they will be poorly probed.
In a future paper, we will explore this effect in more detail by trying to constrain voids.

Given our aim of using clusters as constraints, this circumstance is actually quite fortunate.
One might, however, want to modify the algorithm to better suit void-region constraints.
One possibility is to sample directly in Eulerian (volume) space $\bx$, as opposed to the Lagrangian (mass) space.
This poses problems with the non-uniqueness of the translation from Eulerian to Lagrangian coordinates.
Another option might be to use $\sigma(\bx)$ to compensate for the voids.
One could lower the variance in those regions, thus making the constraints tighter and more likely to be held in a random sample.
However, it still leaves the possibility of intermediate fluctuations, given that the amount of particles ending up in voids will be low.
The effects of this will also be tested in a future study.
This approach is consistent with that of \citet{wang13}, who define the variance as
\begin{equation}
    \sigma(\bx) = \mu \delta^\mathrm{obs}(\bx) \,,
\end{equation}
with $\mu$ a constant; i.e.\ simply a linear function of the density.

\subsection{Future steps towards real data}
\label{sec:discussion_steps_towards_real_data}
A number of improvements may be identified that would enable our formalism to improve the accuracy, reliability and level of realism of our reconstructions.
For future work, especially for applying the formalism on real observational data, we here list the most significant ideas towards improvement.

\subsubsection{Rank ordered density value correction}
\label{sec:discussion_rank_ordering}
When one wants to compare true density fields to fields derived with LPT, one of the first aspects encountered is the difference of one-point density distributions (PDFs).
Perturbation theory approaches do not prevent particles from pursuing their initial track.
Beyond shell crossing, it leads to the overshooting of the position they would take if the mutual gravitational interaction would have been taken into account in the calculations.
True gravitational clustering will give much higher peak density values than LPT, where this artificial shell crossing artifact creates ``fuzzy'' clusters.
A simple 1st order correction to this would be to apply a rank ordered substitution of density values.
A transfer function from a ``LPT Eulerian space'' density to, for instance, a ``$N$-body Eulerian space'' density can be constructed.
\citet{leclercq13} did this by comparing the rank order of density values in LPT and $N$-body densities at $z=0$ (given the same initial conditions) and reassigning densities in one or the other in such a way that the density value PDFs match after the procedure.
This way, at least one is comparing apples to apples for as far as 1st order statistics are concerned.

\subsubsection{Accuracy of gravitational modeling}
We have shown that in the perfect situation of a model that exactly describes our (mock) reality, our algorithm almost perfectly reconstructs the desired mean field and its statistics.
Although the used models describe cosmic structure well on large scales, they are far from perfect for describing non-linear regime structures.
There are a few models that would give a more complete picture of gravitational evolution.

The current LPT (Zel'dovich) based framework is easily adapted to using second order LPT (2LPT) or ALPT \citep{kitaurahess13}.
2LPT has the advantage of matching even better on the largest scales.
At cluster scales 2LPT is more affected by artificial shell crossing than Zel'dovich, leading to ``puffy'' structures.
The latter can be fixed by combining 2LPT with a spherical collapse model on small scales.
This is what ALPT accomplishes.
Both models are fully analytical, so that they can be implemented in the same manner as the Zel'dovich approximation described in this work.
In Appendix~\ref{ap:2lpt_sc} we work out the equations necessary to extend \barcode\ with 2LPT and ALPT structure formation models.
A similar option would be to apply the Multiscale Spherical Collapse Evolution (MUSCLE) model \citep{2016MNRAS.455L..11N}.
This analytical model was shown to perform slightly better than ALPT, especially when combined with 2LPT on large scales.

It would be even better if we could use an $N$-body simulation as our structure formation model.
\citet{wang14} indeed for the first time successfully invoked a particle mesh $N$-body algorithm in this context.
The particle mesh equations are analytical, so every single particle mesh evolution step can be derived to the signal.
By writing the derivative in the form of a matrix operator and combining subsequent particle mesh time steps by means of matrix multiplications, the full likelihood force can be analytically derived.
By means of adjoint differentiation, the large matrices can be efficiently multiplied and the computational cost of this method stays within reasonable limits.
The resulting model accuracy using only 10 particle mesh steps is remarkably high.
When one needs high accuracy on small scales, this seems like the way forward.
More recently, this approach was also adopted by \citet{2018arXiv180611117J}.

Possibly, the method by \citet{wang14} could be augmented by using an $N$-body solver that also has \emph{baryonic particles}.
Whether this is analytically tractable within an HMC framework remains to be investigated.
Another interesting extension might be to employ the COLA method \citep{tassev13,tassev15} as an alternative gravitational solver.
COLA combines the LPT and particle mesh methods, trading accuracy at small scales for computational efficiency.
It yields accurate halo statistics down to masses of $10^{11} \msun$, which would be more than sufficient for cluster studies.
In fact, the COLA method has already found uses in the context of Bayesian reconstruction \citep{leclercq15,leclercq15thesis}, but in these cases COLA was applied after the Bayesian reconstruction step, not self-consistently within the model like in the work of \citet{wang14}.

\subsubsection{Galaxy biasing}
As, in practice, observations concern galaxies, stars and gas, instead of dark matter, it is of key importance to address the issue of in how far the galaxy distribution reflects the underlying dark matter distribution \citep[see][for a recent review]{2018PhR...733....1D}.
For the application to clusters observed in X-ray, SZ or via weak lensing, this is not directly necessary, since for clusters the biasing problem is far less.
\citet{schaller14} showed that the ratio of halo mass in $N$-body simulations with and without baryons is $\simeq 1$ from mass $\simeq 10^{13.5} \msun$ upwards.
It drops off towards $\simeq 0.8$ for galaxies.
Similarly, the halo abundance in cluster-sized haloes was shown to be similar in simulations with and without baryons.
However, we might want to extend the algorithm towards a formalism capable of processing the galaxy distribution \citep[like e.g.][]{wang13,leclercq15b}.

A natural way to solve this would be to incorporate gas particles in the structure formation model.
However, this would still be a mere approximation, due to the complexities of baryonic astrophysical processes.
A more statistical approach would be to explicitly incorporate biasing prescriptions \citep[see e.g.][and references therein]{ata15,kitaura15}.
We have implemented such models in \barcode\ and will explore their use further in an upcoming work.

\subsubsection{Masking and selection functions}
One of the more observationally oriented issues is that of masking and selection functions.
In our general Gaussian likelihood (equation~\ref{eqn:log_likelihood}), we included a weight factor $w_i$.
This can be set to zero if nothing is known about cell $i$ in the data and to one if there is.
Another option is to set it to zero if the area is observed, but was found to be empty (at least up to the depth of the observation).
Either of these constitute a very basic form of masking.

Given the nature of observations, one could think of more advanced selection mask treatments.
For instance, we should take into account the depth of a null-observation and the fact that this implies a \emph{lower limit} to the density, not an absolute absence of mass.
One could implement this by setting $w_i=0$ if $\rho^\obs < \rho^\mathrm{cut}$ for some cut-off density (per cell).
A theoretically more sound model may be to use a Heaviside step function convolved with a Gaussian kernel (i.e.\ an error function, plus one, divided by two) with a step at the cut-off density $\rho^\mathrm{cut}$.
This reflects the fact that there is some uncertainty about whether the density is above the cut-off value that we derived from the depth of our observation.
It further tells us that we are quite sure it is not much higher and that it is most likely lower than the cut-off value.

Such a selection function will give an additional $\delta(\bq)$ dependent term in the Hamiltonian force.
The derivative is simply a 1D Gaussian distribution.
This implies that it should be straight-forward to derive and should give only minimal extra overhead in the algorithm.

These masking treatments will be explored further in an upcoming work.

\subsubsection{Statistical aspects of the implementation}
Another parameter in our Gaussian model we hardly gave any attention is the $\sigma_i$ cell-wise dispersion of the data.
It is highly likely that a more suitable value can be found than the one we used ($1$ for all cells).
In general, the statistical model in the form of prior and likelihood that we use in our formalism might be improved upon.
It seems like the prior is well defined and encapsulates what it must in a proper way.
The likelihood is a different story.

Other authors have mainly employed Poissonian data models \citep{kitaura08,wang13,jaschewandelt13} instead of the Gaussian likelihood we used.
The way to find out what model should be used is to meticulously map out the errors and systematics involved in the observations.
One particular example of this is that of defining a likelihood model for an observational dataset of X-ray clusters.
In this case, we have to take into account the following aspects:
\begin{itemize}
  \item X-ray photon counts, which have Poissonian errors;
  \item X-ray photon energies, which are also Poissonian due to CCD photon counts;
  \item the propagation of photon errors into a $\chi$-squared fit for the temperature and density profile parameters, e.g.\ using a $\beta$-model \citep{king72,sarazin86,mulchaey00}; if the confidence ranges of such parameters are estimated with bootstrapping, that could lead to complicated likelihood shapes for the density profile parameters;
  \item many systematics arising from the use of certain fitting models for density estimation, like cool cluster cores which are not included in simple $\beta$-models \citep[e.g.][]{1996A&A...305..756S};
  \item the combination of the density/mass profile errors with errors in the distance estimators, e.g.\ from redshifts of the brightest cluster galaxies or some weighted average of all clusters' galaxies.
\end{itemize}
Our Gaussian likelihood was chosen as a general error model that is likely to be close to (but not equal to) many true data likelihood functions.
This approach is, in fact, not uncommon in the literature for X-ray cluster density error estimation.
In \citet{eckert11} an error in the density profile of a cluster is estimated using a Monte Carlo approach, by simulating $10^8$ realizations (assuming Poisson statistics) and taking the variation in values as errors.
In \citet{samsing12}, an MCMC sampling of the parameter space is done to characterize the probability function around the best fitting model.
Subsequently, the width of the projected PDFs for every parameter were characterized by the root mean square, meaning that essentially they are approximating the error on the parameters as Gaussian, i.e.\ they leave out higher order PDF characteristics.
For the NORAS survey, \citet{boeringer00} assume statistically independent, Gaussian distributed errors, even though they state this is not 100\% correct.
Finally, many authors use the $\chi^2$-statistic for determining the goodness-of-fit in their analyses.
This statistic is also based on a Gaussian distribution of noise.

The HMC algorithm is quite sensitive to deviations of the data model from the actual data errors.
Using a wrong model can lead to poor reconstructions, as we found in \citet{thesis}.
In that work, we found that the power spectrum was not properly reconstructed, deviating from expected values by as much as 10\% in some ranges.
For this work, to show the potential of the redshift space correction model, we accommodated for the Gaussian likelihood by allowing negative densities in the mock density fields produced by adding Gaussian noise to the Eulerian redshift space density field derived from the ``true'' Lagrangian initial density field.
In other use-cases, when reconstructions deviate significantly from expectations, one could use this as an indicator of model deficiencies.

\subsubsection{RSDs and LPT}
\label{sec:discussion_rsds_and_lpt}
The use of Lagrangian Perturbation Theory imposes two major limitations on a redshift space distortion treatment.
Because of the inaccuracy of the non-linearities in the velocity field, Fingers of God and triple value regions are not accurately represented.
However, just as clustering in density was augmented in the ALPT model, one could try to think of ways to augment LPT to better represent these non-linear density features.

The virialized motion of FoGs is \changed{prohibitively} hard to model as a direct function of the signal (which is necessary for the derivative of $P(s)$).
It is possible to fix this to some extent by adding a dispersion term to the velocity field model \citep{hess13,kitaura14}.
However, this implementation needs an extra stochastic step to sample from a Gaussian velocity distribution (the mean and dispersion of which depend on the local density).
This breaks the philosophy of \barcode, in which we sample from one posterior that integrates all the available models.
It might be possible to find a way to add a similar model to our posterior.
This is a topic of ongoing work.

Another possible way would be to split up very high density particles into a distribution of smaller particles with a virialized velocity distribution.
An easy way to identify these particles would be to use the spherical collapse criterion of the $\divdisp = -3$ lower limit.
To simplify matters, one could collect clustered particles and collectively split them into one big virialized kernel.
However, this would then destroy any remaining substructure and shape.

The triple value regions may require a similar augmentation as the virialized motion, but maybe a simpler form is sufficient.
It is possible that in ALPT these regions are already partly modeled.
However, whether the velocities are also correct in ALPT should be further investigated.
If so, this would indeed solve the triple value region problem.

\subsubsection{Theoretical considerations}
In the relativistic limit, the correspondence between our theoretical redshift space definition and observational redshifts breaks down.
It might be worth looking into whether a relativistic redshift space mapping could be fashioned.
One would ideally also include in this the proper cosmological distance instead of the Hubble approximation that we usually make.
Such a mapping may lead to rather small corrections on small scales.
However, on the largest possible scales, quadratic and higher order terms must be taken into account, as they become dominant.
Furthermore, one should start thinking about relativistic structure formation equations.

Within the limits of the framework described in this work, the above cosmological distance arguments are moot.
A far more important consideration would be that at some point the data that is used can no longer be assumed to be in the same evolutionary state.
Conceivably, LPT should be adaptable to this.
The calculation of a field's evolved state at one point in time is just as easily done as another point in time.
One could then imagine that each particle in the field is simply evolved to a different point in time, depending on where it ends up.
The MCMC nature of the model should be able to iteratively find optimal values for this.

In Section~\ref{sec:method}, we derived the redshift space equations necessary for an application in a fully non-plane parallel redshift space.
Subsequently, we neglected these terms, due to our plane-parallel approximation.
We have not implemented nor tested the additional terms that the full treatment entails.
However, this can easily be done.
The algorithm will then become fractionally slower, but not significantly.
For real galaxy redshift catalogs the plane-parallel approximation can not be used.
When one wants to use these to reconstruct real Universe densities, one must implement the full redshift space equations.

%% file: inc/sec_conclusions.tex

\section{Conclusions}
\label{sec:conclusions}

Galaxy surveys provide information of the luminous matter distribution, such as galaxy, or cluster catalogs. They are affected by selection effects, yielding an incomplete biased and noisy sample of the underlying nonlinear matter distribution. In addition, the tracers are affected by their peculiar motions causing redshift-space distortions (RSDs). A proper analysis of the cosmological large scale structure should account for all these effects. To this end, Bayesian methods have been developed in the past years connecting the observed data to the primordial density fluctuations, which summarize all the cosmic information for a given set of cosmological parameters and structure formation model.

In this work, we have presented \barcode, an algorithm for the analysis of the large scale structure, which for the first time self-consistently solves for coherent RSDs within an analytical Bayesian framework.
The contribution of this work in the context of Bayesian techniques relies on the analytical derivation of the Hamiltonian equations taking into account the transformation of the biased tracer to redshift space.
We present its numerical implementation and a number of tests.

This method could be extended to deal with virialized motions by including a random term accounting for it in the equations, or by a prior fingers-of-god collapse, as suggested in a number of works.

Here we have chosen the grid-based Bayesian approach, which can easily account for survey mask and geometry of a typical galaxy survey. 
We have restricted the numerical study to the Zel'dovich approximation, but have made derivations including the tidal field tensor within second order Lagrangian perturbation theory and small scale spherical collapse based corrections.  
This approach could also be extended at the expense of a higher computational cost to particle mesh solvers.

From the detailed analytical derivations and numerical tests here presented, we draw the following main conclusions:
\begin{enumerate}
  \item Using our self-contained redshift space model we can overcome large scale redshift space distortion effects in observations and reconstruct the true densities to a great degree of accuracy.
  \item When our model is not applied, but rather a naive model based purely on Eulerian real space is used, the redshift space distortions create an anisotropic imprint on the real space reconstructions (Appendix~\ref{ap:results_problem}). We propose to call this effect the \emph{Kaiser effect echo} or Kaiser echo for short.
  \item We have demonstrated that \barcode\ yields unbiased initial density fluctuations from a biased tracer of the dark matter density field in redshift space in terms of the cell-to-cell correlation of the true and reconstructed density field, the respective power spectra and the 2D correlation functions.
  \item In particular, we find that the features of the power spectra characterized by the cosmic variance of the considered volume are recovered in detail solving for the Kaiser factor and beyond (as coherent flows have also an impact on non-linear RSDs), and that the 2D correlation functions are isotropized after applying our method.
  \item Component-wise scrutiny of the Hamiltonian likelihood force in redshift space (Appendix~\ref{ap:hamiltonian_force_zspace}) shows that it contains an unexpected non-radial term that points in the direction of the displacement field and has an amplitude that grows inversely with distance to the observer. This component may play a strong role when the plane parallel approximation is abandoned. This remains to be investigated.
  \item Our adaptive leap-frog time step method was shown to give good MCMC chain performance, yielding uncorrelated samples with a high, stable degree of structural variability, showing that the chain successfully explores the posterior parameter space of initial conditions that match the input (mock) data.
  \item The number of leap-frog steps per iteration of 256 is sufficient for good chain performance and may even be reduced slightly for this set of parameters (box size, resolution, etc.).
  \item The redshift space model constrains the sampler less strongly, which means the chain has to progress through the posterior space slightly slower than a regular space model. Our adaptive $\epsilon$ scheme automatically takes care of this given a large enough $N_\epsilon$ value. When $N_\epsilon$ is too small (as in \citet{thesis}) subsequent samples will be more strongly correlated and the sampler will cover a smaller part of the parameter space.
\end{enumerate}

This code can find applications in a large variety of cosmological problems, such as baryon acoustic oscillations reconstruction or cosmic web reconstruction from the Lyman alpha forest, galaxy or cluster distributions. 
The here presented code, which is made publicly available\footnote{
  Our source code under MIT license can be found at \changed{\url{https://ascl.net/1810.002}} and \url{https://github.com/egpbos/barcode}.
},
thus has the flexibility to tackle realistic large-scale structure analysis from galaxy cluster survey data.

%% file: inc/sec_acknowledgments.tex

\section*{Acknowledgments}

We are grateful for the opportunity offered by IAU Symposium 308 ``the Zeldovich Universe'' (June 2014) to present the details of the code and formalism described in this paper\footnote{The talk can be found at \url{https://doi.org/10.5281/zenodo.1451723}.}.
\changed{
  We express our gratitude to the anonymous reviewer for their critical reading of the initial manuscript and their many incisive questions and suggestions, which no doubt have improved this paper significantly.
}
PB and RvdW thank the Leibniz-Institut f\"ur Astrophysik Potsdam for the great hospitality during the various work visits that defined the basis for this publication.
FSK thanks the Karl-Schwarzschild-Fellowship funding for allowing him to regularly invite PB to Potsdam in this period.
PB further thanks his eScience Center managers Jisk Attema and Rob van Nieuwpoort for supporting the final editing of this manuscript during the past four years.
We thank Jelle Kaastra for the encouragement and motivation given during (mainly the early, defining stages of) this project, and Johan Hidding, Maarten Breddels and Bernard Jones for many interesting and encouraging discussions.
PB acknowledges support by the NOVA project 10.1.3.07.
We would like to thank the Center for Information Technology of the University of Groningen for their support and for providing access to the Peregrine high performance computing cluster.
For the analysis and the figures created for this paper we used a number of scientific software libraries, most importantly: Matplotlib \citep{Hunter:2007} and Seaborn \citep{michael_waskom_2018_1313201} for visualization, NumPy and SciPy \citep{scipy} for numerical computation and IPython \citep{PER-GRA:2007} and Jupyter \citep{Kluyver:2016aa} for interactive analysis.
\barcode\ itself was written in C\texttt{++}, making use of the FFTW3 \citep{Frigo05thedesign} and GSL \citep{gsl} libraries.

%% file: inc/ap_barcode_summary.tex

\section{Barcode: Hamiltonian Monte Carlo sampling algorithm}
\label{ap:barcode_summary}

\changed{In this section we briefly recap the basic concepts, quantities and formulas behind \barcode\ that are necessary for this paper.
See Section~\ref{sec:barcode} for a general overview of the components described in more detail in this appendix.
Since the HMC algorithm, specifically in cosmology, has also been explained in great detail in previous works \citep{neal93,jaschekitaura10,jaschewandelt13,wang13,thesis}, we keep it short here.}

\subsection{Hamiltonian Monte Carlo terminology}
\label{sec:hmc_terminology}

The Hamiltonian $\hamHMC$ is the sum of the potential energy $\potHMC$ and the kinetic energy $\kinHMC$,
\begin{equation}
\hamHMC = \potHMC + \kinHMC \,.
\label{eqn:hamiltonian}
\end{equation}
The kinetic term in this is the momentum (row-vector) times the inverse of the mass matrix $\masHMC$ times the momentum (column-vector),
\begin{equation}
  \kinHMC = \frac12 \momHMC^T \masHMC^{-1} \momHMC \,.
  \label{eqn:kinetic_term}
\end{equation}

The potential $\potHMC$ is coupled to the probability distribution function (PDF) of our signal $\delta(\bq_i)$.
We define this PDF in a Bayesian way as the posterior $P$, which is the product of a prior $\prior$ with a likelihood $\likeli$\footnote{
  Note that the notation we use is not necessarily the most common in Bayesian literature.
  For instance, \citet[page 89]{jaynes03} uses a notation that would translate in this work to something like $L(\delta(\bq)) = P(\delta^\obs(\bx) | \delta(\bq))$, where $L$ is the ``likelihood'' of the hypothesis and $P$ is a probability function that describes it in terms of the random variable $\delta^\obs(\bx)$ and conditional variable $\delta(\bq)$ (the hypothesis).
}.
The connection between the potential and the PDF is given as a canonical ensemble distribution.
In logarithmic form, this gives us an equation for $\potHMC$ in terms of the posterior $P$:
\begin{equation}
  \potHMC = \mathrm{constant} - \ln P = \mathrm{constant} -\lnP -\lnL \,,
  \label{eqn:potential_energy}
\end{equation}
so $-\lnP$ and $-\lnL$ are the terms we need our (statistical) model to specify so that we can calculate $\potHMC$, up to a constant.
However, as we will see below, we only need to know $\delta \hamHMC$ or the derivative of the potential.
We can therefore forget about the constants.

In this work we use a Gaussian prior and likelihood:
\begin{equation}
 -\lnP = \mathrm{constant} + \frac12 \sum_{i,j} \delta(\bq_i) {(S^{-1})}_{ij} \delta(\bq_j) \,,
 \label{eqn:log_prior}
\end{equation}
\begin{equation}
 -\lnL = \mathrm{constant} + \frac12  \sum_i \frac{ {\left[T(\rho^\obs(\bx_i)) - \rho(\bx_i)\right]}^2 w_i}{ \sigma_i^2 } \,.
 \label{eqn:log_likelihood}
\end{equation}
These are two different types of Gaussian exponents.
The likelihood one is a simple sum of one dimensional Gaussians; one for each grid cell.
The prior exponent is a little more computationally intensive, due to the matrix inversion.
However, we can approximate the ${(S^{-1})}_{ij}\delta(\bq_j)$ by treating it as a convolution in real space and thus a simple multiplication in Fourier space, where the (inverse) power spectrum (the FT of $S^{-1}$) is known and easy (diagonal) (see Appendix~3.B.3 of \citet{thesis} for computational details).
The same considerations apply for the kinetic term $\kinHMC$, though the mass matrix $\masHMC$ is a different quantity.
Often it is taken to be $\masHMC = S^{-1}$, meaning that indeed a similar procedure can be applied, only now with the non-inverted power spectrum.

\subsection{Leap frog scheme}
\label{sec:leap_frog}
After drawing a random momentum vector, we ``move'' the MCMC chain forward through the posterior's parameter space by solving the Hamiltonian equations of motion using a leap-frog estimation scheme.
Numerically, we'll be solving the following discretized equations for each time step $\tau + \epsilon$ (based on the previous step at $\tau$):
\begin{equation}
\begin{split}
  \momHMC_i(\tau + \frac{\epsilon}{2}) &\simeq \momHMC_i(\tau) - \frac{\epsilon}{2} \pdif{\potHMC}{\posHMC_i}(\posHMC(\tau)) \\
  \posHMC_i(\tau + \epsilon)           &\simeq \posHMC_i(\tau) + \epsilon \sum_j {(\masHMC^{-1})}_{ij} \momHMC_j(\tau + \frac{\epsilon}{2}) \\
  \momHMC_i(\tau + \epsilon)           &\simeq \momHMC_i(\tau + \frac{\epsilon}{2}) - \frac{\epsilon}{2} \pdif{\potHMC}{\posHMC_i}(\posHMC(\tau+\epsilon)) \,.
\end{split}
\label{eqn:leap_frog_discrete}
\end{equation}

We will need the derivative of $\potHMC$ to each component of the signal, i.e.\ every $\delta(\bq_i)$ sampled on a regular grid $\{\bq_i\}$:
\begin{equation}
    F_i \equiv \pdif{\potHMC}{\delta(\bq_i)} = F^\prior_i + F^\likeli_i \,.
    \label{eqn:dE_ddelta_general}
\end{equation}
This $F_i$ is called the Hamiltonian force and its components are called the prior force and likelihood force respectively.
For our Gaussian prior, the prior force is
\begin{equation}
    F^\prior_i = \sum_j {(S^{-1})}_{ij} \delta(\bq_j) \,,
    \label{eqn:dprior_ddelta_gaussian}
\end{equation}
where $S_{ij}$ is the correlation matrix corresponding to the cosmological power spectrum, given an inverse power spectrum mass (Section~\ref{sec:statistical_parameters}).

In general, we can write the likelihood force as
\begin{equation}
\label{eqn:likeli_force_short}
  F^\likeli_k = \sum_m h(\bq_m) \pdif{\divdisp(\bq_m)}{\delta(\bq_k)} \,,
\end{equation}
where the derivative of the divergence of the displacement field $\divdisp(\bq)$ to the signal $\delta(\bq_k)$ encodes the structure formation model and the $h$ term primarily represents the Gaussian model that we use for our likelihood (see Chapter~3 of \citet{thesis} for more details).
For the Zel'dovich model of structure formation, the likelihood force is given by
\begin{equation}
  F^{\likeli,ZA}_i = -h(\bq_i)  \,.
  \label{eqn:dlog_likeli_zeldovich}
\end{equation}
The function $h(\bq)$ is subsequently defined as:
\begin{equation}
\label{eqn:likeli_force_h_real}
    \hat{h}(\bk_l) = - {\left( - \frac{i \bk_l}{k_l^2} \cdot {\left( \hat{\bs{V'}}(\bk_l) \right)}^* \right)}^* = -\frac{i \bk_l}{k_l^2} \cdot \hat{\bs{V}}_l
\end{equation}
or
\begin{equation}
\label{eqn:likeli_force_h}
    h(\bq_m) = \nabla^{-2} \bnabla \cdot \bs{V}_m \,,
\end{equation}
where
\begin{equation}
\label{eqn:likeli_force_V}
\begin{split}
  \bs{V}_i &\equiv \bs{V}(\bx(\bq_i)) \equiv \pdif{\lnL}{\bx(\bq_i)} \\
           &= m_i \sum_j \pdif{\lnL}{\rho(\bx_j)} \pdif{W(\bx_j - \bx(\bq_i); h_s)}{\bx(\bq_i)} \,.
\end{split}
\end{equation}
Here the derivative of the likelihood to the density in Eulerian space $\rho(\bx)$ for a Gaussian likelihood is given by:
\begin{equation}
  \pdif{\lnL}{\rho(\bx_j)} 
  = \frac{\left( T_j( \rho^\mathrm{obs}(\bx_j) ) - \rho(\bx_j) \right) w_j}{\sigma_j^2} \,.
  \label{eqn:dlog_likelihood_gaussian}
\end{equation}
\changed{
  Note that at this point, a mixing of the Lagrangian and Eulerian coordinate systems has been introduced.
  The signal grid denoted here with subscript $i$ corresponds to the Lagrangian space, that of the primordial density field that we are interested in sampling from our PDF.
  The grid denoted by subscript $j$, in turn, corresponds to the discretized Eulerian grid of the observations.
  This $j$ index is summed out.
  Hence, we are left in the end with the Lagrangian grid cells only.
  It is especially recommended to keep this distinction in mind when considering the derivations in Appendix~\ref{ap:hamiltonian_force_zspace}.  
}

\subsubsection{Leap frog time step precision} 
\label{sec:leap_frog_time_step_precision}
The maximum pseudo time step $\epsilon_m$ is probably the most important parameter that determines overall HMC performance.
The pseudo time step $\epsilon$ is drawn randomly from a uniform distribution between 0 and $\epsilon_m$.

If $\epsilon_m$ is too high, the leap-frog errors will be too large, leading to artificial increase in energy.
The acceptance criterion does not allow this, meaning that many steps will be rejected.
This wastes computational resources.

On the other hand, if it is too low, many steps will be accepted.
However, the samples will be very close together, meaning that the probability distribution is densely and slowly sampled.
This is an inefficient use of computational power, since we do not need a too dense sampling, but rather a representative sampling of the overall high probability regions.

Striking a balance between the two is essential.
Further complications arise from the fact that the energy also scales with the dimensionality of the problem $N$.
To keep the acceptance rate at an acceptable value, $\epsilon$ will have to be smaller for larger problems, meaning that the chain will evolve correspondingly slower.
It is therefore impossible to determine a universally optimal $\epsilon$ for every HMC problem.
\citet{neal12} discusses the $\epsilon$ issue to great lengths.

One of the possible solutions is to automatically adapt $\epsilon_m$ to get an optimal acceptance rate \citep{neal12,hoffman12}.
We implemented our own adaptive scheme in \barcode.
In accordance with \citet{neal12}, we set the target acceptance rate to 0.65.
The code will check after a set number of attempted iterations (50 in our case) if the acceptance rate is still within a reasonable range around our optimal rate ($\pm 0.05$ is allowed).
If it is not, $\epsilon_m$ is adjusted an appropriate amount in the right direction.
Furthermore, to accelerate the initial adaptive stage (the first few HMC iterations) the code checks for acceptance every step instead of checking the acceptance rate every 50 steps.

\subsection{Hybrid Monte Carlo acceptance criterion}
\label{sec:hmc_acceptance_criterion}
Like Metropolis-Hastings and Gibbs sampling, Hamiltonian sampling includes a sample rejection step after the candidate sampling step.
Given the difference $\delta \hamHMC$ between the Hamiltonian of the state $(\posHMC,\momHMC)$ (where $\posHMC=\delta(\bq)$) before the dynamical step, $\hamHMC(\posHMC,\momHMC)$, and that of the candidate state $(\posHMC^*,\momHMC^*)$ after dynamics, $\hamHMC(\posHMC^*,\momHMC^*)$, the probability of acceptance of the candidate state conditional on the previous state $P((\posHMC^*,\momHMC^*)|(\posHMC,\momHMC))$ is given by
\begin{equation}
 P((\posHMC^*,\momHMC^*)|(\posHMC,\momHMC)) = \min \left( 1, e^{-(\hamHMC(\posHMC^*,\momHMC^*)-\hamHMC(\posHMC,\momHMC))} \right) \,.
 \label{eqn:acceptance_criterion}
\end{equation}

\subsection{From Lagrangian to Eulerian coordinates}
\label{sec:lag2eul}
A final important aspect of \barcode\ --- which is unrelated to HMC, but crucial to our cosmological application --- concerns the transformation between the Lagrangian space of the sampled primordial density fields and that of the Eulerian comoving coordinates at some $a>0$.
Lagrangian coordinates $\bq$ can, in general, be transformed to their Eulerian counterparts $\bx$ by adding a displacement field $\displace$ (in our case, defined by the Zel'dovich approximation):
\begin{equation}
    \bx(\bq) = \bq + \displace(\bq) \,.
    \label{eqn:lagrangian_to_eulerian}
\end{equation}
Note that this relation cannot be inverted.
Mathematically, this is because the displacement field is dependent on $\bq$.
Physically, it is because multiple fluid elements originally at $\{\bq_i\}$ might end up at one same Eulerian location $\bx$; through gravitational clustering, this will in fact happen all over the place.
This means that we must take the $\bq_i$ grid as the basis of our analysis, as taking an $\bx$ grid as the basis would make it impossible to find unique corresponding $\bq$ values, which computationally is a problem.
To compute $\displace$ we need $\divdisp$, because in Fourier space, we have:
\begin{align}
  i \bk \tilde{\divdisp} &= - k^2 \tilde{\displace} \\
  &\Rightarrow \tilde{\displace} = - \frac{i\bk}{k^2} \tilde{\divdisp} \label{eqn:psi_k-space} \,.
\end{align}
For the Zel'dovich model of structure formation that we use in this work, this quantity is given by:
\begin{equation}
  \divdisp = -D_1 \Deloneq \,.
  \label{eqn:phi_zeldovich}
\end{equation}

%% file: inc/ap_hamiltonian_likelihood_force.tex

\section{Hamiltonian force in redshift space}
\label{ap:hamiltonian_force_zspace}
\moved{Section~\ref{sec:method}}{The Hamiltonian likelihood force\footnote{
  The Hamiltonian \emph{prior} force is defined in Lagrangian space only.
  We need not take redshift space into account there.
} --- defined for observations in comoving coordinates in equation~\ref{eqn:likeli_force_short} --- is redefined for observations in redshift space as
\begin{equation}
  F^\likeli_i = - \sum_j \pdif{\lnL}{\rho^s(\zspace_j)} \pdif{\rho^s(\zspace_j)}{\posHMC_i} \,,
\label{eqn:likelihood_force_rss}
\end{equation}
where $\rho^s(\zspace_j)$ is the Eulerian density field in redshift space, at grid location $j$, which is found by evolving the Lagrangian density field $\posHMC = \delta(\bq)$ forward (see Appendix~\ref{ap:density_est}),
and $\posHMC_i$ is the Lagrangian density field (the signal) at grid location $i$.
The first right-hand side multiplicative term in equation~\ref{eqn:likelihood_force_rss} is given analogously to equation~\ref{eqn:dlog_likelihood_gaussian} by
\begin{equation}
  \pdif{\lnL}{\rho^s(\zspace_j)} = \frac{w_j}{\sigma_j^2} \left[ T(\rho^\obs(\zspace_j)) - \rho^s(\zspace_j) \right]\,,
\label{eqn:dlikeli_drho}
\end{equation}
where $T(\rho)$ is a transfer function that performs a simple linear density transformation to values in the used structure formation model given values from an $N$-body simulation (which should correspond to the observed quantities).
For the non-RSD models we could use functions from \citet{leclercq13}.
However, the fitted functions from that work were not calibrated for redshift space densities, so we would need to rederive such functions.
We highlight this possibility for future applications (see Section~\ref{sec:discussion_rank_ordering}), but without any loss of generality, we can define $T(\rho) = \rho$ here.
Since the $T$ term in this equation does not depend on the primordial density field, its exact form is not important in our derivation.
}

Note that indices $i$ and $j$ run over completely different regular grids.
The $i$ index here refers to the initial Lagrangian space grid, while the $j$ index is for the present-time Eulerian grid defined by the observations.

The density estimation of $\rho^s(\zspace_j)$ is done with SPH splines (Appendix~\ref{ap:density_est}).
In this case, we replace $\bx$ by $\zspace$ in equation~\ref{eqn:rho_x_kernel_sum}: 
\begin{equation}
  \rho^s(\zspace) = \sum_i m_i W(\zspace - \zspace_i; h_s) \,.
\label{eqn:rho_s_kernel_sum}
\end{equation}
With that we can write the second right-hand side term of equation~\ref{eqn:likelihood_force_rss} as
\begin{equation}
\label{eqn:drho_dsignal}
\begin{split}
  \pdif{\rho^s(\zspace_j)}{\posHMC_i} &= \sum_k m_k \pdif{W(\| \zspace_j - \zspace_k \| )}{\posHMC_i} \\
  &= \sum_k m_k \pdif{W(\| \zspace_j - \zspace_k \| )}{\zspace_k} \cdot \pdif{\zspace_k}{\posHMC_i} \,,
\end{split}
\end{equation}
where the gradient of $W$ is given in equation~\ref{eqn:SPH_kernel_gradient} (replacing $\bx$ by $\zspace$).
The second term is where the real difference with the non-RSD method comes in, because
\begin{equation}
  \pdif{\zspace_k}{\posHMC_i} = \pdif{\displace(\bq_k)}{\posHMC_i} + \pdif{\displacez(\bq_k)}{\posHMC_i}\,,
\label{eqn:ds_dsignal}
\end{equation}
where, besides the derivative of $\displace$, we now have the extra derivative of redshift space displacement $\displacez$ term to deal with.
It is this term that we will derive further in the following subsections.
Putting this back into equation~\ref{eqn:likelihood_force_rss}, we can rewrite as
\begin{equation}
\label{eqn:likeli_force_rss_2}
\begin{split}
  F^\likeli_i
  &=      - \sum_k \bs{V}_k \cdot \left[ \pdif{\displace_k}{\posHMC_i} + \pdif{\displacez_k}{\posHMC_i} \right] \\
  &=      \sum_m h_m \pdif{\divdisp_m}{\posHMC_i} 
          - \sum_k \bs{V}_k \cdot \pdif{\displacez_k}{\posHMC_i} \\
  &\equiv F^{\likeli,\mathrm{reg}}_i + F^{\likeli,\mathrm{rss}}_i \,,
\end{split}
\end{equation}
where the first part is the same as the ``regular'' $F^\likeli$ without redshift space from equation~\ref{eqn:likeli_force_short} and the second part represents the redshift space contribution.

We carry on to derive in detail the multiplicand in this added term, which is the derivative or gradient of the redshift space displacement field $\displacez$ with respect to the signal (the primordial density field $\delta(\bq)$).
For convenience, we multiply by $H a$, yielding:
\begin{equation}
\label{eqn:dz_dsignal}
\begin{split}
  (Ha) \pdif{\displacez(\bq_k)}{\posHMC_i}
  &=       \pdif{\left( (\bv_k - \bv_\obs) \cdot \hatx_k \right) \hatx_k}{\posHMC_i} \\
  &=       \left( \pdif{\bv_k}{\posHMC_i} \cdot \hatx_k \right) \hatx_k + \left( (\bv_k - \bv_\obs) \cdot \pdif{\hatx_k}{\posHMC_i} \right) \hatx_k \\
  &\quad + \left( (\bv_k - \bv_\obs) \cdot \hatx_k \right) \pdif{\hatx_k}{\posHMC_i} \,.
\end{split}
\end{equation}
We can further expand the gradient of $\hatx_k$:
\begin{equation}
\label{eqn:dhatr_dsignal}
\begin{split}
  \pdif{\hatx_k}{\posHMC_i}
  &= \frac{1}{\| \bx_k \|} \pdif{\displace_k}{\posHMC_i} + \bx_k \left[ - \frac12 \frac{2 x_k \pdif{x_k}{\posHMC_i} + 2 y_k \pdif{y_k}{\posHMC_i} + 2 z_k \pdif{z_k}{\posHMC_i}}{{( x_k^2 + y_k^2 + z_k^2 )}^{\frac32}} \right] \\
  &= \frac{1}{\| \bx_k \|} \pdif{\displace_k}{\posHMC_i} + \bx_k \frac{1}{\| \bx_k \|^3} \left( \bx_k \cdot \pdif{\bx_k}{\posHMC_i} \right) \\
  &= \frac{1}{\| \bx_k \|} \left[ \pdif{\displace_k}{\posHMC_i} - \left(\hatx_k \cdot \pdif{\displace_k}{\posHMC_i}\right) \hatx_k \right] \,.
\end{split}
\end{equation}
This is the most explicit form we can give in the general case.
To find a specific solution for our models, we must specify $\bv$.
In LPT, given $\displace$, we know $\bv$:
\begin{align}
  \bv &= a H f^\one \displace^\one  &\mathrm{LPT~(Zel'dovich)} \,, \label{eqn:velocity_zeldovich}\\
  \bv &= a H \left( f^\one \displace^\one + f^\two \displace^\two \right) &\mathrm{2LPT} \,, \label{eqn:velocity_2LPT}
\end{align}

For convenience, in what follows we split $\displacez$ into two parts:
\begin{equation}
\label{eqn:z_split}
\begin{split}
  \displacez  &=        \frac{1}{H a} \left( \sum_o \bv^{(o)} - \bv_\obs \right) \cdot \hatx \hatx \\
              &=        \left( \sum_o f^{(o)} \displace^{(o)} - \frac{\bv_\obs}{H a} \right) \cdot \hatx \hatx 
                \equiv   \sum_o \displacez^{(o)} - \displacez^\obs
              \,,
\end{split}
\end{equation}
where $o$ stands for the LPT order of the $\bv$ and corresponding $\displacez$ terms.
This way we can treat the \changed{observer's velocity} contribution separately from the LPT part(s).
In what follows we work out the \changed{observer's velocity} part and the Zel'dovich first order LPT term.

\subsection{Zel'dovich term}
When we insert equation~\ref{eqn:velocity_zeldovich} into equation~\ref{eqn:dz_dsignal} and combine with equation~\ref{eqn:dhatr_dsignal} we find for the derivative of $\displacez^\one$ that (note that by $\displace$ we really mean $\displace^\one$ here):

\begin{equation}
\label{eqn:dz_dsignal_zeldovich}
\begin{split}
  \frac{1}{f^\one} \pdif{\displacez_k^\one}{\posHMC_i}
  &=         \left( \pdif{\displace_k}{\posHMC_i} \cdot \hatx_k \right) \hatx_k + \frac{1}{\| \bx_k \|} \left[ \displace_k \cdot \left( \pdif{\displace_k}{\posHMC_i} - \left( \hatx_k
  \right. \right. \right. 
  \\
  &\qquad \left. \left. \left.
              \cdot~ \pdif{\displace_k}{\posHMC_i} \right) \hatx_k \right) \right] \hatx_k +  \frac{1}{\| \bx_k \|} (\displace_k \cdot \hatx_k) \left[ \pdif{\displace_k}{\posHMC_i}
  \right.
  \\
  &\qquad \left.
            - \left( \hatx_k \cdot \pdif{\displace_k}{\posHMC_i} \right) \hatx_k \right] \\
  &=        \left( \pdif{\displace_k}{\posHMC_i} \cdot \hatx_k \right) \hatx_k + \frac{1}{\| \bx_k \|} \left[ \displace_k \cdot \pdif{\displace_k}{\posHMC_i} \hatx_k - 2 (\displace_k
  \right. 
  \\
  &\qquad \left.
       \cdot~ \hatx_k) \left( \pdif{\displace_k}{\posHMC_i} \cdot \hatx_k \right) \hatx_k + (\displace_k \cdot \hatx_k) \pdif{\displace_k}{\posHMC_i} \right] \,.
\end{split}
\end{equation}
To facilitate insight into this expression we introduce a few auxiliary variables:
\begin{equation}
  \pdif{\displacez_k^\one}{\posHMC_i} \equiv f^\one \left[ (A^\one_{ki} + B^\one_{ki} - 2 C^\one_k A^\one_{ki}) \hatx_k + C^\one_k \pdif{\displace^\one_k}{\posHMC_i} \right] \,, 
\label{eqn:dz_dsignal_zeldovich_short}
\end{equation}
following the definition:
\begin{align}
  A^\one_{ki} &\equiv \pdif{\displace^\one_k}{\posHMC_i} \cdot \hatx_k \label{eqn:dz_dsignal_zeldovich_aux_A} \\
  B^\one_{ki} &\equiv \frac{1}{\| \bx_k \|} \displace^\one_k \cdot \pdif{\displace^\one_k}{\posHMC_i} \label{eqn:dz_dsignal_zeldovich_aux_B} \\
  C^\one_k    &\equiv \frac{1}{\| \bx_k \|} \displace^\one_k \cdot \hatx_k \,. \label{eqn:dz_dsignal_zeldovich_aux_C} 
\end{align}
The $A$ and $B$ terms have units $\mpch$ and $C$ is unitless.
It is interesting that, while the first three terms represent a contribution in the radial direction, the last term corresponds to a contribution in the direction of the displacement field.
Depending on the density field this may be an arbitrary direction.
The $C$ term in front of it contains a division by the distance to the particle, meaning that the closer you look, the stronger this component will be.
The same is true for the $B$ term.
The derivative of the redshift space displacement field is thus quite different from what one might naively expect from regular (radial) RSD effects, especially near the observer.
The $A$ term is the most straight-forward, simply representing the component of the derivative of the displacement field in the radial direction.

\subsection{\changed{Term involving observer's velocity}}
Elaborating on the remaining part of equation~\ref{eqn:dz_dsignal}, in combination with equation~\ref{eqn:dhatr_dsignal}, yields the \changed{term involving the observer's velocity}:
\begin{equation}
  \displacez^\obs \equiv \frac{\bv_\obs}{H a} \cdot \hatx \hatx \,.
\end{equation}
This term is valid for any structure formation model encoded in a displacement field $\displace$.
\begin{equation}
\label{eqn:dz_obs_dsignal}
\begin{split}
  (H a) \pdif{\displacez_k^\obs}{\posHMC_i}
  &=      \bv_k^\obs \cdot \pdif{\hatx_k}{\posHMC_i} \hatx_k + \bv_k^\obs \cdot \hatx_k \pdif{\hatx_k}{\posHMC_i} \\
  &=      \frac{1}{\| \bx_k \|} \left[ \bv_k^\obs \cdot \left( \pdif{\displace_k}{\posHMC_i} - \left( \hatx_k \cdot \pdif{\displace_k}{\posHMC_i} \right) \hatx_k \right) \right] \hatx_k \\
  &\quad+ \frac{1}{\| \bx_k \|} (\bv_k^\obs \cdot \hatx_k) \left[ \pdif{\displace_k}{\posHMC_i} - \left( \hatx_k \cdot \pdif{\displace_k}{\posHMC_i} \right) \hatx_k \right] \\
  &= \frac{1}{\| \bx_k \|} \left[ \bv_k^\obs \cdot \pdif{\displace_k}{\posHMC_i} \hatx_k \right. \\
  &\quad - 2 (\bv_k^\obs \cdot \hatx_k) \left( \pdif{\displace_k}{\posHMC_i} \cdot \hatx_k \right) \hatx_k \\
  &\quad \left. + (\bv_k^\obs \cdot \hatx_k) \pdif{\displace_k}{\posHMC_i} \right] \,.
\end{split}
\end{equation}
This expression has the same structure as equation~\ref{eqn:dz_dsignal_zeldovich}, except for the first ``$A$'' term.
The latter disappears because it is constant, yielding:
\begin{equation}
\label{eqn:dz_obs_dsignal_short}
\pdif{\displacez_k^\obs}{\posHMC_i} \equiv
\frac{1}{H a} \left[ \left( B^\obs_{ki} - 2 C^\obs_k A^\obs_{ki} \right) \hatx_k + C^\obs_k \pdif{\displace_k}{\posHMC_i} \right]
\,,
\end{equation}
where we define
\begin{align}
  A^\obs_{ki} &\equiv \pdif{\displace_k}{\posHMC_i} \cdot \hatx_k \label{eqn:dz_obs_dsignal_aux_A} \\
  B^\obs_{ki} &\equiv \frac{1}{\| \bx_k \|} \bv^\obs_k \cdot \pdif{\displace_k}{\posHMC_i} \label{eqn:dz_obs_dsignal_aux_B} \\
  C^\obs_k    &\equiv \frac{1}{\| \bx_k \|} \bv^\obs_k \cdot \hatx_k \,. \label{eqn:dz_obs_dsignal_aux_C}
\end{align}
The similarity between the expressions reveals that it has the same behavior of non-radial contributions near the observer as the Zel'dovich term.

\subsection{Plane parallel approximation}
\label{sec:plane_parallel_approximation}
The above two terms can be greatly simplified by taking a plane parallel approximation of redshift space, effectively putting the observer at an infinite distance.
The $\frac{1}{\| \bx \|}$ will then make the $B$ and $C$ terms go to zero, as well as the full \changed{observer's velocity term} of the derivative of $\displacez$ in equation~\ref{eqn:dz_obs_dsignal_short}.

For the Zel'dovich model in plane parallel redshift space, we are left with
\begin{equation}
  \left. \pdif{\displacez_k}{\posHMC_i} \right|_{\stackon[1pt]{$\scriptscriptstyle\mathrm{parallel}$}{$\scriptscriptstyle\mathrm{plane}$}}
  = \pdif{\displacez^\one_k}{\posHMC_i} = f^\one \left( \pdif{\displace_k}{\posHMC_i} \cdot \hatx_k \right) \hatx_k \,.
\label{eqn:dz_dsignal_zeldovich_planepar}
\end{equation}
It is good to point out that $\displacez$ and $\displacez^\one$ are intrinsically different.
Only in this specific case are their derivatives equal, as the \changed{observer's velocity term} part is equal to zero.

Without loss of generality we will restrict our studies to the plane parallel approximation for simplicity.

Finally, we insert this back into the $F^{\likeli,\mathrm{rss}}$ term of equation~\ref{eqn:likeli_force_rss_2}:
\begin{equation}
\label{eqn:likeli_force_rss_part}
\begin{split}
  F^{\likeli,\mathrm{rss}}_i
  &=      - f^\one \sum_k ( \bs{V}_k \cdot \hatx_k ) \left( \pdif{\displace_k}{\posHMC_i} \cdot \hatx_k \right) \\
  &=      - f^\one \sum_k ( \bs{V}_k \cdot \hatx_k ) \frac{1}{N} \sum_l e^{i \bk_l \cdot \bq_k} \left( - \frac{i \bk_l}{k_l^2} \right) \\
    &\qquad \cdot~ \hatx_k \sum_m e^{-i\bk_l \cdot \bq_m} \pdif{\divdisp_m}{\posHMC_i} \\
  &=      - \frac{f^\one}{N} \sum_m \sum_l \sum_k e^{-i\bk_l \cdot \bq_m} e^{i \bk_l \cdot \bq_k} ( \bs{V}_k \cdot \hatx_k ) \left( - \frac{i \bk_l}{k_l^2} \right) \\
    &\qquad \cdot~ \hatx_k \pdif{\divdisp_m}{\posHMC_i} \\
  &\equiv - \frac{f^\one}{N} \sum_m \sum_l \sum_k e^{-i\bk_l \cdot \bq_m} e^{i \bk_l \cdot \bq_k} \left( - \frac{i \bk_l}{k_l^2} \right) \cdot \bs{V}^r_k \pdif{\divdisp_m}{\posHMC_i} \,,
\end{split}
\end{equation}
where we define $\bs{V}^r_i$ as
\begin{equation}
  \bs{V}^r_i \equiv (\bs{V}_i \cdot \hatx_i) \hatx_i \,.
\label{eqn:likeli_force_V_rss}
\end{equation}
We are left with almost the same result as in equation~3.27 of \citet{thesis}, the only difference being the replacement of $\bs{V}$ with $f^\one \bs{V}^r$.
We can then conclude that in order to reuse equation~\ref{eqn:dlog_likeli_zeldovich}, we need to merely replace $h(\bq_m)$ by $h^\mathrm{rss}(\bq_m)$, defined as
\begin{equation}
  h^\mathrm{rss}(\bq_m) \equiv \nabla^{-2} \bnabla \cdot (\bs{V}_m + f^\one \bs{V}^r_m) \,.
\label{eqn:likeli_force_h_rss}
\end{equation}
This results in:
\begin{equation}
  F^\likeli_i = \sum_m h^\mathrm{rss}_m \pdif{\divdisp_m}{\posHMC_i} \,.
\label{eqn:likeli_force_short_rss}
\end{equation}

In the implementation of this algorithm it is convenient to choose the direction of the observer parallel to one of the coordinate axes.
In this way, $\hatx$ only changes one component of $\bs{V}$.
This modification can easily be implemented in a pre-existing HMC sampling code.

%% file: inc/ap_no_rsd_model_problem.tex

\section{Illustration of the redshift space problem}
\label{ap:results_problem}

In this appendix we illustrate the problems that one encounters when redshift space distortions are not modeled.
In summary: redshift space distortions introduce an apparent anisotropic clustering in the data, which must be dealt with to reconstruct the real-space (initial and final) density field.
When one does not correct for these distortions, the resulting reconstructions will be similarly distorted.

Note that we explicitly differentiate between the Kaiser effect and the effects we illustrate in this Appendix.
Both in the Kaiser effect and in the effects in our ``uncorrected redshift space'' runs, the main issue is deformation of structures along the line of sight.
However, the term \emph{Kaiser effect} pertains specifically to the generated structure \emph{in redshift space}.
In our case, the Kaiser effect is only present in the mock observations that we put into \barcode.
It is not present in the resulting reconstructed realizations, because these are either in Lagrangian space or in Eulerian (real) space, not in redshift space!
The imprint of the Kaiser effect in these runs is visible because the wrong assumption was made that the observations are in real space, not in redshift space.
The algorithm then tries to find a set of initial conditions that reproduce the anisotropies purely through gravitational evolution, without any redshift space assumptions.
To clearly distinguish this imprint of the Kaiser effect on the reconstructions from the Kaiser effect itself in the future, we propose to call this effect the \emph{Kaiser effect echo}.

\subsection{RSD artifacts in reconstructions}
To show the problem with RSDs in observations we obtained a set of regular Zel'dovich model samples with \barcode.
As with real data, the mock observational input is in redshift space, \emph{but we do not account for this in the model}.
These are the ``obs\_rsd'' type runs that we mentioned in Section~\ref{sec:astronomical_parameters}.

\begin{figure*}
\centering
\includegraphics[width=\textwidth]{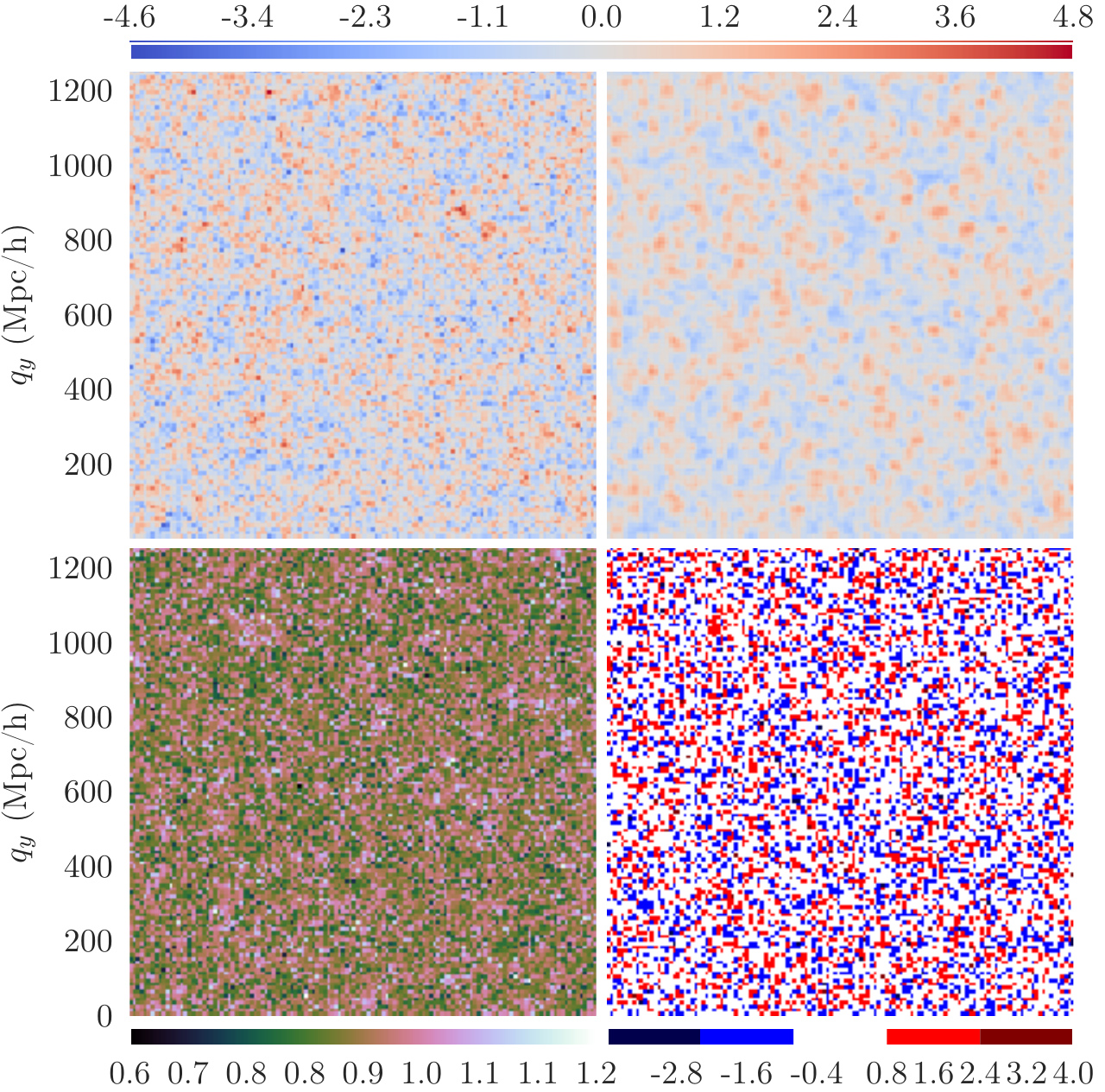}
\caption{Mean sample \emph{Lagrangian} densities (top right) compared to true (top left), together with std (bottom left) and the difference of the true and mean fields (bottom right).}
\label{fig:den_vstrue_lag_obs}
\end{figure*}

In figure~\ref{fig:den_vstrue_lag_obs}, we first look at a slice of the Lagrangian density field.
By eye, the results are strikingly similar to those with the redshift space model.
Structures are again well reconstructed in the mean of the reconstructions (top right) and the amount of variation around the mean is at a similar level (bottom panels).

\begin{figure*}
\centering
\includegraphics[width=\textwidth]{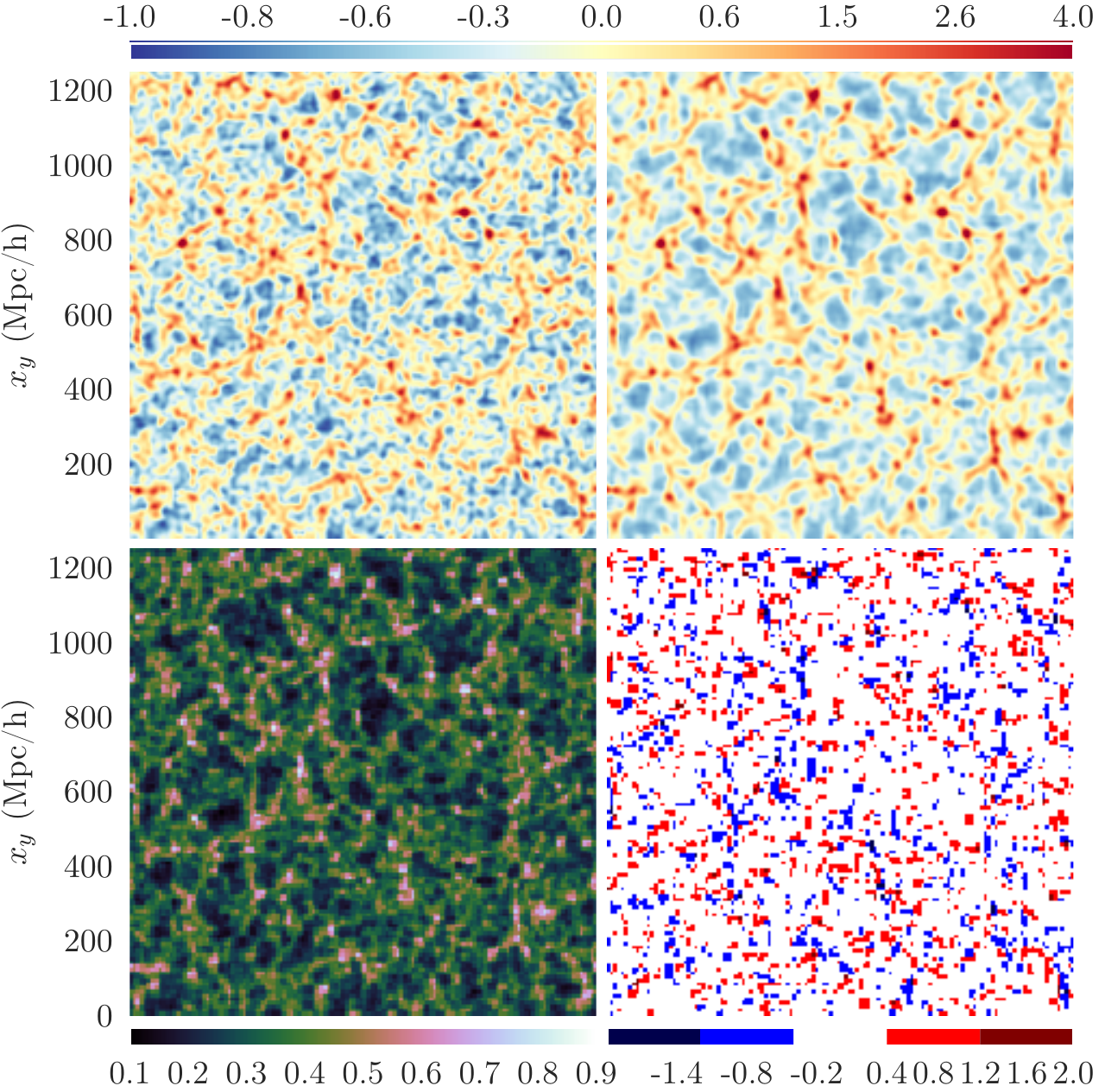}
\caption{Mean sample \emph{Eulerian} densities (top right) compared to true (top left), together with std (bottom left) and the difference of the true and mean fields (bottom right).}
\label{fig:den_vstrue_eul_obs}
\end{figure*}

We next look at the match of the corresponding ensemble average Eulerian density field to the true field in figure~\ref{fig:den_vstrue_eul_obs}.
The differences between the true (top left) and mean-reconstructed (top right) fields again are subtle and hard to discern by eye on paper.
The best way to see the difference is to alternate between the mean reconstruction image of this figure and that of the RSS-model results in figure~\ref{fig:den_vstrue_eul_rsd}, laid on top of each other in an animation\footnote{An animated GIF showing this is provided at \url{http://egpbos.nl/rsd-eul__obs-eul/}.}.
In this way, it becomes immediately apparent that a strong anisotropic imprint is imparted on the reconstructions: along the line of sight direction (horizontal in the image), the voids are stretched and the overdense structures are compressed.
In fact, when one compares this figure to the the RSS density field of figure~\ref{fig:img_results_01_lag_eul_rss_nobs} the problem becomes apparent\footnote{Animated GIF at \url{http://egpbos.nl/rsd-rss__obs-eul/}}.
\emph{When one does not model RSS, the algorithm converges on the true RSS field instead of the true Eulerian space field.}

Our intention is to \emph{remove} redshift space distortions, not reproduce them.
In the latter case, the corresponding Lagrangian density is quite incorrect as well.
When one does not include redshift space distortions and one uses data that is itself affected by redshift space distortions, the model is simply incorrect.
One assumes the redshift space density distribution to equal the Eulerian space density.
This begets a Lagrangian density that reproduces the observed redshift space distortions as if they were true physical features of the large scale structure.
Instead, they are optical illusions and must be treated accordingly.

The difference plot (bottom right) shows the shift of many features along the $z$-axis.
A bi-polar pattern along this line-of-sight axis is apparent for most clusters.

\subsection{RSD artifacts in two-point statistics}

\begin{figure}
  \centering

  \includegraphics[width=\columnwidth]{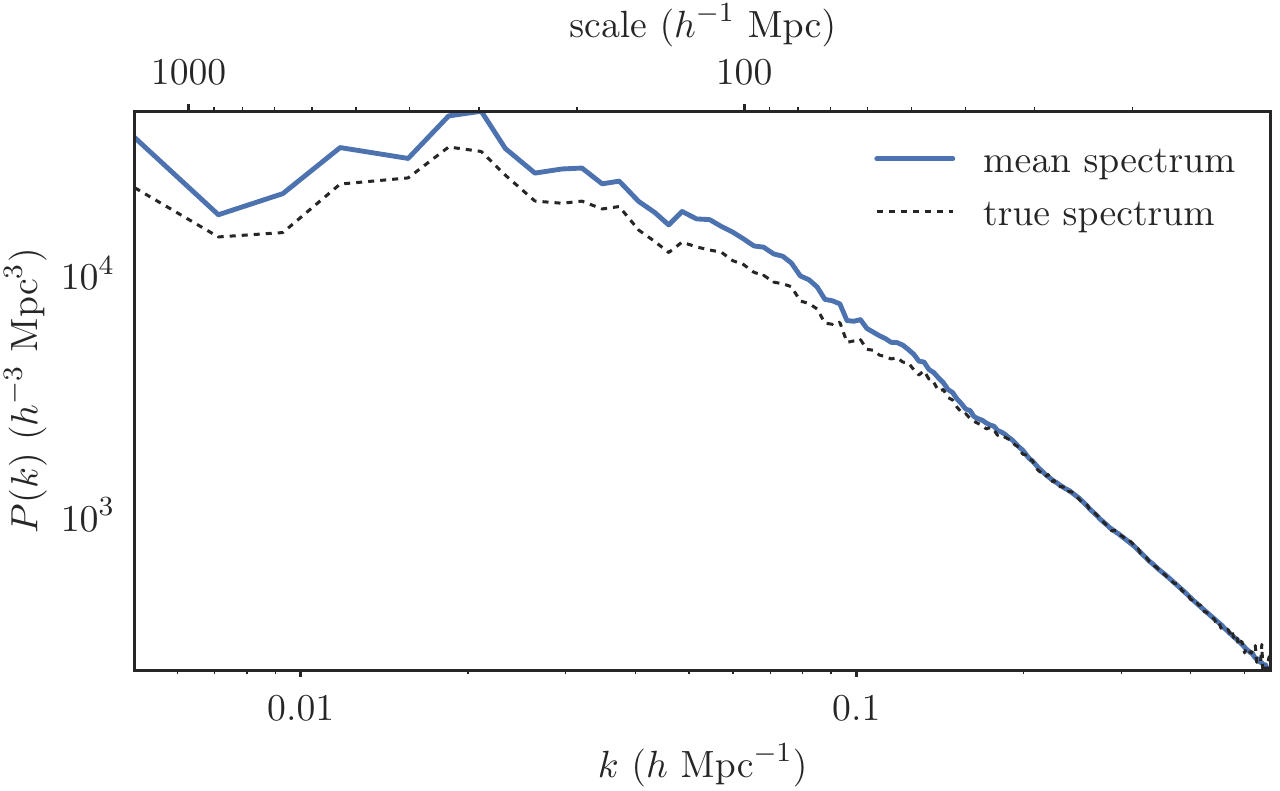}

  \includegraphics[width=\columnwidth]{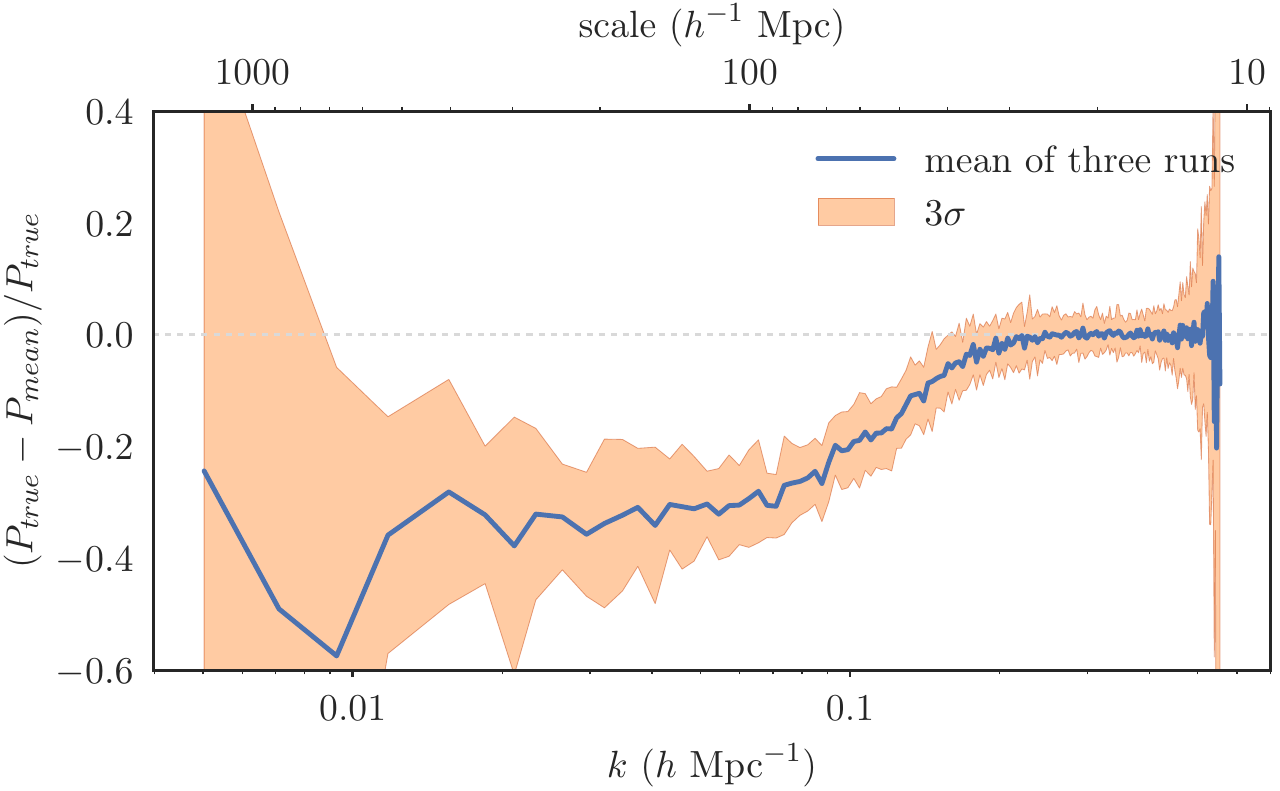}

  \caption{
    Sample Lagrangian power spectra compared to true; true compared to the mean and standard deviation of the sampled densities.
    Note that the difference plot (bottom panel) uses samples from three different runs with different true fields, while the power spectrum plot itself (top panel) is only from one true field.
  }
\label{fig:ps_vstrue_lag_obs}
\end{figure}

The reconstruction problems are visible in the power spectrum as well, as can be seen in figure~\ref{fig:ps_vstrue_lag_obs}.
This shows the boost in power at low $k$ that is known as the Kaiser effect.
In the difference plot (bottom panel of figure~\ref{fig:ps_vstrue_lag_obs}) this reveals itself as a downwards turn.
We want to get rid of this effect as well, since we are trying to reconstruct RSD-less density fields.

In figure~\ref{fig:2DCF_vstrue_mean_obs} we show 2D correlation functions (see Appendix~\ref{ap:2D_corr_fct}) that further illustrate the problem that is caused by having RSDs in the observations.
What one would like to see is in the left panel: the isotropic $\twodcf$ of the true density field.
The central panel shows the ensemble average of the samples that the algorithm obtains when RSDs are present in the mock observations, but are not modeled.

The reconstructions show large anisotropy (non-circular $\twodcf$).
Note again that we only see here the effect of the coherent flow RSDs, which cause ``great walls''.
This is seen in the horizontal (perpendicular to line of sight) stretching of the otherwise circular function.

\begin{figure}
\centering
\includegraphics[width=\columnwidth]{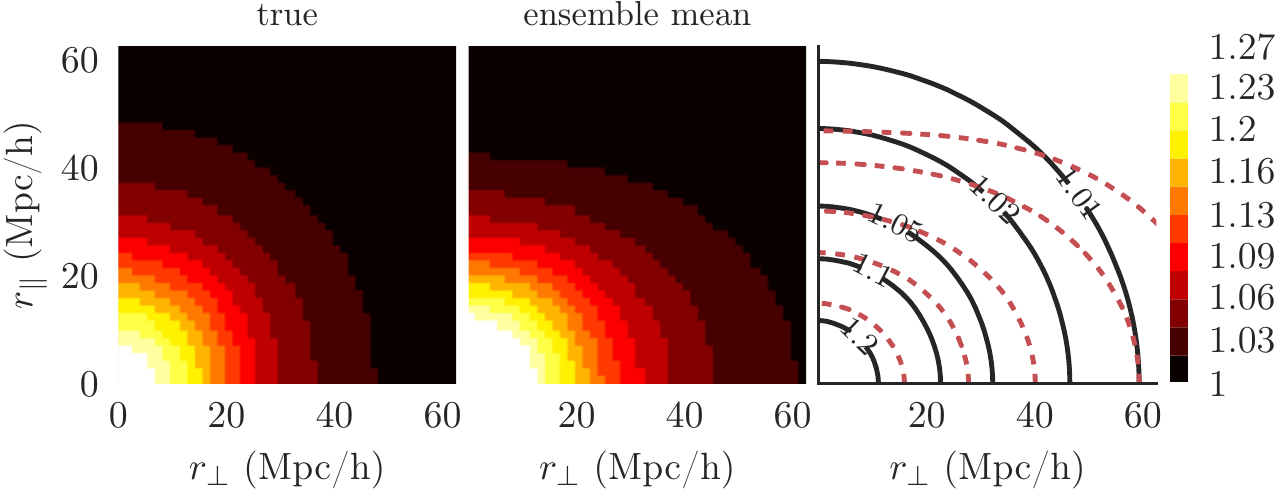}
\caption{
  Eulerian space 2D correlation functions: true (left) vs ensemble mean (center).
  The right-hand panel shows the two functions in contours: solid black for the true function and dashed red for the ensemble mean.
}
\label{fig:2DCF_vstrue_mean_obs}
\end{figure}

%% file: inc/ap_mcmc_performance.tex

\section{MCMC chain performance}
\label{ap:mcmc_performance}
\moved{Previously Section~5}{
  Some questions remain regarding performance of the algorithm given our new model, in terms of both speed and accuracy.
  Here we formulate some answers to the questions of how the chain evolves compared to the case without RSDs presented in \citet{thesis} \changed{and in Appendix~\ref{ap:results_problem}} and whether we see any remaining imprints of the RSDs.
}

\subsection{Burn-in phase}

\begin{figure}
\centering
\includegraphics[width=\columnwidth]{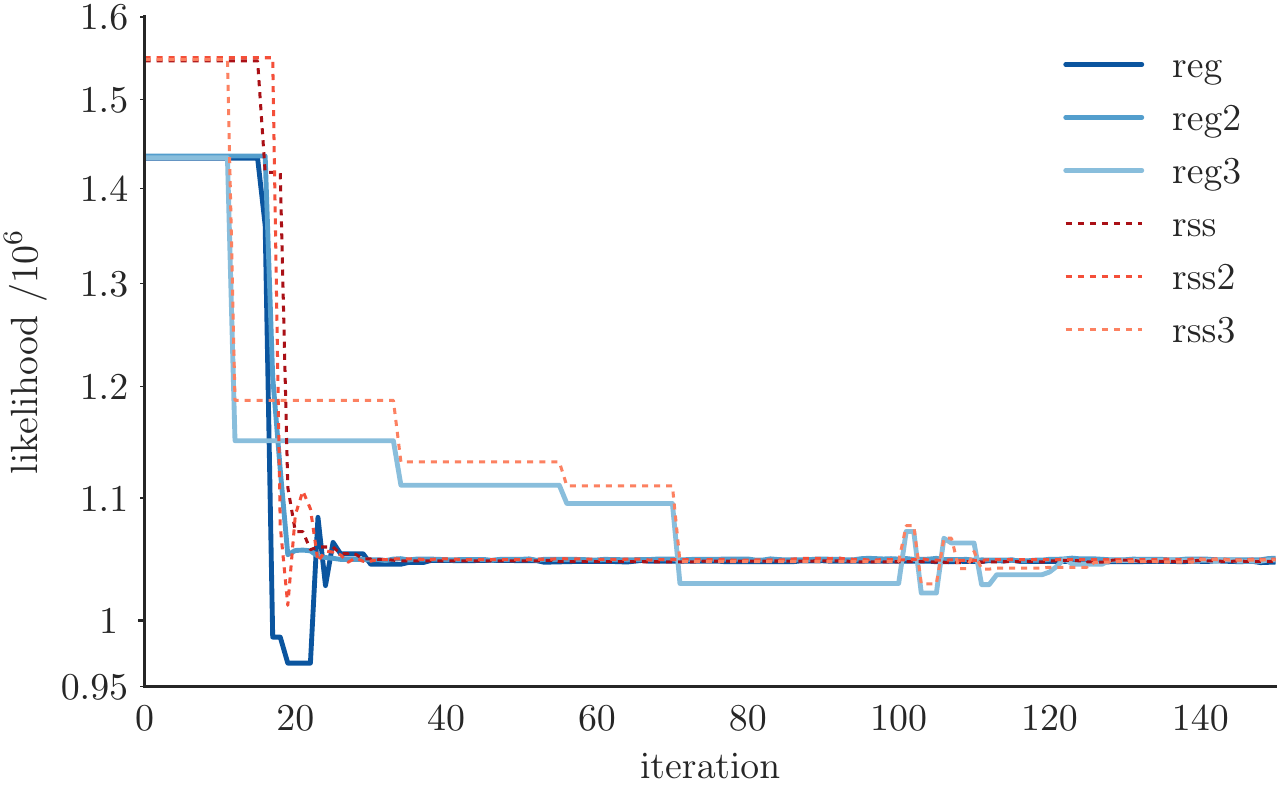}
\caption{
  Likelihood during the burn-in phase of six chains.
  Results are shown for the ``regular'' Eulerian space model in solid blue lines and the ``rss'' model in redshift space in dashed red lines.
  For each model, the three lines correspond to runs with three different true Lagrangian initial fields, generated with three different random number generator seeds.
}
\label{fig:burnin_reg_rsd}
\end{figure}

When trying to analyze chain mixing speed, one obvious place to begin is the burn-in phase.
In figure~\ref{fig:burnin_reg_rsd} we compare the likelihood of three different chains for each of the two models, the Zel'dovich model in Eulerian and in redshift space.
From this we conclude that burn-in does not seem at all slower.
The runs seem equally fast in Eulerian and in redshift space.
That also means that defining step 150 as the cut-off for burn-in is acceptable for the redshift space model as well.

\subsection{Chain performance after burn-in and adaptive $\epsilon$ evaluation}

\begin{figure}
\centering
\includegraphics[width=\columnwidth]{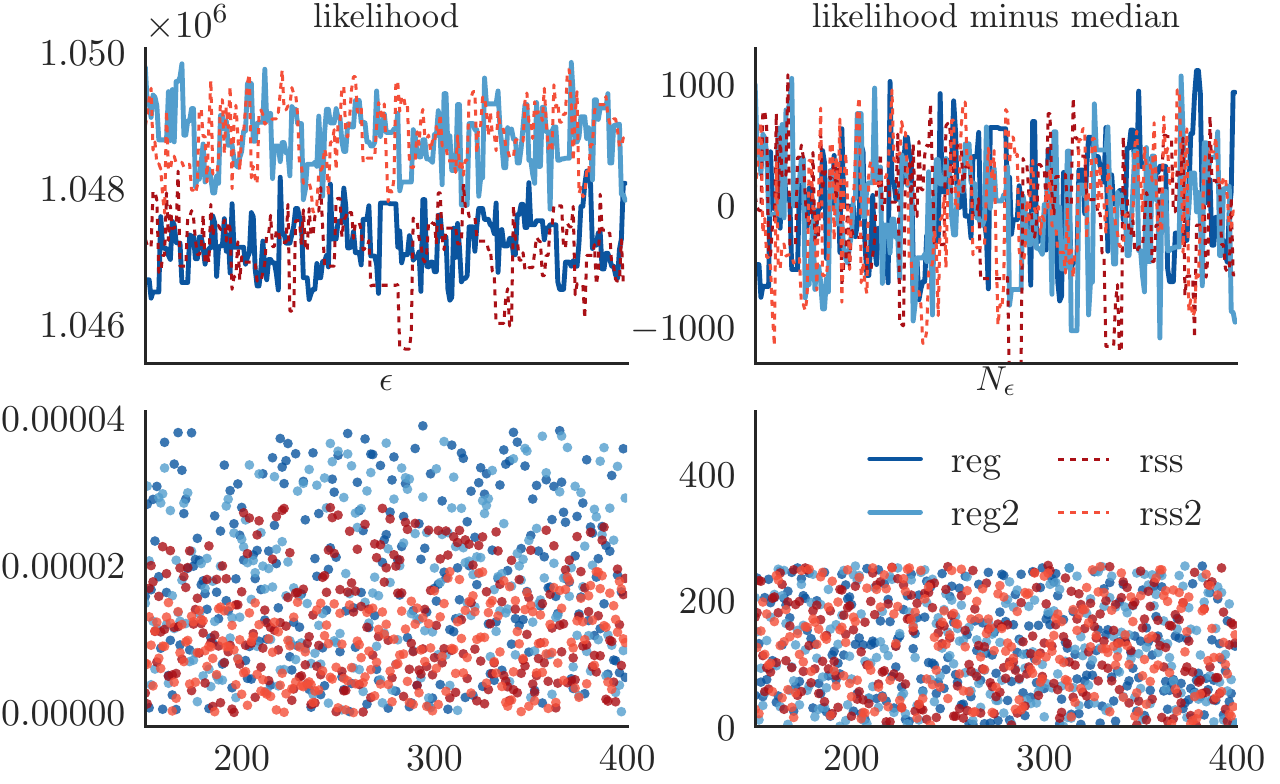}
\caption{
  Chain parameter evolution vs iteration number (horizontal axis).
  The iteration number includes both accepted and rejected steps.
  The top panels show the likelihood (right panel with the median of the likelihood subtracted for easier comparison), bottom left panel shows $\epsilon$, bottom right shows $N_\epsilon$.
  Regular runs in solid blue shades, redshift model runs in dashed red shades.
}
\label{fig:perform_reg_rsd}
\end{figure}

In figure~\ref{fig:perform_reg_rsd} we show chain parameters for the whole chain (after burn-in).
Note that the iteration number on the horizontal axis includes both accepted and rejected steps.
The variation in the likelihood can be used as a measure of the variability of the chain, as it compares the sample to the input field.
Two Eulerian space and two redshift space chains are shown in these figures.

The variation in likelihood is nearly identical for the regular and redshift space chains, with a standard deviation of slightly over 400 for both chains.
This is more directly illustrated in the top right panel, where we subtract the median value of the chain, so that the actual variation around the median can be compared for the different chains.
The random patterns are statistically indistinguishable.

In the bottom panel we additionally show $\epsilon$ (bottom left) and $N_\epsilon$ (bottom right).
Here finally the expected difference in the chain performance between the regular and rss runs shows.
The leap-frog step-size adapts to smaller values in the redshift space model runs.
This means that for some reason the orbit through the posterior's parameter space accrues more errors, i.e.\ its Hamiltonian is more sensitive to the exact path taken.
The redshift space model constrains the sampler less tightly, since it allows for more possible configurations along the line of sight than when using a regular space model.
This means that it is easier for the sampler to end up in a location that fits less well with the true field than in the regular model case.
This explains why a shorter leap-frog path may be necessary in the redshift space model.

\subsection{Optimizing number of leap-frog steps}
\label{sec:optimizing_number_leap_frog_steps}

In \citet{thesis}, we found that for runs with $L=200\hmpc$ and $N_x=64$, a higher number of leap-frog steps $N_\epsilon$ than $256$ improved the performance of the chain in that its samples covered a larger part of the posterior's parameter space.
We tried a higher value in this work as well to test whether we have reached optimal chain performance.

\begin{figure}
\centering
\includegraphics[width=\columnwidth]{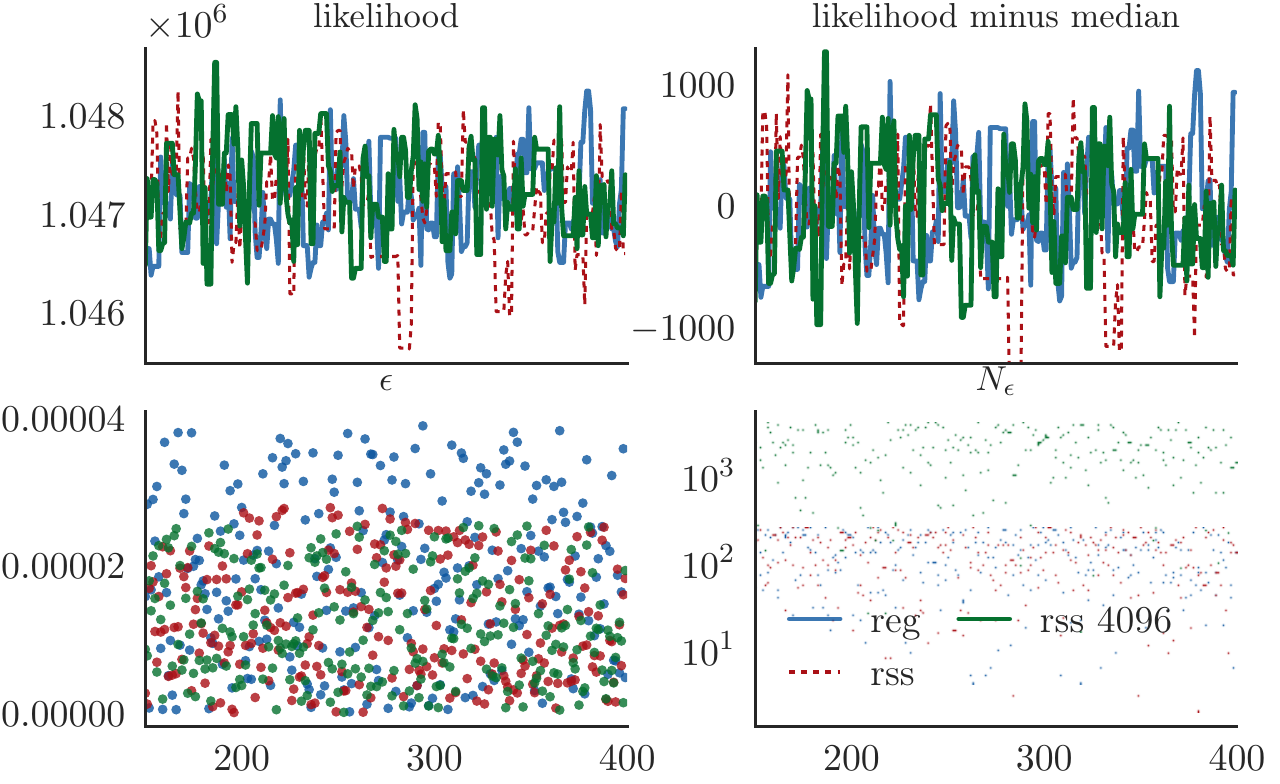}
\caption{
  Chain parameter evolution vs iteration number (horizontal axis) for different $N_\epsilon$ values.
  The iteration number includes both accepted and rejected steps.
  The top panels show the likelihood (right panel with the median of the likelihood subtracted for easier comparison), bottom left panel shows $\epsilon$, bottom right shows $N_\epsilon$.
  Regular run in solid blue, redshift model runs (including the two with a higher $N_\epsilon$) in dashed red shades.
}
\label{fig:perform_reg_rsd_Neps}
\end{figure}

In figure~\ref{fig:perform_reg_rsd_Neps}, we show the results of another redshift space model runs with a higher $N_\epsilon$ value of $4096$.
This significantly higher value does not seem to make any noticeable difference in the chain evolution parameters.
It does makes the code slower, as it scales with $N_\epsilon$.
We can conclude that 256 leap-frog steps per iteration are plenty for proper chain performance.

%% file: inc/ap_2D_corr_fct.tex

\section{2D correlation functions}
\label{ap:2D_corr_fct}

The stronger clustering implied by the Kaiser effect results in a boost of the contrast of walls and filaments perpendicular to the line of sight.
This boost is reflected in the higher amplitude of the power spectrum.
However, it does not differentiate between the line-of-sight and the other two, unaffected directions.
A common statistic that does take this into account is the two dimensional (sky --- redshift) correlation function.
We give a short recap on this statistic.

In general, the correlation function $\xi(\bx_A, \bx_B)$ of point A with point B in a homogeneous field is dependent only on the distance vector $\bs{r}_{AB}$ between the two points, i.e.
\begin{equation}
  \xi(\bx_A, \bx_B) = \xi(\bx_A - \bx_B) \equiv \xi(\bs{r}_{AB}) \,.
\label{eqn:correlation_function_homogeneous}
\end{equation}

Often, one also assumes the field to be isotropic, which further reduces the dependence of the correlation function to only the (absolute) distance scalar between two points:
\begin{equation}
  \xi(\bs{r}_{AB}) = \xi(|\bs{r}_{AB}|) \equiv \xi(r_{AB}) \,.
\label{eqn:correlation_function_homogeneous_isotropic}
\end{equation}
The assumptions of homogeneity and isotropy form the basis of the FLWR\footnote{
  Friedmann, Lema\^{\i}tre, Walker and Robertson.
}
metric of space-time.

When dealing with correlation functions of fields with RSDs, the latter assumption no longer holds.
The RSDs create a marked anisotropy between the line of sight and the sky.
The isotropy in the plane of the sky remains preserved.
This means that for correlation functions of density fields affected by RSDs, we can reduce the correlation function to a function of two scalar variables instead of one: one in the direction of the line of sight, and one in the remaining two (transverse) directions.
In our simplified case, where the observer is far away, the line of sight direction can be (and is) chosen to be along the $z$-axis.
This gives us:
\begin{equation}
  \xi(\bs{r}) = \xi(|\bs{r} - r_z \hat{z}|, r_z) \equiv \twodcf \,,
\label{eqn:correlation_function_homogeneous_RSD}
\end{equation}
where $r_z = \bs{r} \cdot \hat{z}$ and for simplicity we drop the $AB$ subscripts.

In the literature, often $r_\parallel$ is called the radial pair separation $\pi$ and $r_\perp$ is called the transverse pair separation $\sigma$ \citep{peacock01}.
Note that the parallel (to the line-of-sight) axis, which in our case is the $z$-axis, is vertical in the $\twodcf$ plots, but horizontal in the density slice plots.
The former was chosen to correspond to the way $\twodcf$ plots are usually shown in the literature \citep[see e.g.]{hawkins03}.

The Kaiser effect is reflected in the 2D correlation functions.
In this case, only in the transverse direction.
The increased correlation on large scales corresponds to the ``great walls'' that are enhanced in the observations \citep{sargentturner77,ryden96,thomas04}.
While the difference appears to be in the transverse direction, the cause of this flattening is the compression along the line of sight direction.
Due to this compression, both a modest amount of flattening in the radial direction and the more apparent widening into the transverse direction are brought about.
The compression also produces a generally higher level of correlation.

Each quadrant contains exactly the same information, since $\twodcf$ is symmetric along its two axes.
In this paper, we plot the upper left quadrant only, which offers greater detail at the expense of making it slightly harder to see the shape.

%% file: inc/ap_density_est.tex

\section{Kernel density estimation}
\label{ap:density_est}

The derivative of the density field $\pdif{\rho(\bx_i)}{\bx(\bq)}$ pops up in the derivative of the log-likelihood of equation~\ref{eqn:likelihood_force_rss}.
This must be treated very carefully, because a mixing of coordinate systems takes place here.
We take the derivative of the Eulerian density field on a regular grid defined by the observations, but we derive \emph{to} an irregular $\bx(\bq)$ coordinate.
Even though $\bq$ is on a regular (initial Lagrangian density field) grid, $\bx(\bq)$ is certainly not.
This means we cannot simply put the density on a grid and then take a derivative (easily done numerically in Fourier space by multiplying by $i\bk$).
One might think that by interpolating the regularized field $\rho(\bx_i)$ on locations $\bx(\bq)$ one could also overcome this, but this is not the case.
In fact, this introduces severe instabilities in the algorithm.

As always for HMC algorithms, the only solution is to keep every calculation as explicit as possible.
We do this by formulating the density field in terms of particles with a kernel that distributes their mass over space.
This allows us to analytically express the derivative to the particle kernel necessary in the likelihood force.
The following general expression for density $\rho(\bx)$ then applies:
\begin{equation}
\label{eqn:rho_x_kernel_sum}
  \rho(\bx) = \sum_{\mathclap{\mathrm{particles}~i}} m_i W(\bx-\bx_i) \,,
\end{equation}
given a kernel $W(\bx_i,\bx)$ that gives the density contribution of the particle at $\bx$ given a particle $i$ at $\bx_i$, and a particle mass $m_i$ to the continuous density field $\rho$ at $\bx$.\footnote{
  To transform from the $\bq$ to the $\bx$ grid, we start by evolving particles forward in time to the desired redshift using the structure formation model we chose (Zel'dovich in this work), given the input field $\delta(\bq)$.
  The resulting particle distribution in $\bx$ comoving coordinate space is then converted to a new regularly gridded density field $\rho(\bx)$ by means of an SPH field estimation scheme.
}
Our mass will be defined by the critical density $\rho_c$, box volume $V$ and number of grid cells $N$, as we are only concerned with equal mass Lagrangian particles:
\begin{equation}
  m_i = \rho_c \frac{V}{N} \,.
  \label{eqn:SPH_kernel_mass}
\end{equation}
This kernel formulation conveniently integrates the particle nature of the transformation from Lagrangian to Eulerian space; this is, after all, a hydrodynamical representation of matter fields, based on particles.

One could in principle convolve fields with these kernels to smooth them out, or deconvolve fields with them by dividing by them in Fourier space.
\citet{wang13} suggest that the latter is necessary when using kernel-based density estimation and/or interpolation methods.
We do not follow them in this, as we discuss in \citet{thesis}.

For this work, we chose to use an SPH spline kernel \citep{monaghan92,springel10} for each particle at $\bx(\bq)$.
The advantages of this kernel are:
\begin{itemize}
  \item smoothness (it is a second order Taylor expansion of a Gaussian), which also gives us nicely behaving derivatives (first and second), and
  \item boundedness within a few neighboring cells, which makes it far more computationally efficient than e.g.\ a Gaussian which has infinite extent.
\end{itemize}
Other options, like Gaussian, NGP, CIC and TSC kernels only have one of the above properties, not both.
We can also not use the more precise DTFE scheme \citep{schaap00,weygaert11,cautun11a}, because it is not kernel-based, making analytical derivatives a lot harder to calculate.

The SPH spline kernel is defined in 3D following \citet{monaghan92}:
\begin{equation}
  W(q;h_s) = \frac{1}{\pi h_s^3}
  \begin{cases}
    1 - \frac{3}{2} q^2 + \frac{3}{4} q^3       &\mbox{ if } 0 \leq q < 1 \\
    \frac{1}{4} {(2 - q)}^3                     &\mbox{ if } 1 \leq q < 2 \\
    0                                           &\mbox{ otherwise} \,,
  \end{cases}
  \label{eqn:SPH_kernel}
\end{equation}
where $h_s$ is the scale parameter of the kernel which defines its size and $q$ is given by:
\begin{equation}
  q \equiv \frac{|\bx - \bx_i|}{h_s} \,.
  \label{eqn:SPH_kernel_q}
\end{equation}
An important property of the kernel thus normalized is that, independent of the choice of $h_s$, we have
\begin{equation}
  \int_V W(q;h_s) d\bx = 1 \,,
  \label{eqn:SPH_kernel_integral_1}
\end{equation}
which is necessary for equation~\ref{eqn:rho_x_kernel_sum} to be correct.
Note that the integral is unitless, as $W$'s units are $\mpc^{-3}$; $W$ is effectively a number density that spreads the mass of the Lagrangian tracer particle i at $\bx_i$ over its surroundings.

We need the gradient of this kernel with respect to $\bx_i = \bx(\bq_i)$ to calculate the likelihood force term.
This gradient is given by:
\begin{equation}
  \pdif{W(\bx - \bx_i; h_s)}{\bx_i} = \frac{1}{\pi h_s^5} \frac{\bx_i - \bx}{q}
  \begin{cases}
    \frac{9}{4} q^2 - 3 q      &\mbox{ if } 0 \leq q \leq 1 \\
    - \frac{3}{4} {(2 - q)}^2  &\mbox{ if } 1 \leq q \leq 2 \\
    0                          &\mbox{ otherwise}
  \end{cases} \,,
  \label{eqn:SPH_kernel_gradient}
\end{equation}
where in the $F^\likeli$ equations $\bx_i$ is replaced with $\bx(\bq_i)$ and $\bx$ with $\bx_j$, which are the terms we'll actually use in the formulas (though it works for any combination of $\bx$ positions).

%% file: inc/ap_2lpt_sc.tex

\section{Second order Lagrangian perturbation theory extension}
\label{ap:2lpt_sc}

In the implementation used in this paper, \barcode\ uses the Zel'dovich or first order Lagrangian perturbation theory model as its structure formation model.
This model has the advantage of giving a decent match to reality over a range of scales; from large to semi-linear scales.
On large scales, however, the second order Lagrangian perturbation theory (2LPT) model of structure formation is more accurate, reproducing higher order structural statistics better than for instance the Zel'dovich approximation \citep{neyrinck13}.
Also, it takes into account the second order tidal terms, which is important for overall large scale structure formation.
In this appendix (Section~\ref{sec:2lpt}), we derive the equations necessary to include the 2LPT model in \barcode.

While 2LPT works well at large scales, it fails badly at smaller scales.
Its main downside is that it produces even stronger overshoot in clustered regions than the Zel'dovich model.
Figure~6 from \citet{neyrinck13} illustrates this fact succinctly.

To correct 2LPT on small scales, we adopt the \emph{Augmented Lagrangian Perturbation Theory} model of \citet{kitaurahess13}.
This model adds a spherical collapse (``SC'') component to the 2LPT model.
Through this combination, the massive overshoot is removed completely.
Small scales are corrected for in a way that is still analytically tractable and thus usable in our Hamiltonian sampler.
We will refer to this model by the more explicit acronym ``2LPT+SC'' instead of the previously used ALPT.\@
In Section~\ref{sec:2lpt_sc}, we derive the equations necessary to implement 2LPT+SC in \barcode.

This model, though a great improvement to bare 2LPT, is not without its shortcomings.
One of particular interest is the fact that it is, by nature, spherical.
However, structure formation as a process is driven by anisotropic collapse \citep{zeldovich70,icke73}.
Since this is what we study, we need to assess the impact of the addition of a spherical collapse model to anisotropic measures of large scale structure.
The validity of these models with respect to important aspects of the large scale structure will be tested in a future study.

\subsection{Notation}
Most of this appendix uses the same definitions and the same notation as the rest of this paper.
However, we must make some additional remarks on index notation.
Normal indices usually start from $i$ in the alphabet and indicate grid cells.
We use these indices when talking about Fourier transforms, for instance.
Underlined indices start from $a$, i.e.\ $\ua$.
These are used when talking about components of regular vectors like coordinate vectors $\bx$, velocities $\bv$ or a displacement vector $\displace$.
Some quantities have a combination of these two types of indices, e.g.\ when talking about (Fourier transforms of) discretized vector fields, like the $i$-th cell of the $\ua$-th component of a displacement field $\displace$: $\displacescalar_{i\ua}$.

When we use matrices, two indices are necessary.
The same goes for the Kronecker delta $\krodel_{ij}$.
When, instead of indices, the full coordinates are written in the subscript of the Kronecker delta, e.g.\ $\krodel_{\bq_i,\bq_j}$, a comma is put in between for clarity.
We clarify this, since, in other cases, a comma in the subscript means a derivative with respect to the coordinates corresponding to the indices (for instance in equation~\ref{eqn:delta_two}).

\subsection{Second order Lagrangian perturbation theory}
\label{sec:2lpt}

2LPT adds a second order term to the equation for $\divdisp$ \citep[equation 94]{bernardeau02}:
\begin{equation}
  \divdisp^{(2)}(\bq) = D_2 \Deltwo \,,
  \label{eqn:phi_2lpt_part}
\end{equation}
so that the full equation for $\divdisp$ in 2LPT reads
\begin{equation}
  \divdisp = -D_1 \Delone + D_2 \Deltwo \,.
  \label{eqn:phi_2lpt}
\end{equation}
The second order spatial term is dependent on the first order term from equation~\ref{eqn:phi_zeldovich}.
This allows us to take the derivative to $\delta(\bq_i) = \delone(\bq_i)$, obtaining
\begin{equation}
  \pdif{\divdisp(\bq)}{\delta(\bq_i)} = -\krodel_{\bq,\bq_i} + D_2 \pdif{\Deltwo}{\delta(\bq_i)} \,.
  \label{eqn:dphi_ddelta_2lpt}
\end{equation}

The second order term of the density in LPT, $\Deltwo$, is given by
\begin{equation}
 \Deltwo \equiv \sum_{\ua>\ub} \left( \phi^{(1)}_{,\ua\ua}(\bq) \phi^{(1)}_{,\ub\ub}(\bq) - {\left( \phi^{(1)}_{,\ua\ub}(\bq) \right)}^2 \right) \,,
 \label{eqn:delta_two}
\end{equation}
where the comma-$\ua$/$\ub$-subscripts denote the derivative to the $\ua$th/$\ub$th component of the coordinates ($x$, $y$, $z$).
$\phi^{(1)}$ is the first order term in the LPT gravitational potential expansion, which is found by solving the Poisson equation for the first order terms:
\begin{equation}
  \nabla_\bq^2 \phi^{(1)}(\bq) = \Delone \,.
\end{equation}
For the derivative of $\Deltwo$ in equation~\ref{eqn:dphi_ddelta_2lpt}, one then obtains:
\begin{multline}
\label{eqn:d_deltwo_d_delta_q}
 \pdif{\Deltwo}{\delta(\bq_i)} = \sum_{\ua>\ub} \left( \phi^{(1)}_{,\ua\ua}(\bq) \pdif{\phi^{(1)}_{,\ub\ub}(\bq)}{\delta(\bq_i)} + \phi^{(1)}_{,\ub\ub}(\bq) \pdif{\phi^{(1)}_{,\ua\ua}(\bq)}{\delta(\bq_i)} \right. \\
 \left. - 2 \phi^{(1)}_{,\ua\ub}(\bq) \pdif{\phi^{(1)}_{,\ua\ub}(\bq)}{\delta(\bq_i)} \right) \,.
\end{multline}

The derivative of $\phi^{(1)}_{,\ua\ub}$ to $\delta(\bq_i) = D_1 \Delone(\bq_i)$ can be solved in Fourier-space.
The Poisson equation in Fourier-space can be written as
\begin{equation}
    \hat{\phi}^{(1)} (\bk) = - \frac{1}{k^2} \hat{\Delta}^{(1)}(\bk) \,,
\end{equation}
which gives us this relation for $\phi^{(1)}$:
\begin{equation}
  \label{eqn:phi_one_poisson_fourier}
  \phi^{(1)}(\bq) = - \frac{1}{N} \sum_j e^{-i \bk_j \cdot \bq} \frac{1}{k_j^2} \hat{\Delta}^{(1)}(\bk_j) \,.
\end{equation}
Taking the derivatives to the $\ua$th and $\ub$th coordinate components:
\begin{equation}
  \label{eqn:phi_second_deriv}
  \begin{split}
    \phi^{(1)}_{,\ua\ub}(\bq)
    &=  \partial_{\ua} \partial_{\ub} \phi^{(1)}(\bq) \\
    &=  \frac{1}{N} \sum_j e^{i \bk_j \cdot \bq} \frac{k_{j\ua} k_{j\ub}}{k_j^2} \left[ \sum_{\bq'} e^{-i \bk_j \cdot \bq'} \Delta^{(1)}(\bq') \right] \,,
  \end{split}
\end{equation}
and the derivative of this to $\delta(\bq_i)$, we end up with:
\begin{equation}
\label{eqn:d_phi,lm_d_delta}
  \begin{split}
    \pdif{\phi^{(1)}_{,\ua\ub}(\bq)}{\delta(\bq_i)}
    &=  \frac{1}{N} \sum_j e^{i \bk_j \cdot \bq} \frac{k_{j\ua} k_{j\ub}}{k_j^2} \left[ \sum_{\bq'} e^{-i \bk_j \cdot \bq'} \pdif{\Delta^{(1)}(\bq')}{D_1 \Delta^{(1)}(\bq_i)} \right] \\
    &=  \frac{1}{D_1 N} \sum_j e^{i \bk_j \cdot \bq} \frac{k_{j\ua} k_{j\ub}}{k_j^2} \left[ \sum_{\bq'} e^{-i \bk_j \cdot \bq'} \krodel_{\bq',\bq_i} \right] \\
    &=  \frac{1}{D_1 N} \sum_j e^{i \bk_j \cdot \bq} \frac{k_{j\ua} k_{j\ub}}{k_j^2} e^{-i \bk_j \cdot \bq_i} \,. 
  \end{split}
\end{equation}
The first exponential we can use, in the end, to get rid of the sum in equation~\ref{eqn:likeli_force_short}, by turning them into a DFT.\@
The second exponential will be used with the sum over $j$ in this equation.
Note that for this equation to be non-zero, the $\bq_i$ and $\bq$ grids need to be the same!
If they are not, the Kronecker delta will only equal zero.
This final step thus necessarily couples the $\bq_i$-grid of the signal, the $\delta(\bq)$ field sample, and that of the $\bx(\bq)$ grid used in the likelihood $\likeli$.

Now, to simplify equation~\ref{eqn:d_deltwo_d_delta_q}, we can write each of the terms in the $\ua>\ub$ sum in a general way, writing them as specific instances of a general four-index quantity.
Including the multiplication by $h(\bq_m)$ and the sum over index $m$ from equation~\ref{eqn:likeli_force_short}, and using the result from equation~\ref{eqn:d_phi,lm_d_delta}, we then get for each generalized part (note the indices $\ua$, $\ub$, $\uc$ and $\ud$) of the full term:
\begin{equation}
\label{eqn:four_index_term_2lpt}
\begin{split}
   \sum_m & h(\bq_m) D_2 \phi^{(1)}_{,\uc\ud}(\bq_m) \pdif{\phi^{(1)}_{,\ua\ub}(\bq_m)}{\delta(\bq_i)} \\
   \qquad&=       \sum_m h(\bq_m) \frac{D_2}{D_1} \frac1N \phi^{(1)}_{,\uc\ud}(\bq_m)  \sum_j e^{i\bk_j\cdot\bq_m} \frac{k_{j\ua} k_{j\ub}}{k_j^2} e^{-i\bk_j\cdot\bq_i} \\
  \qquad&\equiv  \frac{D_2}{D_1} \frac1N \sum_j \hat{\alpha}^*_{\uc\ud}(\bk_j) \frac{k_{j\ua} k_{j\ub}}{k_j^2} e^{-i\bk_j\cdot\bq_i} \\
  \qquad&\equiv  \frac{D_2}{D_1} {\xi^*}^{\ua\ub}_{\uc\ud}(\bq_i) \\
  \qquad&=  \frac{D_2}{D_1} \xi_{\uc\ud}^{\ua\ub}(\bq_i) \,,
\end{split}
\end{equation}
where we define the convenience functions $\alpha(\bq)$ and $\xi(\bq)$ as
\begin{equation}
  \alpha_{\uc\ud}(\bq) \equiv {\left( h(\bq) \phi^{(1)}_{,\uc\ud}(\bq) \right)}^* = h(\bq) \phi^{(1)}_{,\uc\ud}(\bq) \,,
  \label{eqn:convenience_alpha_2lpt}
\end{equation}
\begin{equation}
\label{eqn:convenience_xi_2lpt}
\begin{split}
  \hat{\xi}_{\uc\ud}^{\ua\ub}(\bk) &\equiv {\left( \frac{k_{\ua} k_{\ub}}{k^2} \hat{\alpha}^*_{\uc\ud}(\bk) \right)}^* = \frac{k_{\ua} k_{\ub}}{k^2} \hat{\alpha}_{\uc\ud}(\bk) \\ 
  &\Rightarrow \xi^{\ua\ub}_{\uc\ud}(\bq) = \partial_{\ua} \partial_{\ub} \nabla^{-2} \left( h(\bq) \phi^{(1)}_{,\uc\ud}(\bq) \right)\,,
\end{split}
\end{equation}
where we get rid of the conjugations because $\phi^{(1)}_{,\uc\ud}$ is real, as it is a physical term, $h(\bq)$ is also real, as previously noted, and the $\bk$'s are real scale vectors.
The double conjugation on $\hat{\alpha}_{\uc\ud}$ cancels out as well.

We will see $\xi^{\ua\ub}_{\uc\ud}$ again in Section~\ref{sec:2lpt_sc}, but in a slightly different form. This prompts us to use a functional instead, which we can use again later on and which simplifies our code by only writing the general form of the function once:
\begin{equation}
  \chi^{\ua\ub}_{\uc\ud}(f(\bq); \bq) \equiv \partial_{\ua} \partial_{\ub} \nabla^{-2} \left( f(\bq) \phi^{(1)}_{,\uc\ud}(\bq) \right) \,.
  \label{eqn:convenience_chi}
\end{equation}

Then, finally, we can write down a worked out 2LPT version of the likelihood force (equation~\ref{eqn:likeli_force_short}), combining the first and second order results:
\begin{equation}
  F^{\likeli,2LPT}_i = - h(\bq_i) + \frac{D_2}{D_1} \sum_{\ua>\ub} \left( \chi_{\ub\ub}^{\ua\ua} + \chi_{\ua\ua}^{\ub\ub} - 2\chi_{\ua\ub}^{\ua\ub} \right)(h(\bq_i); \bq_i) \,.
  \label{eqn:loglikeli_2lpt}
\end{equation}
The argument $(h(\bq_i); \bq_i)$ applies to all three $\chi$s.

\subsection{2LPT and spherical collapse}
\label{sec:2lpt_sc}

For the 2LPT+SC model, the divergence of the displacement field $\divdisp(\deloneq)$ is approximated by the weighted sum of a 2LPT part on large scales and a spherical collapse part on small scales, where the scales are split using Gaussian convolution (meaning that there is not an absolute splitting of scales, there will be overlap):
\begin{equation}
\label{eqn:phi_alpt}
\begin{split}
  \divdisp
  &\equiv  \left[K_G \ast \left( \bnabla \cdot \displace^\mathrm{2LPT} \right) \right] + (\bnabla \cdot \displace^\mathrm{SC}) - \left[K_G \ast ( \bnabla \cdot \displace^\mathrm{SC} ) \right] \\
  &=       \left[K_G \ast \left( -D_1 \Delone + D_2 \Deltwo \right) \right] + 3 \left( \sqrt{1 - \frac23 \delta} - 1 \right) \\
  & \qquad  {} - \left[K_G \ast \left( 3 \left( \sqrt{1 - \frac23 \delta} - 1 \right) \right) \right] \,,
\end{split}
\end{equation}
where $K_G(\bq)$ is a Gaussian kernel at some characteristic scale $R_G$.
The convolution (denoted by $\ast$) with the 2LPT part preserves mainly the scales above $R_G$ and the other with ``$1-K_G$'' preserves scales below that for the spherical collapse part.
All terms are defined on the Lagrangian $\bq$ grid (we left out the function arguments for brevity).

The derivative we need can be identified as
\begin{equation}
\label{eqn:dphi_ddelta}
\begin{split}
 \pdif{\divdisp(\bq)}{\delta(\bq_i)} &=  \left[ K_G \ast \left( -K^K_i + D_2 \pdif{\Deltwo}{\delta(\bq_i)} \right) \right](\bq) \\
 &\quad+ {\left(1 - \frac23 \delta(\bq)\right)}^{-\frac12} K^K_i(\bq) \\
 &\quad- \left[ K_G \ast \left( {\left(1 - \frac23 \delta(\bq)\right)}^{-\frac12} K^K_i \right) \right](\bq) \,.
\end{split}
\end{equation}
Here we use an unusual notation for Kronecker delta:
\begin{equation}
  K^K_i(\bq) \equiv \krodel_{\bq,\bq_i} \,.
\end{equation}
This helps us to neatly write the Kronecker delta in convolution notation.
The 2LPT part of this equation has been worked out in Section~\ref{sec:2lpt}, so we recycle those results below.
The convolutions with the Gaussian are also most efficiently done in (discrete) Fourier space.
Note that the conversion between continuous and discrete FTs causes the following relation to contain a small error (a convolution of a Gaussian with a general function $f(\bq)$):
\begin{equation}
  \label{eqn:gaussian_convolution}
  \begin{split}
    [K_G \ast f](\bq)
    &=  \frac{1}{N} \sum_j e^{i \bk_j \cdot \bq} \hat{K}_G(\bk_j) \hat{f}(\bk_j) \\
    &=  \frac{1}{N} \sum_j e^{i \bk_j \cdot \bq} e^{-\frac{k_j^2 R_G^2}{2}} \hat{f}(\bk_j) \,.
  \end{split}
\end{equation}

In what follows, we will, term by term, reduce equation~\ref{eqn:likeli_force_short}, combined with the 2LPT+SC prescription of this section, to a series of Fourier transforms of several intermediate convenience functions.
The numbered terms that form
\begin{equation}
  F^{\likeli,2LPT+SC}_i = \circled{1} + \circled{2} + \circled{3} + \circled{4} \,,
\end{equation}
are
\begin{equation}
\begin{split}
    \circled{1} &= \sum_m h(\bq_m) \left[ K_G \ast \left( -\krodel_{\bq_m,\bq_i} \right) \right]  \notag \\
    \circled{2} &= \sum_m h(\bq_m) \left[ K_G \ast \left( D_2 \pdif{\Delta^{(2)}(\bq_m)}{\delta(\bq_i)} \right) \right]  \equiv  \sum_m h(\bq_m)  \left[ \circled{$\ast$} \right] \notag \\
    \circled{3} &= \sum_m h(\bq_m) {\left(1 - \frac23 \delta(\bq_m)\right)}^{-\frac12} \krodel_{\bq_m,\bq_i} \notag \\
    \circled{4} &= - \sum_m h(\bq_m) \left[ K_G \ast \left( {\left(1 - \frac23 \delta(\bq_m)\right)}^{-\frac12} \krodel_{\bq_m,\bq_i} \right) \right] \,.
\end{split}
\end{equation}

\subsubsection{Term 1}
This term contains, on first hand, a somewhat unintuitive convolution with a Kronecker delta function. After some manipulation, however, this term will quickly conform to something more in line with our comfortable view of the world.

Using the DFT of the Kronecker delta:
\begin{equation}
 \dft\left[ \krodel_{\bx,\bx'} \right] = \sum_{\bx} e^{-i\bk\cdot\bx} \krodel_{\bx,\bx'} = e^{-i\bk\cdot\bx'} \,,
 \label{eqn:DFT_kronecker_delta}
\end{equation}
we reshape term \circled{1} into a simple convolution:
\begin{equation}
\label{eqn:f_likeli_2lpt_sc_term_1}
\begin{split}
  \circled{1}
        &=       \sum_m h(\bq_m) \left[ K_G \ast \left( -K^K_i \right) \right](\bq_m) \\
        &=       -\frac{1}{N} \sum_m h(\bq_m) \sum_j e^{i\bk_j\cdot\bq_m} \hat{K}_G(\bk_j) e^{-i\bk_j\cdot\bq_i} \\
        &=       -\frac{1}{N} \sum_j \sum_m h(\bq_m) e^{i\bk_j\cdot\bq_m} \hat{K}_G(\bk_j) e^{-i\bk_j\cdot\bq_i} \\
        &\equiv  -\frac{1}{N} \sum_j \hat{c}^*(\bk_j) \hat{K}_G(\bk_j) e^{-i\bk_j\cdot\bq_i} \\
        &=       -{\left[ K_G \ast c \right]}^* (\bq_i) \\
        &=       -\left[K_G \ast c\right](\bq_i)   \,,
\end{split}
\end{equation}
where the convenience function $c(\bq)$ is defined as
\begin{equation}
\label{eqn:convenience_c}
    c(\bq) \equiv h^*(\bq) = h(\bq) \,,
\end{equation}
and the conjugation has no effect because $h(\bq)$ is real (and $K_G\ast c$ is as well).

\subsubsection{Term 2}
This term itself consists of several terms, coming from the terms in the sum of equation~\ref{eqn:d_deltwo_d_delta_q}.
In particular, the convolution part $\left[ \circled{$\ast$} \right]$, in full, reads
\begin{equation}
\begin{split}
       \left[ \circled{$\ast$} \right]
        &=       \frac{D_2}{N} \sum_j e^{i\bk_j\cdot\bq_m} \hat{K}_G(\bk_j) \sum_k e^{-i\bk_j\cdot\bq_k} \\
        &\qquad      \sum_{\ua>\ub} \left( \phi^{(1)}_{,\ua\ua}(\bq_k) \pdif{\phi^{(1)}_{,\ub\ub}(\bq_k)}{\delta(\bq_i)} + \phi^{(1)}_{,\ub\ub}(\bq_k) \pdif{\phi^{(1)}_{,\ua\ua}(\bq_k)}{\delta(\bq_i)} \right. \\
        &\qquad      \left. {} \qquad - 2 \phi^{(1)}_{,\ua\ub}(\bq_k) \pdif{\phi^{(1)}_{,\ua\ub}(\bq_k)}{\delta(\bq_i)} \right) \,.
\end{split}
\end{equation}
Like for equation~\ref{eqn:four_index_term_2lpt}, we can write the three different parts (as defined by the three summands between parentheses) of this equation as four-index quantities:
\begin{equation}
\begin{split}
  \frac{D_2}{D_1} {\eta}^{\ua\ub}_{\uc\ud}(\bq_i)
  &\equiv        \sum_m h(\bq_m) \frac{D_2}{N} \sum_j e^{i\bk_j\cdot\bq_m} \hat{K}_G(\bk_j) \\
  &\qquad               \sum_k e^{-i\bk_j\cdot\bq_k} \phi^{(1)}_{,\uc\ud}(\bq_k) \pdif{\phi^{(1)}_{,\ua\ub}(\bq_k)}{\delta(\bq_i)} \\
  &=             \sum_m h(\bq_m) \frac{D_2}{D_1} \frac{1}{N^2} \sum_j e^{i\bk_j\cdot\bq_m} \hat{K}_G(\bk_j) \\
  &\qquad               \sum_k e^{-i\bk_j\cdot\bq_k} \phi^{(1)}_{,\uc\ud}(\bq_k) \sum_l e^{i \bk_l \cdot \bq_k} \frac{k_{l\ua} k_{l\ub}}{k_l^2} e^{-i \bk_l \cdot \bq_i} \\
  &\equiv        \frac{D_2}{D_1} \frac{1}{N^2} \sum_j \hat{c}^*(\bk_j) \hat{K}_G(\bk_j) \sum_k e^{-i\bk_j\cdot\bq_k} \phi^{(1)}_{,\uc\ud}(\bq_k) \\
  &\qquad               \sum_l e^{i \bk_l \cdot \bq_k} \frac{k_{l\ua} k_{l\ub}}{k_l^2} e^{-i \bk_l \cdot \bq_i} \\
  &\equiv        \frac{D_2}{D_1} \frac{1}{N} \sum_k b^*(\bq_k) \phi^{(1)}_{,\uc\ud}(\bq_k) \\
  &\qquad               \sum_l e^{i \bk_l \cdot \bq_k} \frac{k_{l\ua} k_{l\ub}}{k_l^2} e^{-i \bk_l \cdot \bq_i} \\
  &\equiv        \frac{D_2}{D_1} \frac{1}{N} \sum_l \hat{a}^*_{\uc\ud}(\bk_l) \frac{k_{l\ua} k_{l\ub}}{k_l^2} e^{-i \bk_l \cdot \bq_i} \\
  &\equiv        \frac{D_2}{D_1} {\eta^*}^{\ua\ub}_{\uc\ud}(\bq_i) \\
  &=             \frac{D_2}{D_1} {\eta}^{\ua\ub}_{\uc\ud}(\bq_i) \,,
\end{split}
\end{equation}
where the conjugation trivially drops out and we further define convenience functions $b(\bq)$, $a_{\uc\ud}(\bq)$ and $\eta_{\uc\ud}^{\ua\ub}(\bk)$ as
\begin{align}
  \hat{b}(\bk) &\equiv {\left( \hat{c}^*(\bk) \hat{K}_G(\bk) \right)}^* = \hat{h}(\bk) \hat{K}_G(\bk) \,, \notag \\
  &\Rightarrow b(\bq) = \left[ K_G * h \right](\bq) \label{eqn:convenience_b} \\
  a_{\uc\ud}(\bq)  &\equiv {\left( b^*(\bq) \phi^{(1)}_{,\uc\ud}(\bq) \right)}^* = b(\bq) \phi^{(1)}_{,\uc\ud}(\bq) \,, \\
  \hat{\eta}^{\ua\ub}_{\uc\ud}(\bk) &\equiv {\left( \frac{k_{\ua} k_{\ub}}{k^2} \hat{a}^*_{\uc\ud}(\bk) \right)}^* =  \frac{k_m k_n}{k^2} \hat{a}_{\uc\ud}(\bk) \label{eqn:convenience_eta} \notag \\
  &\Rightarrow \eta^{\ua\ub}_{\uc\ud}(\bq) = \partial_{\ua} \partial_{\ub} \nabla^{-2} \left( \left[ K_G * h \right](\bq) \phi^{(1)}_{,\uc\ud}(\bq) \right)\,.
\end{align}
Here we can again easily get rid of the conjugations.
Apart from obvious term-wise double conjugations, the first term contains the Gaussian kernel, which is real, $\phi^{(1)}_{,\uc\ud}$ is also real, as it is a physical term, and the $\bk$ terms are real scale vectors.
Finally, we replace $\eta$ by $\chi$:
\begin{equation}
  \eta^{\ua\ub}_{\uc\ud}(\bq) = \chi^{\ua\ub}_{\uc\ud}(b(\bq); \bq) \,.
  \label{eqn:convenience_eta_as_chi}
\end{equation}

Putting it all together, term \circled{2}, in full, reads
\begin{equation}
     \circled{2} = \frac{D_2}{D_1} \sum_{\ua>\ub} \left( {\chi}^{\ua\ua}_{\ub\ub} + {\chi}^{\ub\ub}_{\ua\ua} - 2 {\chi}^{\ua\ub}_{\ua\ub} \right) (b(\bq_i); \bq_i) \,,
\end{equation}
which is almost the same as the previous 2LPT part in equation~\ref{eqn:loglikeli_2lpt}, but with a Gaussian smoothed version of $h$, namely $b$.

\subsubsection{Term 3}
The simplest term, a matter of a sum over an expression with a Kronecker delta that changes the $\bq_m$'s to $\bq_i$'s:
\begin{equation}
  \circled{3} = \frac{h(\bq_i)}{\sqrt{1 - \frac23 \delta(\bq_i)}} \,. 
  \label{eqn:loglikeli_2lpt+sc_term3}
\end{equation}

\subsubsection{Term 4}

\begin{equation}
\begin{split}
    \circled{4}
    &=      -\sum_m h(\bq_m) \left[ K_G \ast \left( {\left(1 - \frac23 \delta\right)}^{-\frac12} K^K_i \right) \right](\bq_m) \\
    &\equiv -\frac{1}{N} \sum_m h(\bq_m) \sum_j e^{i \bk_j\cdot\bq_m} \hat{K}_G(\bk_j) \\
    &\qquad     \sum_k e^{-i\bk_j\cdot\bq_k} \krodel_{\bq_k,\bq_i} \Delta(\bq_k) \\
    &=      -\frac{1}{N} \sum_j \hat{c}^*(\bk_j) \hat{K}_G(\bk_j) \sum_k e^{-i\bk_j\cdot\bq_k} \krodel_{\bq_k,\bq_i} \Delta(\bq_k) \\
    &=      -\sum_k {\left[ K_G*c \right]}^*(\bq_k) \Delta(\bq_k) \krodel_{\bq_k,\bq_i} \\
    &=      -\Delta(\bq_i) {\left[ K_G*c \right]}^*(\bq_i) \\
    &=      -\Delta(\bq_i) \left[ K_G*c \right](\bq_i) \,,
\end{split}
\end{equation}
where we used the convenience function $\Delta(\bq)$
\begin{equation}
    \Delta(\bq) \equiv {\left(1- \frac23 \delta(\bq)\right)}^{-1/2} \,.
    \label{eqn:convenience_delta}
\end{equation}

It may, on first hand, seem counter-intuitive that while we started with a convolution of the Gaussian with one function ($\Delta(\bq) \krodel_{\bq,\bq_i}$), we end up convolving another function ($c(\bq)$) and do not convolve with the original one at all.
This is actually easily understandable when you consider that what we were calculating is a double convolution.
In fact, we could derive the term in another way, without going to Fourier space, but instead only using the definition of a discrete convolution:
\begin{equation}
\label{eqn:loglikeli_2lpt+sc_term4_alt_deriv}
\begin{split}
  \circled{4}
  &=      -\sum_m h(\bq_m) \left[ K_G \ast \left( {\left(1 - \frac23 \delta(\bq_m)\right)}^{-\frac12} \krodel_{\bq_m,\bq_i} \right) \right] \\
  &=      -\sum_m h(\bq_m) \sum_j K_G(\bq_m-\bq_j) \Delta(\bq_j) \krodel_{\bq_j,\bq_i} \\
  &=      -\sum_j \left[ K_G \ast h \right](\bq_j) \Delta(\bq_j) \krodel_{\bq_j,\bq_i} \\
  &=      -\left[ K_G \ast h \right](\bq_i) \Delta(\bq_i) \,,
\end{split}
\end{equation}
where in the third step we used that $K_G(\bq) = K_G(-\bq)$.
Term \circled{4} is a convolution of $h(\bq)$ with the convolution of $K_G$ with the other term, and convolutions are associative.
However, because that latter term contains a Kronecker delta, things get a little bit funky, as they always do when delta functions turn up.
With this derivation, we also immediately see that the $c(\bq)$ function is superfluous, as we already concluded when we first encountered it in term \circled{1} (equation~\ref{eqn:f_likeli_2lpt_sc_term_1}).
Actually, term \circled{1} could also easily have been derived with this double convolution method, circumventing the complication of the Fourier transform of the Kronecker delta function.

\subsubsection{All terms together}
Putting it all together, we obtain for \emph{minus} the derivative of the likelihood in a 2LPT+SC model:
\begin{equation}
\label{eqn:loglikeli_2lpt+sc}
\begin{split}
  F^{\likeli,2LPT+SC}_i &= -\left[ K_G \ast h \right] (\bq_i) \\
  &\quad+ \frac{D_2}{D_1} \sum_{\ua>\ub} \left( {\chi}^{\ua\ua}_{\ub\ub} + {\chi}^{\ub\ub}_{\ua\ua} - 2{\chi}^{\ua\ub}_{\ua\ub} \right) (b(\bq_i); \bq_i) \\
  &\quad + \left( \frac{h(\bq_i) - \left[ K_G*h \right](\bq_i)}{\sqrt{1-\frac23 \delta(\bq_i)}} \right) \,,
\end{split}
\end{equation}
with $\chi^{\ua\ub}_{\uc\ud}(f(\bq); \bq)$ in equation~\ref{eqn:convenience_chi} and $b(\bq)$ in equation~\ref{eqn:convenience_b}. There are no conjugations left and the remaining $c$ functions have been be replaced by $h$.

\emph{Note} that in the last term, the square root has an upper bound: $\delta(\bq_i) \leq \frac32$.
If it is larger, the square root will become imaginary.

In fact, this is supposed to happen; the lower bound for the stretching parameter $\divdisp^\mathrm{SC}$ is $-3$ \citep{neyrinck13}.
This means that the given formula is only valid up to the given upper bound.
After that, the entire term (i.e.\ the sum of terms \circled{3} and \circled{4}) becomes zero --- the derivative of $-3$ --- i.e.:

\begin{equation}
\begin{split}
  F^{\likeli,2LPT+SC}_i &= -\left[ K_G \ast h \right] (\bq_i) \\
  &\quad + \frac{D_2}{D_1} \sum_{\ua>\ub} \left( {\chi}^{\ua\ua}_{\ub\ub} + {\chi}^{\ub\ub}_{\ua\ua} - 2{\chi}^{\ua\ub}_{\ua\ub} \right) (b(\bq_i); \bq_i) \\
  &\quad +  
  \begin{cases}
    \left( \frac{h(\bq_i) - \left[ K_G*h \right](\bq_i)}{\sqrt{1-\frac23 \delta(\bq_i)}} \right) \,, & \text{if}\ \delta(\bq_i) \leq \frac32 \,, \\
    0, & \text{otherwise.}
  \end{cases}
\end{split}
\end{equation}